\documentclass[%
 reprint,
 amsmath,amssymb,
 aps,
]{revtex4-2}

\usepackage{amssymb}
\usepackage{graphicx}
\usepackage{algorithm}
\usepackage{algorithmicx}
\usepackage[noend]{algpseudocode}
\usepackage{graphicx,subcaption}
\usepackage{amsmath}
\usepackage{url}
\usepackage{dcolumn}
\usepackage{bm}
\usepackage{float}
\usepackage{color,xcolor,soul}

\begin{document}

\preprint{APS/123-QED}

\title{Time-inversion of spatiotemporal beam dynamics using \\ uncertainty-aware latent evolution reversal} 
\thanks{Contact author: mrautela@lanl.gov}%

\author{Mahindra Rautela*}
\author{Alan Williams}%
\author{Alexander Scheinker}%
\affiliation{%
 Applied Electrodynamics Group (AOT-AE), Los Alamos National Laboratory, USA
}%

\date{\today}

\begin{abstract}
Charged particle dynamics under the influence of electromagnetic fields is a challenging spatiotemporal problem. Many high performance physics-based simulators for predicting behavior in a charged particle beam are computationally expensive, limiting their utility for solving inverse problems online. The problem of estimating upstream six-dimensional phase space given downstream measurements of charged particles in an accelerator is an inverse problem of growing importance. This paper introduces a reverse Latent Evolution Model (RLE) designed for temporal inversion of forward beam dynamics. In this two-step self-supervised deep learning framework, we utilize a Conditional Variational Autoencoder (CVAE) to project 6D phase space projections of a charged particle beam into a lower-dimensional latent distribution. Subsequently, we autoregressively learn the inverse temporal dynamics in the latent space using a Long Short-Term Memory (LSTM) network. The coupled CVAE-LSTM framework can predict 6D phase space projections across all upstream accelerating sections based on single or multiple downstream phase space measurements as inputs. The proposed model also captures the aleatoric uncertainty of the high-dimensional input data within the latent space. This uncertainty, which reflects potential uncertain measurements at a given module, is propagated through the LSTM to estimate uncertainty bounds for all upstream predictions, demonstrating the robustness of the LSTM to random perturbations in the input.
\end{abstract}

\maketitle

\section{Introduction}
Particle accelerators are complex high-dimensional systems with hundreds to thousands of components which include radio-frequency (RF) resonant cavities that accelerator the charged particle beams and magnets that focus and steer the beams. The control and optimization of accelerators is not only challenging due to their complexity, but also because both the accelerator components and the initial conditions of their beams drift with time. Furthermore, for any given accelerator condition, the charged particle beam dynamics evolve in a six-dimensional position and momentum phase space $(x,y,z,p_x,p_y,p_z)$ and experience complex collective effects such as space charge forces in which the electromagnetic fields of the beam's own particles perturb its shape and energy in a highly nonlinear way. To numerically estimate the spatiotemporal evolution of a beam in an accelerator is computationally demanding because of the large length scales involved. For example, while the individual particles in a charged particle bunch which are experiencing collective space charge effects can be bounded by a sphere with a radius of millimeters the bunch also traverses a kilometer long machine as it is accelerated by the external electromagnetic fields of RF cavities and magnets.


In recent years, machine learning (ML) has shown promising abilities in solving complex problems in physical sciences \cite{vinuesa2022enhancing,huerta2021accelerated,scheinker2023adaptive_PRE,boehnlein2022colloquium,bormanis2024solving,morison2024nonlinear}. However, most of the research has focused on either spatial or temporal dynamics with limited emphasis on spatiotemporal dynamics. Some of the techniques for solving spatiotemporal dynamical problems are Graph Neural Networks (GNNs) \cite{kipf2016semi}, Convolutional Long Short-Term Memory (ConvLSTM) \cite{shi2015convolutional},Deep Convolutional Generative Adversarial Networks (DCGAN) \cite{cheng2020data}, three-dimensional Convolutional Neural Networks (3DCNN) \cite{wandel2021teaching}. 

Latent evolution models have recently gained traction for solving challenging spatiotemporal dynamics problems. In these machine learning models, the learning process is decomposed into two subproblems for training, which are later integrated to make predictions \cite{wiewel2019latent}. First, a dimensionality reducer captures spatial correlations by projecting high-dimensional images into a lower-dimensional latent space. Then, another model learns the temporal correlations within this latent space. Latent evolution models become computationally efficient by working in lower dimensional space. The encoder-decoder framework is effective at extracting representative features, which improves the learning of temporal correlations using recurrent neural networks. In \cite{scheinker2021adaptive_JOI}, adaptive feedback acting on the latent embedding of a convolutional autoencoder (AE) is proposed for time-varying beam dynamics in particle accelerators. In \cite{montes2021accelerating}, principal component analysis and LSTM are employed for learning phase-field-based microstructure evolution. In  \cite{wiewel2019latent, nakamura2021convolutional, maulik2021reduced, vlachas2022multiscale}, AE and LSTM are utilized for addressing fluid flow problems \cite{wiewel2019latent, nakamura2021convolutional, maulik2021reduced, vlachas2022multiscale}. In \cite{solera2024beta}, $\beta$-variational autoencoders (VAE) and transformers are used for reduced-order modelling of fluid flow. In \cite{rautela2024conditional,rautela24towards}, conditional VAE with autoregressive LSTMs are proposed for charged particle beam dynamics in accelerators.

Deep learning methods are used to solve forward and inverse problems in beam physics and diagnostics. In \cite{scheinker2018demonstration}, a model-independent feedback coupled with a neural network is demonstrated for automatic control of longitudinal phase space (LPS) of electron beams. In \cite{emma2018machine}, a ML-based virtual diagnostic was developed to predict LPS distribution. In \cite{ivanov2020physics}, a physics-based ML approach is discussed where polynomial neural networks with symplectic regularization is proposed to represent Taylor maps of particle dynamics. Deep Lie map networks were developed to identify magnetic field errors based on beam position monitor measurements in synchrotrons \cite{caliari2023identification}. 
In \cite{zhu2021high}, an encoder-decoder based virtual diagnostics were developed  to model the longitudinal phase-space-diagnostic beamline at the photoinector of the European XFEL. In \cite{mayet2022predicting}, a fully connected neural network was developed to evaluate measured phase advance scan data of space charge-dominated beams. In \cite{cropp2023virtual}, ML methods have been combined with multilinear regression to create virtual time of arrival and beam energy diagnostics at HiRES. In \cite{tran2022predicting}, a convolutional autoencoder compresses the data into lower dimension and used to relate the phase-space information of 4D phase space distribution to the beam loss at four different locations.

Deep learning is also implemented to solve inverse problems of calibration, fault analysis, tuning and optimization. Bayesian optimization (BO) is popularly used for accelerator tuning \cite{duris2020bayesian,jalas2023tuning,kirschner2022tuning,breckwoldt2023machine,ji2024multi}. Deep reinforcement learning is another widely used method for control and tuning of different accelerator facilities \cite{kain2020sample,pang2020autonomous,meier2022optimizing}. Inverse problem of fault analysis in accelerators using machine learning is a critical problem of growing interest. In \cite{obermair2022explainable}, explainable ML-based models using SHapley Additive exPlanation (SHAP) values were developed for predicting breakdowns in high-gradient cavity. In \cite{rajput2024robust}, supervised machine learning based conditional siamese neural network
(CSNN) and conditional variational auto encoder (CVAE) models to predict errant beam pulses under different system configurations. In \cite{tennant2020superconducting}, manual feature engineering based ML for classifying superconducting radio-frequency (SRF) cavity faults. In Ref. \cite{li2018genetic,wan2019improvement,bellotti2021fast}, genetic algorithm based multi-objective optimization is demonstrated for accelerator problems.  In \cite{li2023time}, time series forecasting methods and their applications to particle accelerators are discussed.  

One of the key inverse problems in accelerators is estimating the upstream phase space of charged particles from downstream measurements. This estimation is crucial for characterizing the beam in earlier sections of the accelerator, which aids in understanding beam dynamics, minimizing beam losses, and tuning the accelerator for optimal performance. Towards this direction, an adaptive ML approach is proposed to map output beam measurements to input beam distributions for rapidly changing systems \cite{scheinker2021adaptive_SciRep}. In \cite{wolski2022transverse}, a method was developed for 4D transverse phase space tomography. In \cite{roussel2023phase}, neural networks are integrated with differentiable particle tracking to learn mappings from measurements to initial phase space distributions.

In the accelerator physics community, numerous research efforts are focused on quantifying two distinct types of uncertainties: aleatoric (also referred to as parametric or stochastic) and epistemic (also known as model-form, systematic, or approximation) \cite{smith2013uncertainty,zhang2019quantifying}. Aleatoric uncertainty emerges from inherent randomness or variability in the system whereas epistemic uncertainty arises from insufficient knowledge or incomplete information about the system being modeled. In simpler words, probability distributions can be placed over outputs or model parameters to quantify epistemic and aleatoric uncertainties, respectively \cite{acharya2023learning}. In Ref. \cite{adelmann2019nonintrusive}, the author has investigated aleatoric uncertainties associated with charged particle beam parameters in a cyclotron particle accelerator using a non-intrusive polynomial-chaos expansion (PCE) based approach. In Ref. \cite{convery2021uncertainty}, two different techniques i.e., deep ensembles (bagging) and deep quantile regression (DQR) are utilized to quantify epistemic uncertainty in predicting current profile or longitudinal phase space images of the electron beam. If Ref. \cite{mishra2021uncertainty}, Bayesian neural networks (BNN) are employed for epistemic uncertainty and compared with bootstrapped ensembles for three different accelerator problems. Various other techniques like Siamese neural network (SNN), Monte Carlo dropout (MCD), DQR, and deep Gaussian process approximation (DGPA) are utilized for epistemic uncertainty quantification for anomaly detection problems \cite{blokland2022uncertainty,schram2023uncertainty}. 

More recently, adaptive latent space tuning of autoencoder network is presented to compensate for aleatoric uncertainty for different accelerator problems \cite{scheinker2021adaptive_JOI, scheinker2021adaptive_SciRep, scheinker2023adaptive_PRE}. Authors have also studied aleatoric uncertainty while predicting RMS beam dynamics where a variance-based uncertainty quantification approach is used while minimize the negative log-likelihood function of a Gaussian distribution \cite{garcia2024machine}. 

\subsection{Summary of main results}
In this paper, we propose a reverse latent evolution model (RLE) designed to predict 2D projections of the upstream 6D phase space from downstream phase space measurements. This two-step self-supervised framework operates as follows: first, a conditional VAE learns the low-dimensional latent distribution of the 15 unique projections of the charged particle beam's phase space. Next, a Long Short-Term Memory (LSTM)-based recurrent neural network is trained autoregressively to capture the reverse the forward temporal dynamics in the latent space. The coupled network is used to predict phase space projections in upstream accelerating sections based on limited phase space measurements from downstream sections. Our model effectively captures aleatoric uncertainty from the image space and translates it into the latent space. It also demonstrates robustness against in-distribution variations in the input data. While the proposed technique can address general spatiotemporal problems, we have implemented it specifically for a charged particle beam in the linear accelerator at the Los Alamos Neutron Science Center (LANSCE) at Los Alamos National Laboratory (LANL). 
 
A latent space-based approach, like CVAE, can be leveraged to quantify aleatoric uncertainty because it minimizes the negative log-likelihood of the dataset ($\log{p_{\theta}(X)}$), assuming a Gaussian distribution in the latent space. Similar to VAEs, other explicit generative models, such as the Neural Autoregressive Distribution Estimator (NADE) \cite{uria2016neural}, Masked Autoencoder for Distribution Estimation (MADE) \cite{papamakarios2017masked}, normalizing flows \cite{papamakarios2021normalizing}, and diffusion models \cite{ho2020denoising}, have similar capabilities for capturing uncertainty. However, while this approach effectively captures aleatoric uncertainty in the latent space, it does not account for epistemic uncertainty, which arises from limited training data and modeling procedures \cite{acharya2023learning}.

Our current research is inspired by previous work \cite{rautela2024conditional}, where a latent evolution model (LEM) is used to solve the forward problem of forecasting the spatiotemporal dynamics of charged particles. However, the inverse problem of estimating upstream states given downstream measurement is more complex problem. In the literature \cite{scheinker2021adaptive_SciRep,wolski2022transverse,roussel2023phase}, separate models are often trained to address this inverse problem. In a LEM, spatial and temporal dynamics are learned independently, enhancing its effectiveness for inverse problem-solving. Contrary to \cite{rautela2024conditional}, our objective is to invert the spatiotemporal dynamics of charged particles. We achieve this by training the LSTM in the latent space in a reverse autoregressive setting while keeping the spatial learner (CVAE) intact. The main contributions of this research include (a) solving the inverse problem of estimating upstream states from downstream measurements using a RLEM, and (b) studying uncertainty by (i) capturing aleatoric uncertainty in the input space within the latent space, and (ii) propagating aleatoric uncertainty in the latent space for enhanced robustness.

\section{Methods}
\subsection{Charged particles beam dynamics}\label{ssec:beamdynamics}

Accurate measurement of a beam's six-dimensional (6D) phase space is one of the most significant challenges in accelerator physics. While such measurements are essential for precise control and optimization of beam parameters, they remain exceptionally difficult to perform. The first iterative method for reconstructing a beam’s 6D phase space, developed six years ago, required 32 hours and over 5.7 million individual beam measurements \cite{cathey2018first}. Although advancements have reduced this time to approximately 18 hours, the approach is still impractical for real-time applications. Fast, single-shot measurement techniques provide limited 2D projections, such as (x,y) and (z,E), but these are typically destructive and unsuitable for high-intensity proton beams due to potential damage to diagnostic devices. Further challenges include limited resolution for short beam pulses, restricted operation times, and temporal shifts in the beam distribution caused by variations in accelerator components or initial conditions. These constraints underscore the need for robust theoretical and numerical frameworks to analyze beam dynamics and predict phase space evolution.

The beam dynamics is governed by the interaction between charged particles and electromagnetic fields. At the core of this interaction lies the Vlasov-Maxwell equations, a set of coupled equations describing the self-consistent evolution of charged particle systems in phase space \cite{wiedemann1994particle}. The state of the system in phase space is represented by both the position $(x,y,z)$ and momentum $(p_x,p_y,p_z)$ of the particles.

The relativistic \textit{Vlasov equation} governs the evolution of the particle distribution function \(f(\mathbf{x}, \mathbf{p}, t)\), encapsulating the density of particles in phase space:

\[
\frac{\partial f}{\partial t} + \mathbf{v} \cdot \nabla_{\mathbf{x}} f + \frac{\mathbf{F}}{m} \cdot \nabla_{\mathbf{p}} f = 0,
\]

where the force \(\mathbf{F} = q(\mathbf{E} + \mathbf{v} \times \mathbf{B})\) is determined by the electromagnetic fields \(\mathbf{E}\) and \(\mathbf{B}\). Simultaneously, \textit{Maxwell's equations} govern the evolution of these fields:

\[
\nabla \cdot \mathbf{E} = \frac{\rho}{\epsilon_0}, \quad \nabla \cdot \mathbf{B} = 0,
\]
\[
\nabla \times \mathbf{E} = -\frac{\partial \mathbf{B}}{\partial t}, \quad \nabla \times \mathbf{B} = \mu_0 \mathbf{J} + \mu_0 \epsilon_0 \frac{\partial \mathbf{E}}{\partial t}.
\]

Here, the charge density \(\rho = q \int f \, d^3\mathbf{p}\) and current density \(\mathbf{J} = q \int \mathbf{v} f \, d^3\mathbf{p}\) are derived from the particle distribution function. Together, the Vlasov and Maxwell equations form a self-consistent framework, where the particle distribution evolves under the influence of electromagnetic fields, and the fields themselves are shaped by the particles' charge and current densities. 

Electromagnetic fields in accelerators are beam-based sources, which are self-generated fields acting on particles via space charge forces and distorting the 6D phase space, and external sources, such as resonant accelerating structures and magnets, which accelerate, guide, and focus the beam. In high-intensity accelerators, such as the Los Alamos Neutron Science Center (LANSCE), the strong self-fields of the beam result in pronounced nonlinear effects. Nonlinear space charge forces further distort and filament the beam’s phase space, creating highly complex and nonuniform distributions. These intricate dynamics pose significant analytical and computational challenges for accurate modeling and simulation.

\subsubsection{Dataset Generation and Description}
High Performance Simulator (HPSim) is an open-sourced code developed at LANL which enables rapid, online simulations for multiple-particle beam dynamics \cite{pang2014gpu}. It solves the Vlasov-Maxwell equations to compute the impact of external accelerating and focusing forces on the charged particle beam, including space charge forces within the beam. The software is designed to replicate the accelerator and, therefore, provides a realistic representation of the true beam used at the Los Alamos Neutron Science Center (LANSCE). 

We simulate beam dynamics of 1M macroparticles in the LANSCE linear accelerator at Los Alamos National Laboratory. The beam's behavior in each module is governed by two radio-frequency (RF) parameters: amplitude and phase set-points. The accelerator consists of 48 modules through which the beam propagates, including 4 modules of a 201.25 MHz drift tube linac (DTL) and 44 modules of an 805 MHz coupled cavity linac (CCL). 

The dataset was generated using HPSim by varying the RF set points (8 parameters in total) which were randomly sampled from a uniform distribution, while the RF set points for the remaining 44 modules (88 parameters) were kept constant. The resulting 96-dimensional RF set point vector was used as input to HPSim to simulate beam dynamics across the accelerator's 48 modules. For each simulation, the raw 6D phase space output of HPSim was projected onto 15 unique 2D planes, including \(x-p_x\), \(x-y\), \(x-p_y\), \(x-z\), \(x-p_z\), \(y-p_x\), \(y-p_y\), \(y-z\), \(y-p_z\), \(z-p_x\), \(z-p_y\), \(z-p_z\), \(E-\phi\), \(p_x-p_y\), \(p_x-p_z\), and \(p_z-p_y\), for every module. 

In total, 1400 simulations were performed to generate the training dataset, and an additional 700 simulations were conducted for testing. The data output from each single simulation has a size of [48, 15, 256, 256] where at each of the 48 modules there are 15 projections, each at a $256\times 256$ pixel resolution. The module indices (m) and projections (X) were normalized to the range [0, 1]. The dataset has been made publicly available as an open-source resource on Zenodo \cite{rautela_2024_10819001}.

\begin{figure*}[htbp]
\centering
\includegraphics[trim={0 0 0 0},clip, width=0.95\textwidth]{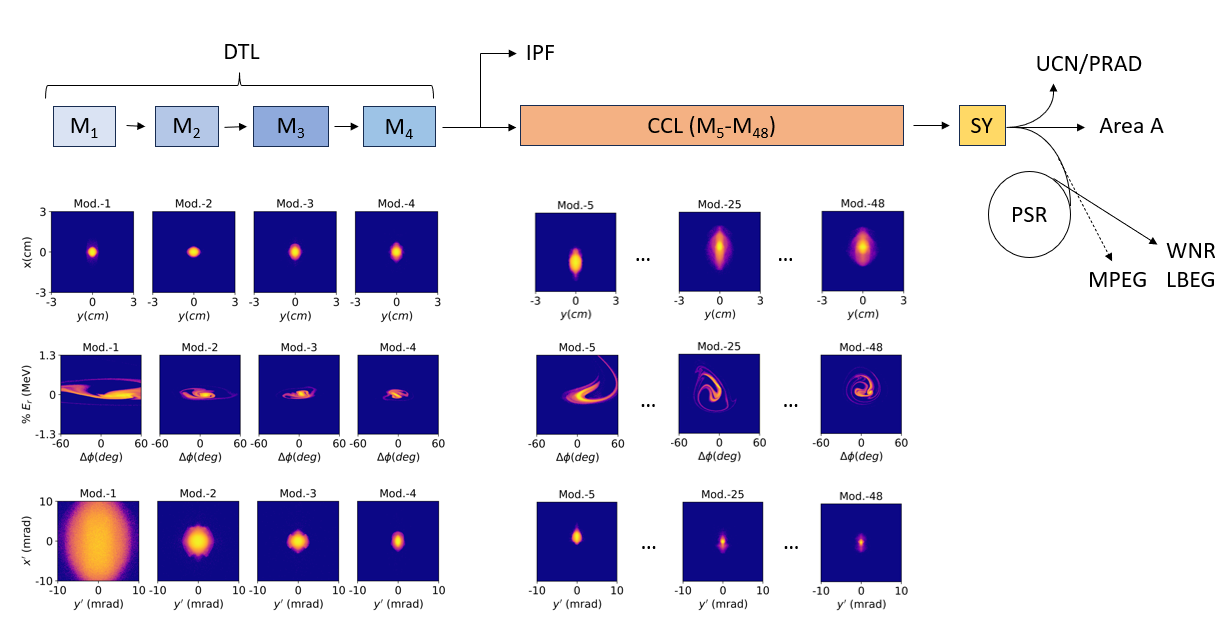}
\caption{Three 2D projections ($x-y$,$E-\phi$,$x'-y'$) out of 15 projections of 6d phase space of charged particle beam in the LANSCE linear accelerator. Accelerating modules - 1 to 4 are 201 MHz drift tube linac (DTL) and 5 to 48 are 805 MHz coupled cavity linac (CCL). The beam serves various scientific areas like isotope production facility (IPF), ultra-cold neutrons (UCN), proton radiography (PRAD), weapons neutron research (WNR), proton storage ring (PSR).}
\label{fig:dataset}
\end{figure*}

A high level overview of LANSCE and RF module locations, as well as a sample of the dataset are presented in Fig.~\ref{fig:dataset}. Three of the 15 projections ($x-y$, $E-\phi$, $x'-y'$) are shown at various locations along LANSCE. The figure highlights the complex, nonlinear, and multi-scale spatiotemporal evolution observed in the projections of the 6D beam phase space.

\subsection{Reverse latent evolution model (RLEM)} \label{ssec:rlem}

The forward discretized spatiotemporal beam dynamics (described above) can be written as $X_{t} = H(X_1, X_2,\dots,X_t,\dots, X_{T-1})$, where $H$ is a unknown nonlinear function that can be learned as $P(X_T|X_1,X_2,\dots,X_t,\dots,X_{T-1})$ with a latent evolution model as presented in \cite{rautela2024conditional}.

Reversing the temporal dynamics to predict the current state $X_t$ given future states $X_{t+1}, X_{t+2}, \dots, X_{T}$ requires inverting the function $H$, which becomes a challenging inverse problem. Similar to \cite{rautela2024conditional}, the inverse problem can be learned by calculating the joint probability distribution $P(X_1, X_2, ... ,X_T)$ by factorising it using the chain rule of probability. The chain rule for probability is temporal direction-invariant, meaning it permits joint probability decomposition in either forward or reverse directions. Therefore, the joint probability distribution can be factored as in reverse direction as $P(X_T)P(X_{T-1}|X_{T}),..,
P(X_0|X_1,X_2,...,X_T)$.

In particle accelerators, $X$ can represent phase space comprising of positions and momentum of billions of particles. However, the high-dimensionality of $X$ makes it difficult to calculate $P(X_{t} | X_{t+1}, X_{t+2}, \dots, X_{T})$. Fortunately, latent variable models (like variational autoencoders \cite{kingma2013auto}) transform the higher-dimensional distribution $P(X_1, X_2, ... ,X_T)$ into a lower-dimensional distribution $P(\mathbf{z}_1, \mathbf{z}_2, ..., \mathbf{z}_{T})$, $p_{\theta}(\mathbf{z}|\mathbf{X}) = p_{\theta}(\mathbf{X}|\mathbf{z})p(\mathbf{z})/p_{\theta}(\mathbf{X})$. The spatial dynamics is learned by minimizing the evidence lower-bound (ELBO) loss, $\mathbb{E}_{\mathbf{z} \sim q_{\phi}(\mathbf{z}|\mathbf{x}, m)} [\log p_{\theta}(\mathbf{x}|\mathbf{z})] 
- D_{KL}(q_{\phi}(\mathbf{z}|\mathbf{x},m)||p_{\theta}(\mathbf{z}|m))$. The first term of the loss function captures the reconstruction error between the original and generated data, while the second term involves the Kullback–Leibler (KL) divergence, which quantifies the difference between the approximate and true posterior distributions \cite{yin2021neural}. 

In lower-dimensions, the temporal dynamics can be learned through
the reverse autoregression $P(z_{t} | z_{t+1}, z_{t+2}, \dots, z_{T})$ to estimate $\mathbf{z}_{t}$. A recurrent neural network (RNN) like an LSTM can be used to invert the forward temporal dynamics in the latent space. The LSTM learns $P(z_{t} | z_{t+1}, z_{t+2}, \dots, z_{T}) = f(z_t|h_{t+1},c_{t+1})$, where the hidden state, $h_{t+1} = g(z_{t+1}, h_{t+2}, c_{t+2})$ is calculated through future hidden and memory states and future observations \cite{toneva2022combining}. A mean squared error based loss can be used to train the network $L(\psi;\mathbf{z}) = \parallel\mathbf{z}_{t} - \hat{\mathbf{z}}_{t}\parallel_2$. Once the inverse temporal dynamics is learned in the latent space, the decoder part of the VAE transforms it back to the high-dimensional space, obtaining $P(X_{t} | X_{t+1}, X_{t+2}, \dots, X_{T})$.

From a learning perspective, the autoregressive procedure used to train the LSTM is a crucial aspect of the latent evolution model. It is a more general formulation which does not assume a strict Markov property, i.e. $P(z_{t} | z_{t+1}, z_{t+2}, \dots, z_{T}) = P(z_{t}|z_{t+1})$, on the temporal dynamics. Since the temporal dynamics are learned in a lower-dimensional latent space rather than the higher-dimensional image space ($n_d \approx 10^6$), 
$T$-step autoregression compensates for the information loss due to spatial dimensionality reduction. A single step autoregression model, using $P(z_{t}|z_{t+1})$, suffers in performance compared to the $T$-step autoregressive model. Additionally, autoregression procedure leverages temporal periodicity (short-term or long-term seasonality) present in the dataset better than Markov property.

\subsubsection{RLEM for beam dynamics in LANSCE accelerator}
\begin{figure*}[htbp]
\centering
\includegraphics[trim={0 0 0 0},clip, width=1.0\textwidth]{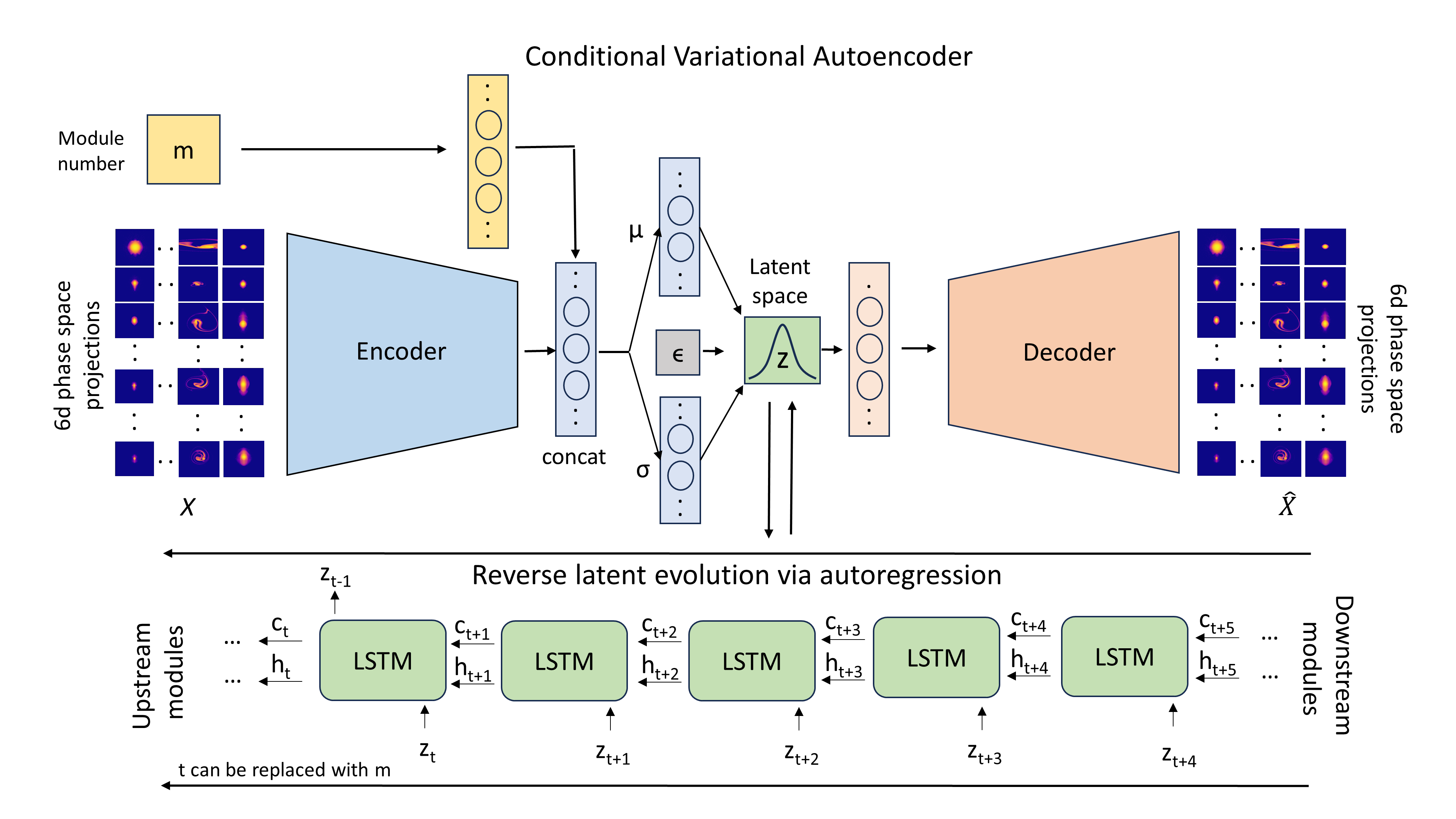}
\caption{Reverse latent evolution model (RLEM): CVAE captures a lower-dimensional latent distribution of a higher-dimensional 6D phase space followed by an autoregressive LSTM learning to reverse temporal dynamics in the latent space.}
\label{fig:model}
\end{figure*}

The above mathematical explanations are more general and uses $t$ as a temporal index with $T$ as the final index where the states $X$ are measured or observed. We apply the RLEM to a dataset representing phase space (X) at 48 different accelerating modules of the LANSCE accelerator. Therefore, all the expressions can be written in terms of $m$, where $M = 48$ is the final module. First, a conditional variational autoencoder (CVAE) projects the 15 unique projections of the 6D phase space of charged particle beams along with the module number into a lower-dimensional latent distribution. Subsequently, a LSTM network learns the reverse temporal dynamics within this latent space. The architecture of the RLEM is represented in Fig.~\ref{fig:model}.

The model learns spatial and temporal correlations independently in two-step self-supervised process. The encoder performs convolution operations to extract features from the images. The module number is passed through dense layers and concatenated with the features vectors out of the encoder. The concatenated vector is the input to the latent space. The LSTM network operates on the learned latent space to predict the previous latent space points (corresponding to projections in the upstream modules) based on future points (corresponding to projections in the downstream modules). The predicted latent points are passed through a decoder which generates phase space projections across different modules of the accelerator. The methodology is outlined in Algorithm~\ref{alg:reverse_prediction}. The input of the algorithm is the user-defined phase space at downstream location of the measurement ($X_{m_{end}}$) and the upstream module/modules until where phase space predictions are desired, denoted by $m_{start}$.  Typically, $m_{start}$ is set to 1, corresponding to the index of the first module, but it can also be any other module number $< m_{end}$. $m_{end}$ can be a single module, or multiple modules together, typically $m_{end} = M = 48$. For the first iteration, the encoder of CVAE projects $X_{M:m}$ into the latent space to obtain $\mathbf{z}_{M:m}$. Subsequent iterations use only the LSTM and decoder of the CVAE. The LSTM acts on the latent representation to predict $\mathbf{z}_{m-1}$. The decoder of the CVAE then reconstructs $X_{m-1}$, providing the phase space projections at the previous module. The predicted latent point $\mathbf{z}_{m-1}$ is then concatenated with input $\mathbf{z}_{M:m}$ to form $\mathbf{z}_{M:m-1}$. This autoregressive procedure is iteratively applied to compute $\mathbf{z}_{m-2}$, followed by $X_{m-2}$, and so forth, until $m_{start}$.

\begin{figure*}[ht]
    \centering
    \begin{minipage}{\textwidth}
        \begin{algorithm}[H]
            \caption{Reverse latent evolution model (RLEM) to invert the beam dynamics}
            \begin{algorithmic}[1]
                \State \textbf{Inputs:} State: \{$X_m$,m\}, Initial index: $m_{start}$, final index: $m_{end}$, pretrained CVAE model, pretrained LSTM model
                \For{$m$ = $m_{end}$ \textbf{to} $m_{start}$}
                    \If {$m = m_{end}$}
                        \State Compute latent encoding $\mathbf{z}_{M:m} = \text{CVAE-encoder}(X_{M:m}, M:m)$
                    \EndIf
                    \State Obtain $\mathbf{z}_{m-1}$ by passing $\mathbf{z}_{M:m}$ through the LSTM: $\mathbf{z}_{m-1} = \text{LSTM}(\mathbf{z}_{M:m})$
                    \State Reconstruct $X_{m-1}$ by decoding $\mathbf{z}_{m-1}$: $X_{m-1} = \text{CVAE-decoder}(\mathbf{z}_{m-1})$
                    \State Update latent sequence $\mathbf{z}_{M:m-1}$ by appending $\mathbf{z}_{m-1}$: $\mathbf{z}_{M:m-1} = \text{concat}(\mathbf{z}_{M:m}, \mathbf{z}_{m-1})$
                \EndFor    
                \State \textbf{Return:} reconstructed sequence $X_{m_{end}:m_{start}}$
            \end{algorithmic}
            \label{alg:reverse_prediction}
        \end{algorithm}
    \end{minipage}
\end{figure*}

\section{\label{sec:results} Results and Discussion}
The training dataset comprising 1400 data objects is partitioned into training and validation sets with an 85:15 ratio. The encoder of the CVAE is composed of five convolutional layers with 32, 64, 128, 256, and 512 filters of size 3×3, each with a stride of 2, followed by a dense layer with 256 neurons. Additionally, a separate dense branch processes the module number through two layers of 32 neurons. The outputs of the primary and branch layers are concatenated and fed into the latent dense layer. We have performed a trial-and-error study balancing dimensions and loss to select the dimensionality of the latent space. This lower-dimensional representation enhances interpretability and reduces training time. The decoder mirrors the encoder's architecture, featuring a dense layer followed by convolutional layers with 512, 256, 128, 64, and 32 filters of size 3×3. Each layer employs the LeakyReLU activation function and batch normalization.

The CVAE network is optimized using the Adam optimizer with a learning rate of 0.001 and a batch size of 32, over 1500 epochs. This configuration proved effective in balancing performance and efficiency.

\subsection{Latent space visualizations}
\begin{figure*}[htbp]
    \centering
    \begin{minipage}[b]{0.132\linewidth}
        \centering
        \includegraphics[trim={0cm 0cm 0cm 0cm},clip, width=1.0\textwidth]{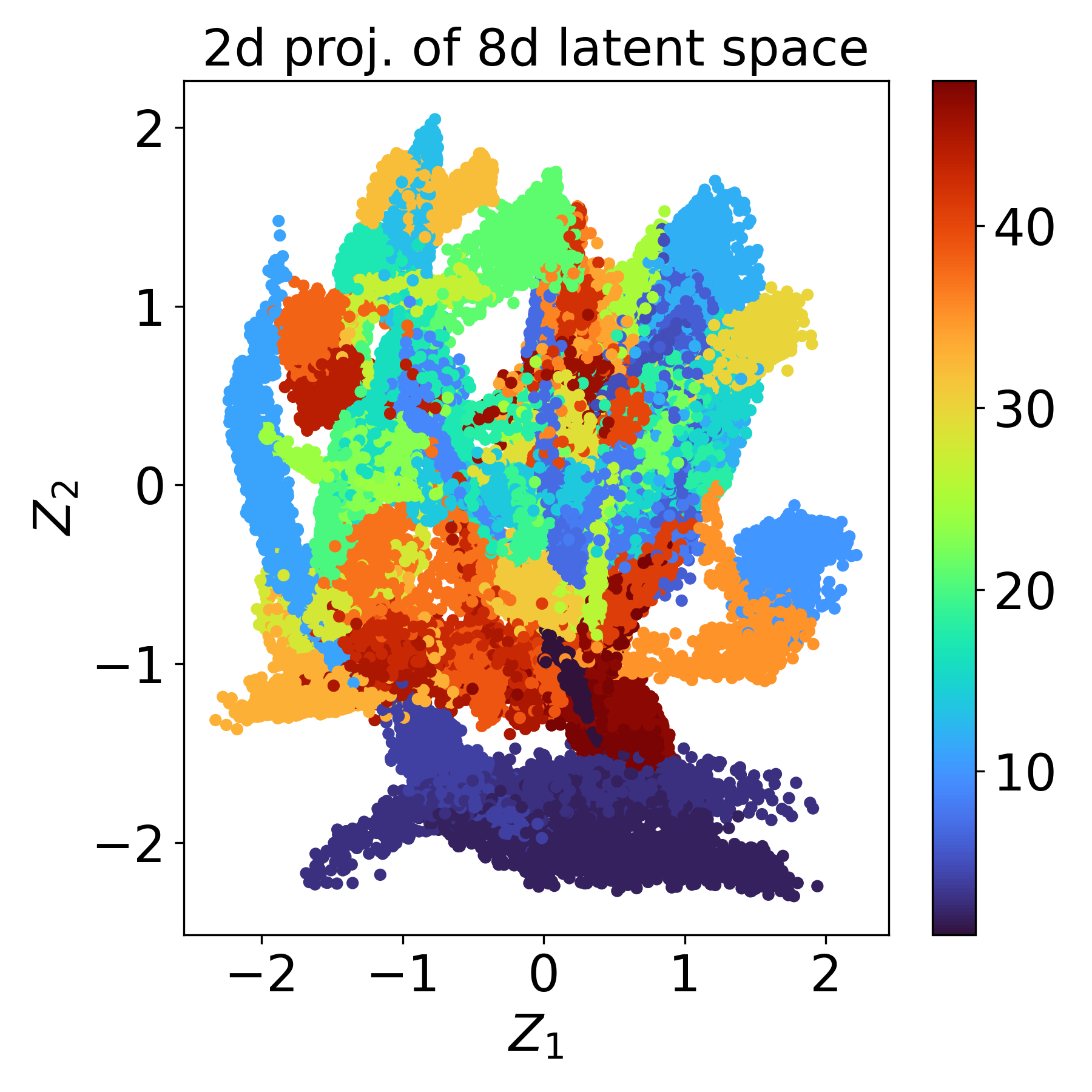}
    \end{minipage}
    \begin{minipage}[b]{0.132\linewidth}
        \centering
        \includegraphics[width=1.0\textwidth]{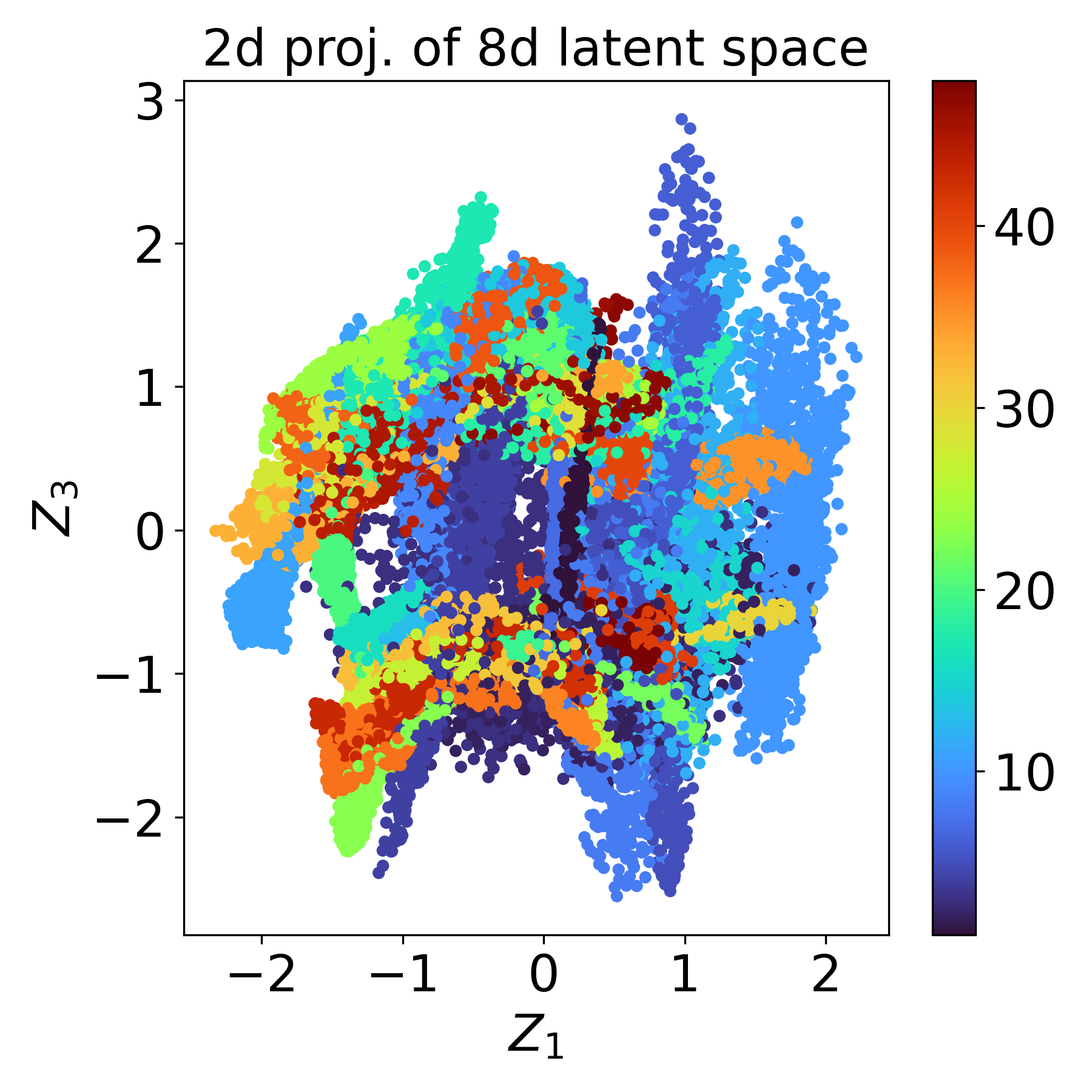}
    \end{minipage}
    \begin{minipage}[b]{0.132\linewidth}
        \centering
        \includegraphics[width=1.0\textwidth]{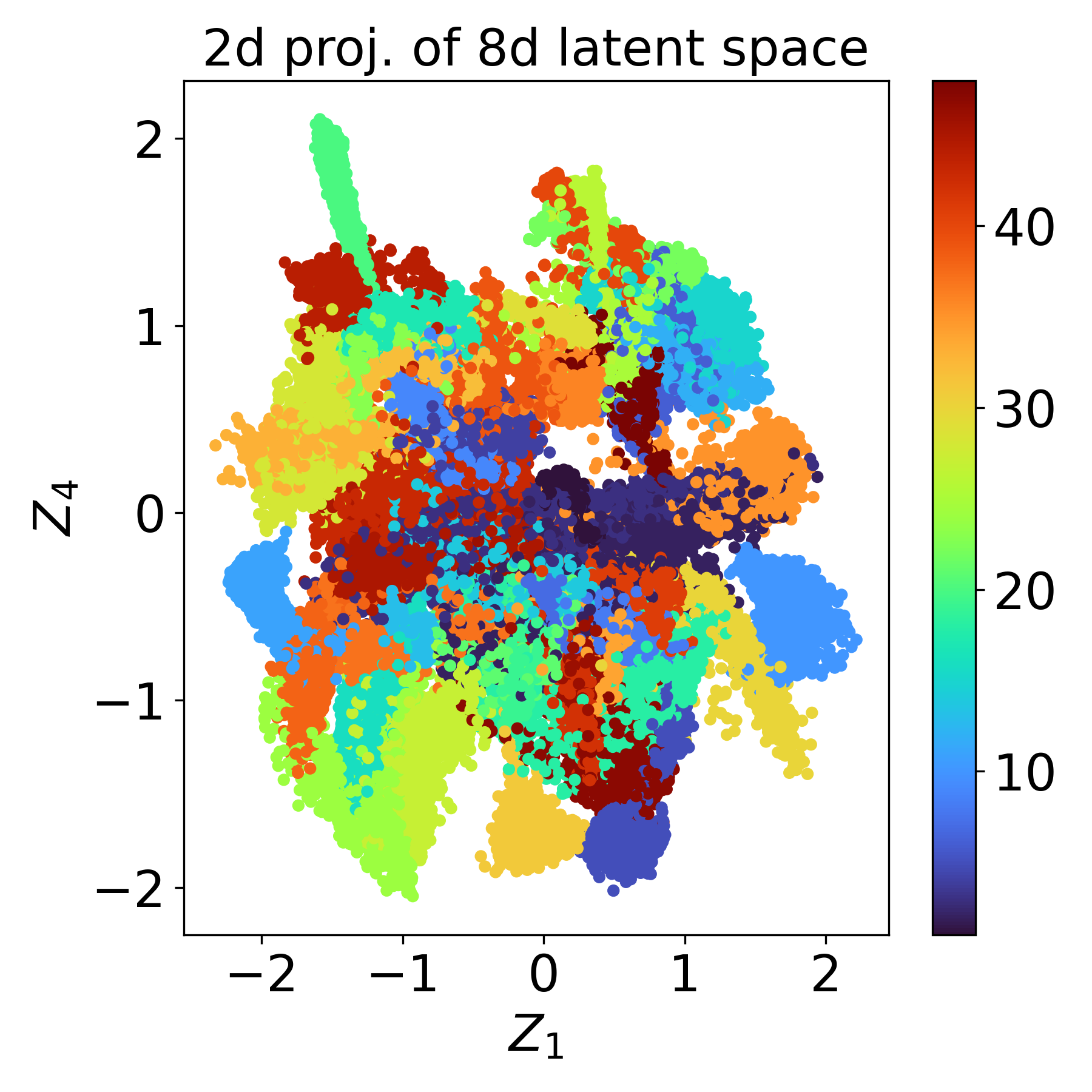}
    \end{minipage}
    \begin{minipage}[b]{0.132\linewidth}
        \centering
        \includegraphics[width=1.0\textwidth]{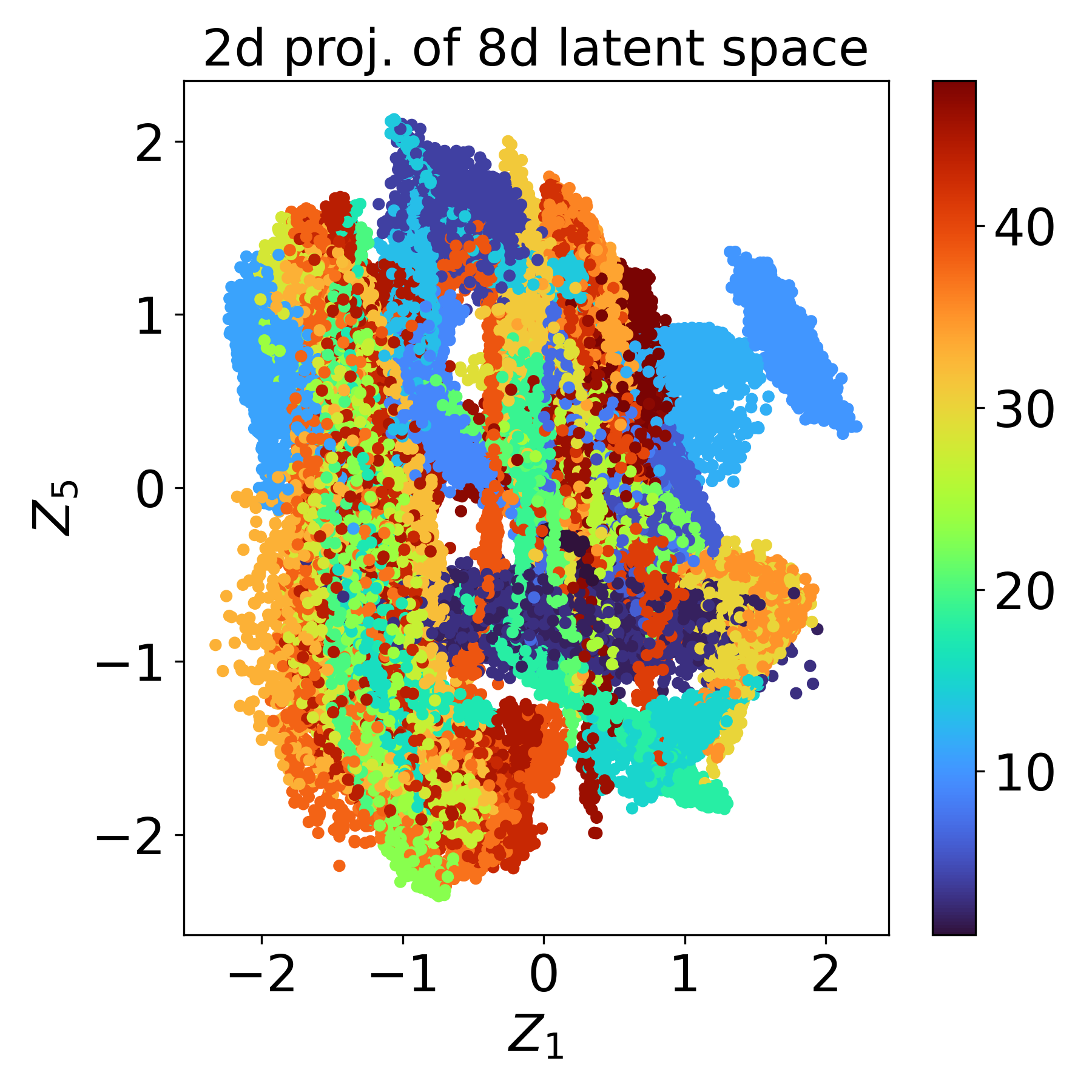}
    \end{minipage}
    \begin{minipage}[b]{0.132\linewidth}
        \centering
        \includegraphics[width=1.0\textwidth]{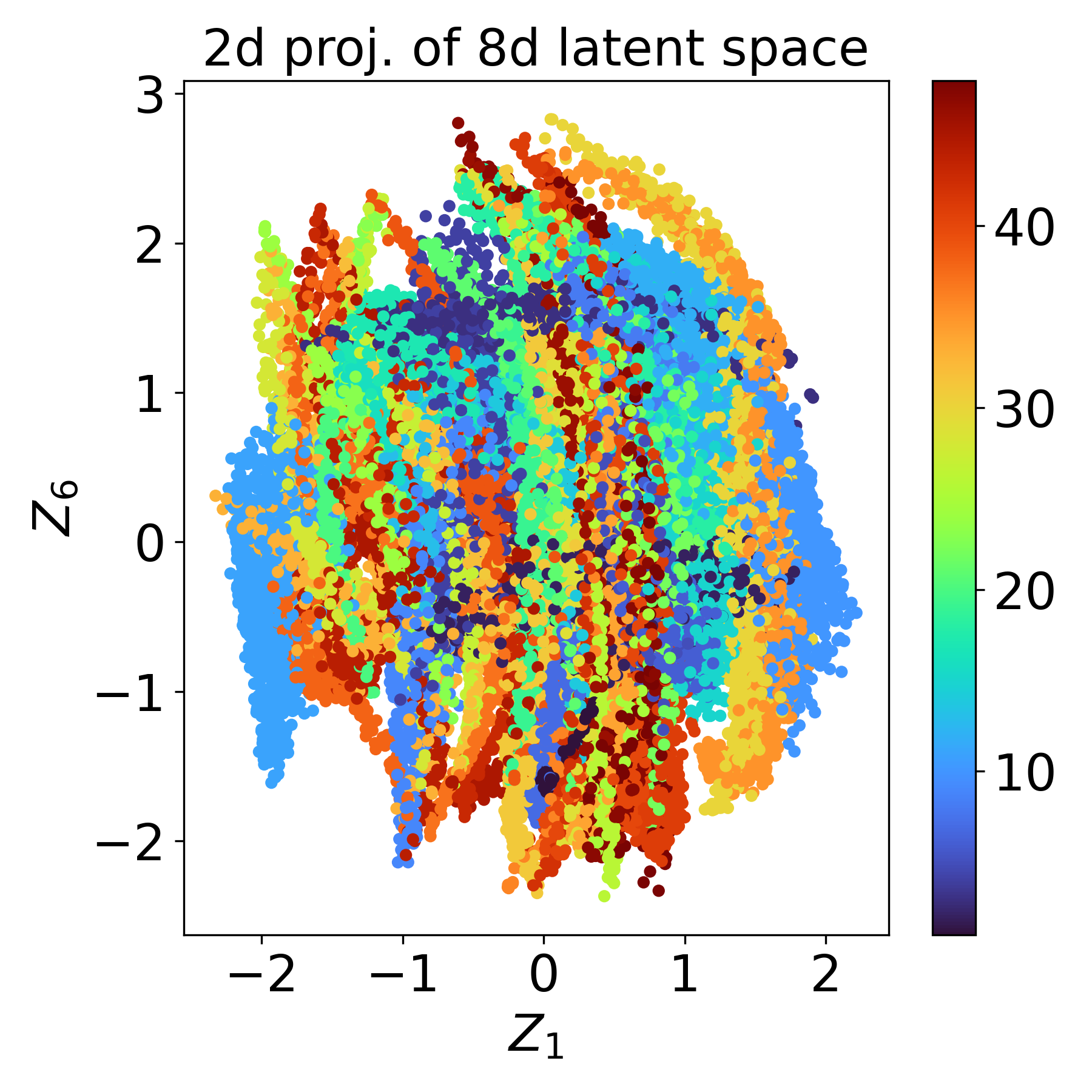}
    \end{minipage}
    \begin{minipage}[b]{0.132\linewidth}
        \centering
        \includegraphics[width=1.0\textwidth]{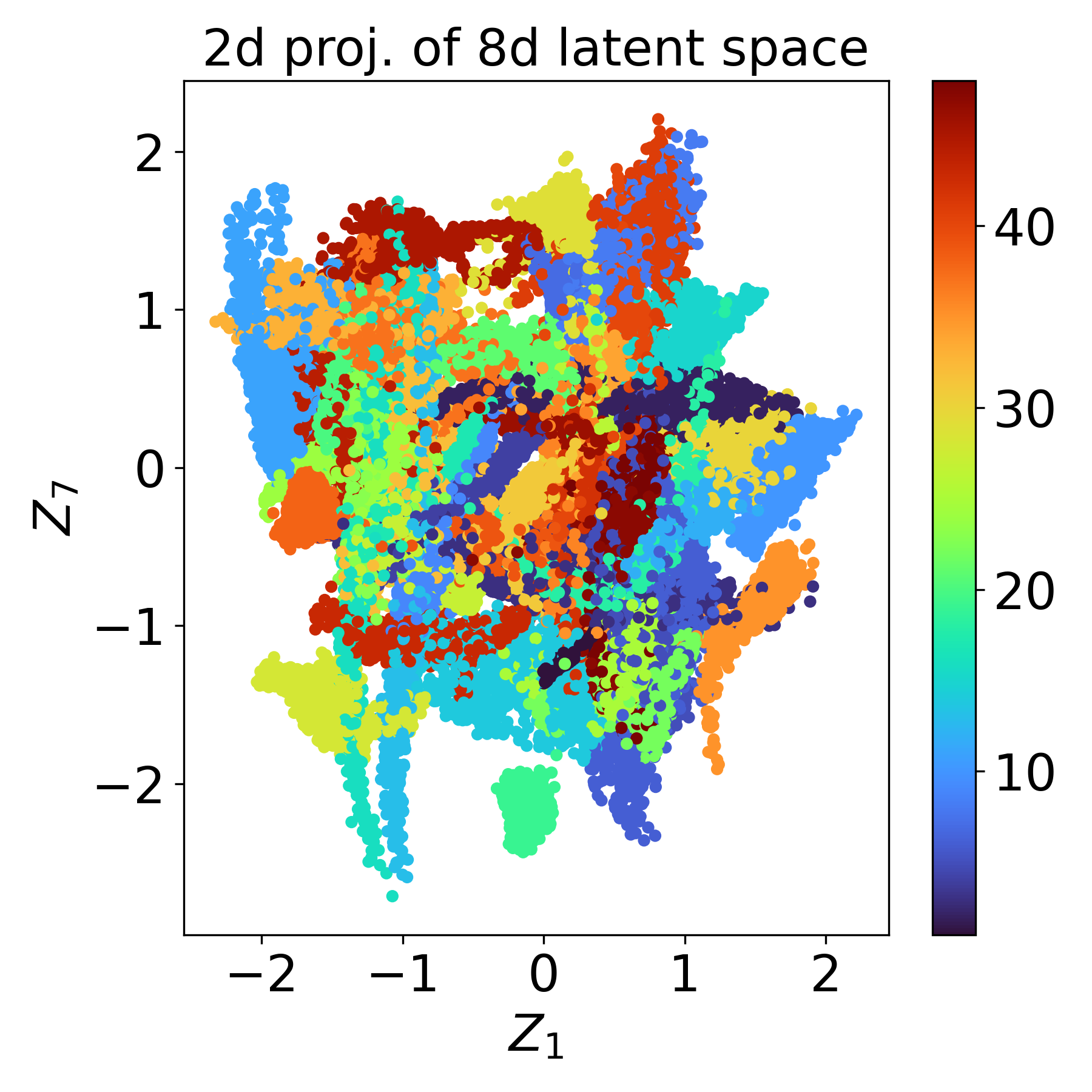}
    \end{minipage}
    \begin{minipage}[b]{0.132\linewidth}
        \centering
        \includegraphics[width=1.0\textwidth]{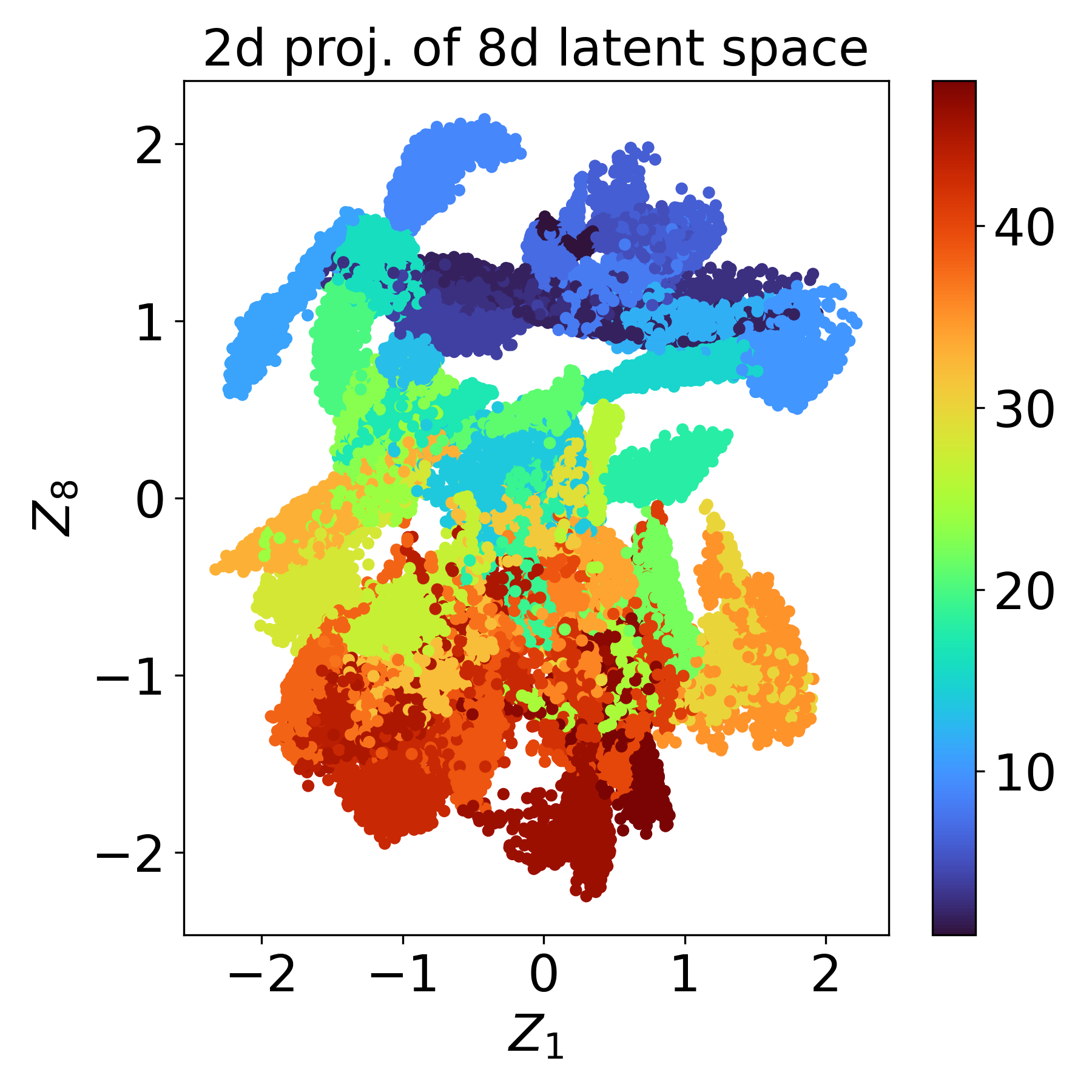}
    \end{minipage}
    \begin{minipage}[b]{0.23\linewidth}
        \centering
        \includegraphics[width=1.0\textwidth]{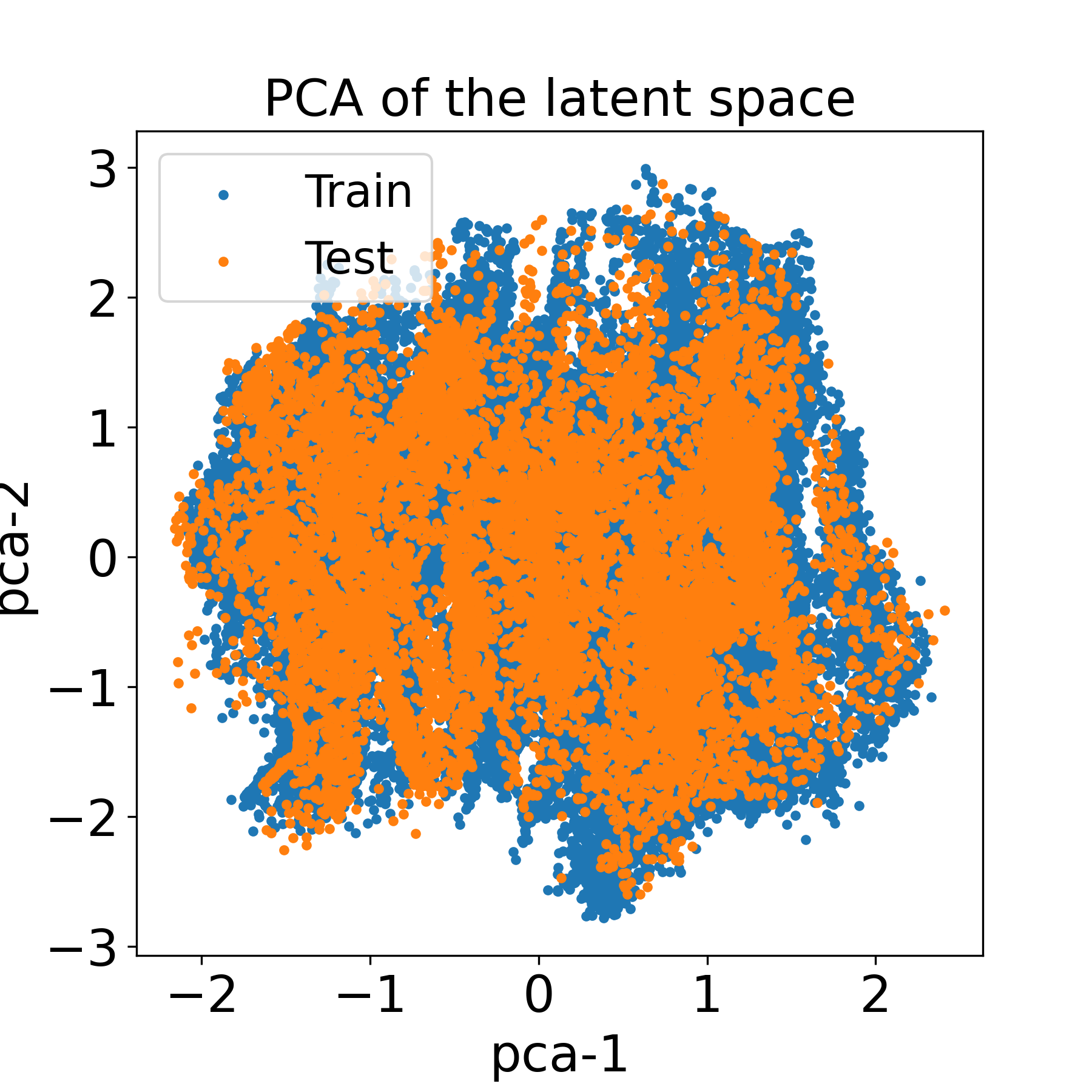}
    \end{minipage}
    \begin{minipage}[b]{0.23\linewidth}
        \centering
        \includegraphics[width=1.0\textwidth]{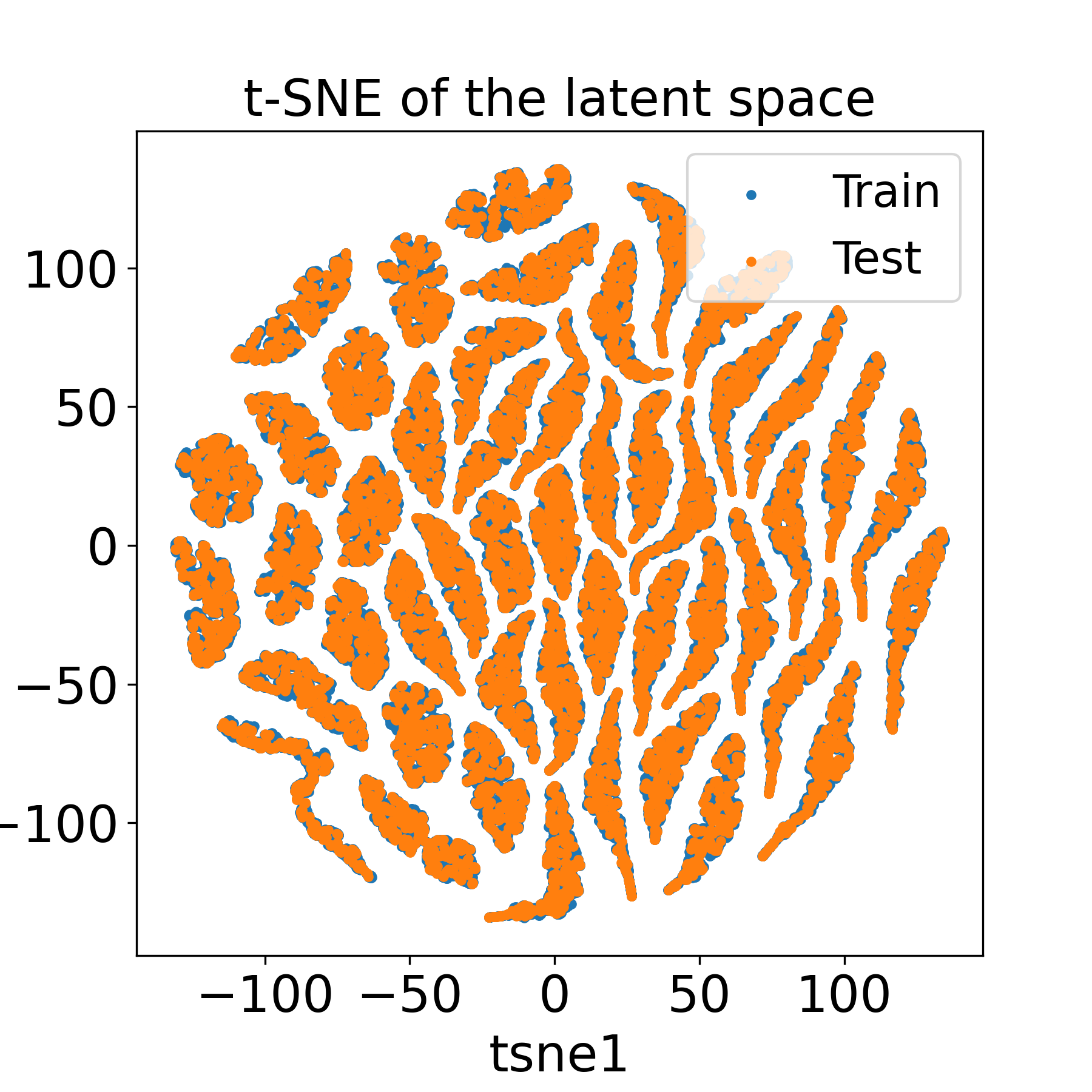}
    \end{minipage}
    \begin{minipage}[b]{0.23\linewidth}
        \centering
        \includegraphics[width=1.0\textwidth]{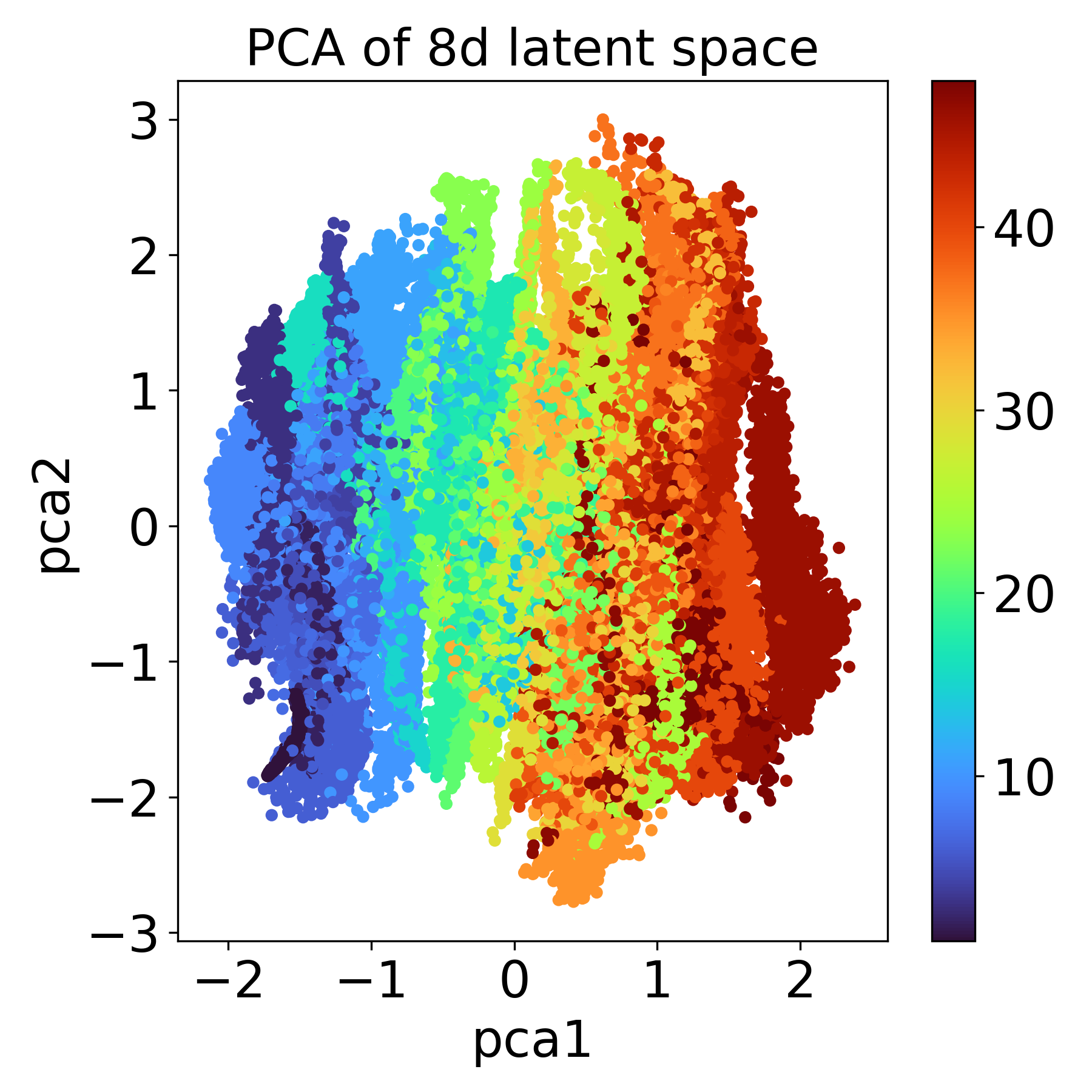}
    \end{minipage}
    \begin{minipage}[b]{0.23\linewidth}
        \centering
        \includegraphics[width=1.0\textwidth]{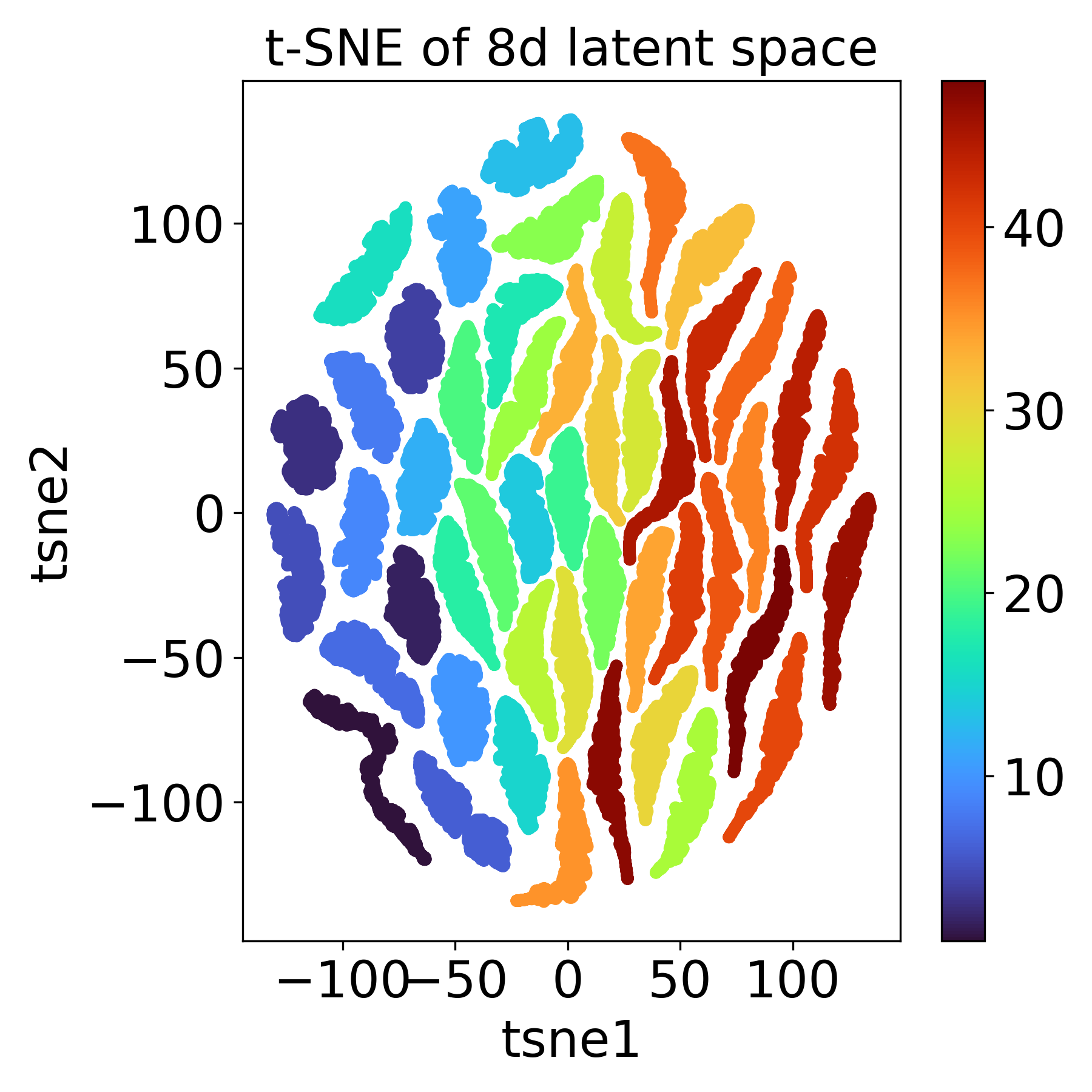}
    \end{minipage}
    \caption{Latent space visualization: Top row shows 7 2d projections ($Z_1 - Z_{2:7}$) of 8d latent space. The bottom row shows 2D PCA and 2D t-SNE of the 8D latent space. The first two figures in the row shows train-test points and the last two figures shows combined train-test colored based on 48 modules.}
    \label{fig:visualization}
\end{figure*}

The latent space of a VAE naturally lead to continuous latent space representations that can be smoothly traversed, making them well suited for probabilistic density estimation and generation \cite{rautela2022towards}. Visualization of the latent space is important because it offers valuable insights into the extracted features, which eventually enhances the network's interpretability. Seven 2D projections of the the 8D latent space is plotted in the top row of Fig.~\ref{fig:visualization}. It can be observed from the 2D projection of the latent space that the phase space in different modules are much more separated in $Z_1-Z_8$ than $Z_1-Z_2$, and $Z_1-Z_4$  and $Z_1-Z_5$ projections. But, visualizing remaining 2D projections is still cumbersome even with 8D latent space. 

We have transformed 8D latent space into 2D space using two of the most popular dimensionality reduction techniques i.e. principal component analysis (PCA) and t-stochastic nearest neighbors (t-SNE) \cite{van2008visualizing}. PCA is a linear reduction technique, whereas t-SNE is a non-linear manifold learning approach. PCA and t-SNE of the 8D latent space is shown in Fig.~\ref{fig:visualization}. The first two plots in the bottom row shows the location of the points corresponding to the training and test set in PCA and t-SNE space. The last two plots show PCA and t-SNE space color coded with different modules.

It is seen from the PCA plot that the points belonging to the initial and end modules are well separated, with a gradual variation in modules as we move along the PCA space. In t-SNE space, similar to PCA, the initial and end modules occupy different ends of the space. It is easily noticeable that there is better separation of different modules into clusters in the t-SNE space. This is to be expected as PCA is a linear method with limited expressive power as compared to nonlinear t-SNE method. Also, PCA is formulated to project high-dimensional data into principal space by maximizing the variance of the data. On the other hand, t-SNE is designed for clustering of points in the lower dimensional space.

The ability to distinguish phase space projections in various modules at different positions demonstrates that the CVAE has effectively learned the spatial correlations of phase space projections across different modules.

\subsection{Prediction results}
For training LSTM, the latent space of trained CVAE is reshaped from 67,200$\times$8 to 1400$\times$48$\times$8, where 1400 signifies the number of samples, 48 denotes the total number of modules, and 8 is the dimension of the latent space. In LSTM terminology, 48 becomes the trajectory length and 8 are number of features. In order for autoregressive prediction, the LSTM is trained with variable-length inputs and single-length output. The 8D trajectory with length 48 is split using input window of sizes varying from 1 to 47 and output window of size of 1. The two windows slides on the 8D trajectory from right to left to split it into different input sizes varying from $1\times8$, $2\times8$, ..., $47\times8$, while the output remains fixed at ${1\times8}$ (representing the upstream module). The variable-length trajectories are made compatible with the LSTM by padding them with zeros to achieve a uniform length of 47. However, the LSTM masks these padded zeros while learning the temporal dynamics through recurrence.

The LSTM training with variable-length input and a single output allows the network to learn the autoregressive temporal patterns of the latent points, which is essential for making autoregressive predictions in the latent space. Our LSTM architecture comprises two layers, each with 64 hidden units. We minimize the mean squared error (MSE) between the true and predicted latent vectors using Adam optimizer. The initial learning rate is set to 0.001 and is halved after 10 epochs if the validation loss plateaus.

A trained RLEM network can be used to autoregressively predict upstream phase space projections based on downstream phase space projections provided as inputs. The prediction procedure is detailed in Algorithm~\ref{alg:predict} and explained in Sec.~\ref{ssec:rlem}. The algorithm operates with a user-defined downstream location (input) where the measurement is specified and the number of upstream locations where predictions are needed (output). It is flexible regarding inputs, accommodating multiple consecutive or non-consecutive downstream locations. Depending on the inputs, it can provide phase space projections for single, multiple, consecutive, or non-consecutive upstream locations.

\begin{figure*}[htbp]
    \centering
    \begin{minipage}[b]{1.0\linewidth}
        \centering
        \includegraphics[trim={4cm 1cm 4cm 0cm},clip, width=1.0\textwidth]{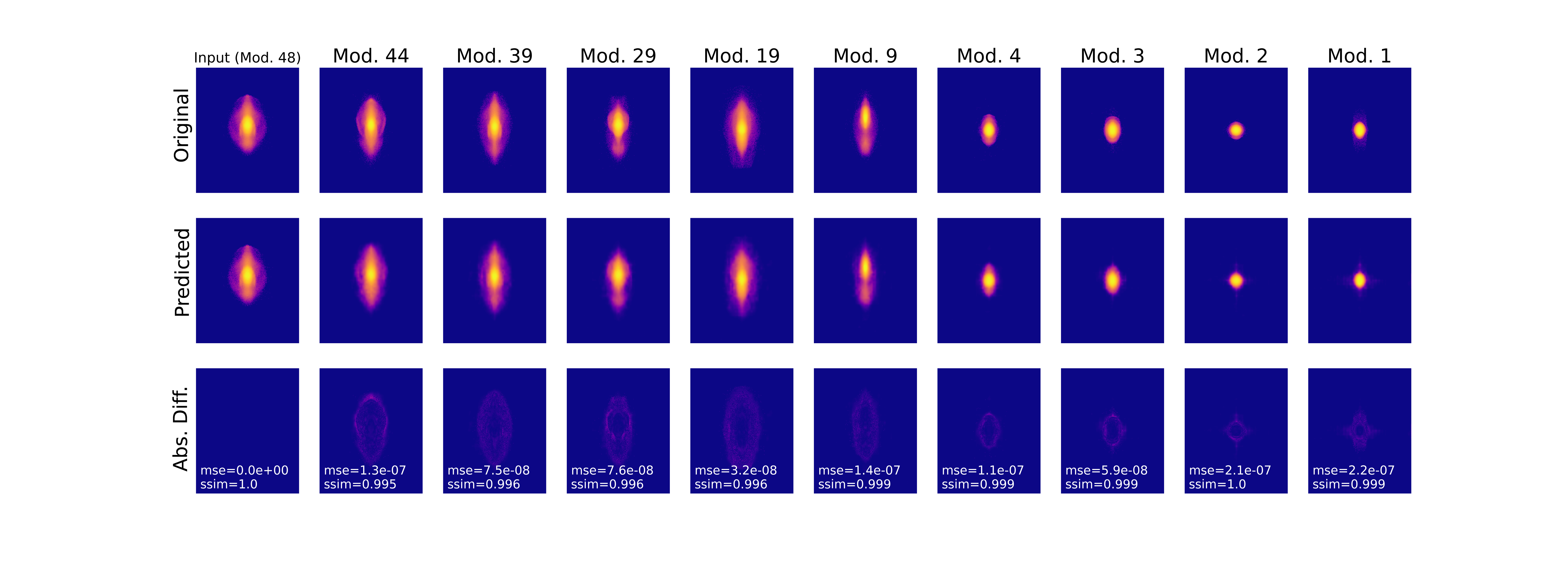}
        \subcaption{$x-y$ projection}
    \end{minipage}
    \begin{minipage}[b]{1.0\linewidth}
        \centering
        \includegraphics[trim={4cm 1cm 4cm 0cm},clip,width=1.0\textwidth]{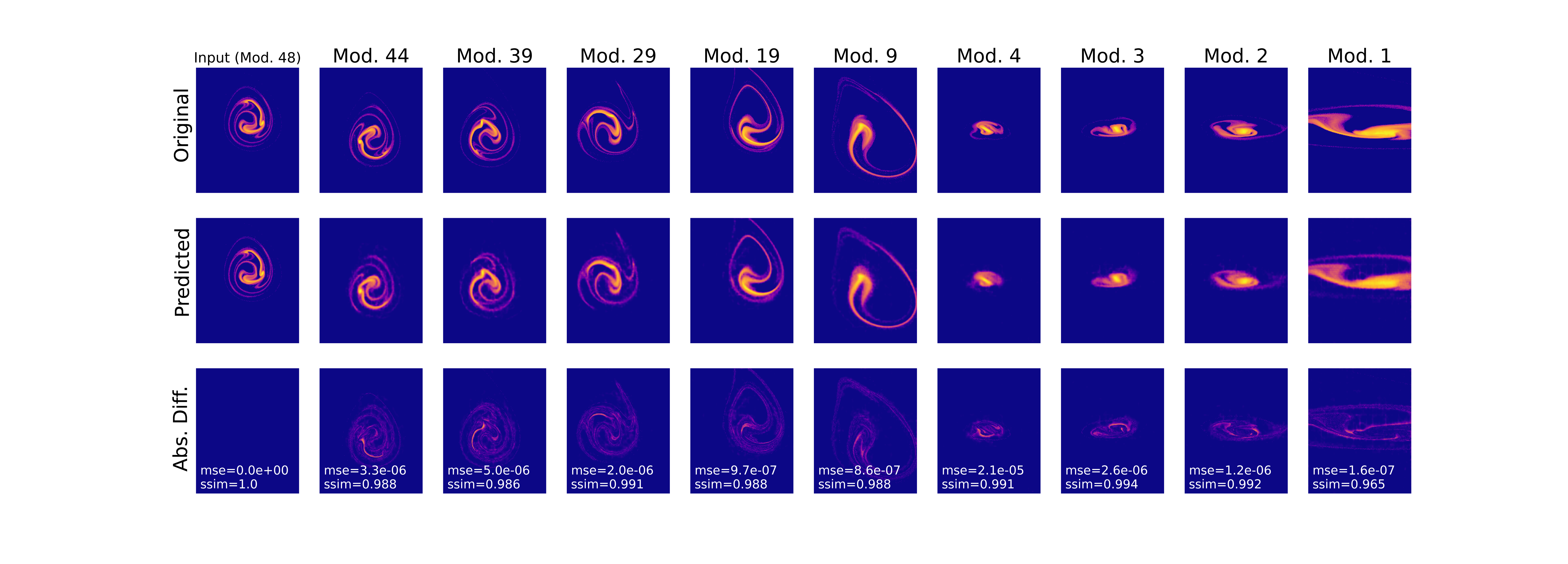}
        \subcaption{$E-\phi$ projection}
    \end{minipage}
    \begin{minipage}[b]{1.0\linewidth}
        \centering
        \includegraphics[trim={4cm 1cm 4cm 0cm},clip,width=1.0\textwidth]{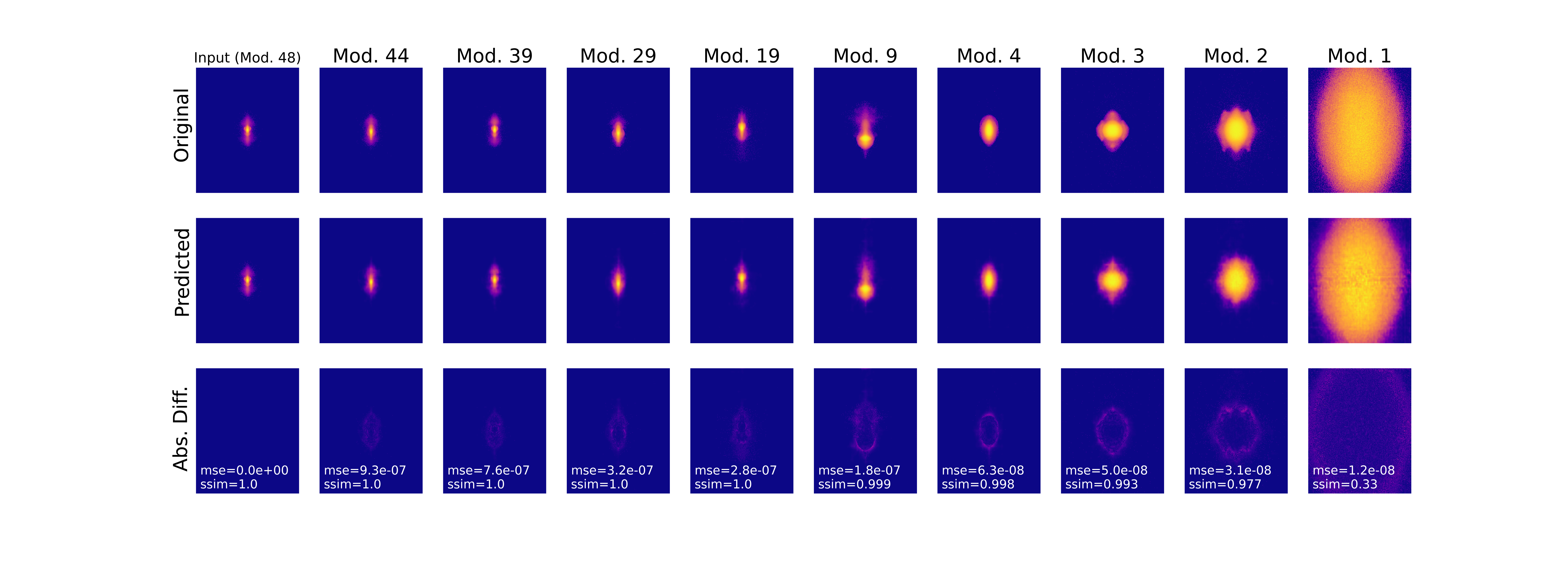}
        \subcaption{$x'-y'$ projection}
    \end{minipage}
    \caption{Predictions from the trained RLE at different upstream modules given module 48 as the input. The original, predicted and the absolute difference are plotted for (a) $x-y$ projection, (b) $E-\phi$ projection and (c) $x'-y'$ projection across nine different modules.}
    \label{fig:prediction_48}
\end{figure*}

Fig.~\ref{fig:prediction_48} shows prediction in all the upstream modules given the last module (module 48) as the input. Similar plots using module 10 and module 25 as inputs are shown in Figs.~\ref{fig:prediction_10} and \ref{fig:prediction_25}. For brevity, three out of fifteen projections are shown across nine different modules. The original and predicted images can be compared using absolute difference plots. The mean squared error (MSE) and structural similarity index (SSIM) are also calculated and displayed alongside the absolute difference plots. It is evident from the figures and comparison metrics that RLE performs well, demonstrating high accuracy prediction capabilities. We achieve an overall MSE and SSIM of ($\approx$ 5e-7, $\approx$ 0.998) and ($\approx$ 1e-6, $\approx$ 0.976) on the training and test set. 

In this work we are only modifying the RF parameters of the first 4 RF modules of LANSCE which comprise the initial 201.25 MHz section of the accelerator. The quadrupole magnetic lattice and the 44 subsequent RF cavities of the 805 MHz section of the accelerator are unchanged and their limited beam acceptance acts as a powerful screening mechanism so that the variance of the beams which do survive to the end of the accelerator cannot be very large. In future work, we plan on adjusting the RF parameters of all 48 RF modules of LANSCE and of the magnetic lattice, which will allow for the survival (particles not being lost as they hit the walls of the accelerator beam pipe along the way) of a much more diverse set of beams to the end of the accelerator. 

Despite the screening process of the accelerator, upon a detailed inspection one can see significant differences in the detailed phase space distributions of the various beams on our training and test sets, and these differences are easily quantified via metrics such as mean squared error (MSE). In order to clearly demonstrate the ability of the LSTM to distinguish between different beams, we show 2 examples in Fig.~\ref{fig:compare_prediction_mse_48} where we show two beam examples side by side ($A_{true}$ and $B_{true}$), their differences from each other at various modules ($A_t - B_t$), and their differences from the LSTM-based autoregressive predictions ($A_t - A_p$ and $B_t - B_p$). It is shown that the LSTM is able to accurately track them with a much lower MSE than the difference between the two beams, which is what would happen if one example was simply used as a guess attempting to predict the other.

Generating predictions for all 15 projections in all 47 modules takes less than one second with RLE, whereas the simulator requires about 10 minutes on similar computing infrastructure. This results in a speedup factor of approximately 600.

\subsection{Uncertainty analysis results}
\subsubsection{Aleatoric uncertainty in the latent space}
\begin{figure*}[htbp]
    \centering
    \includegraphics[trim={0cm 0cm 0 0cm},clip, width=0.85\textwidth]{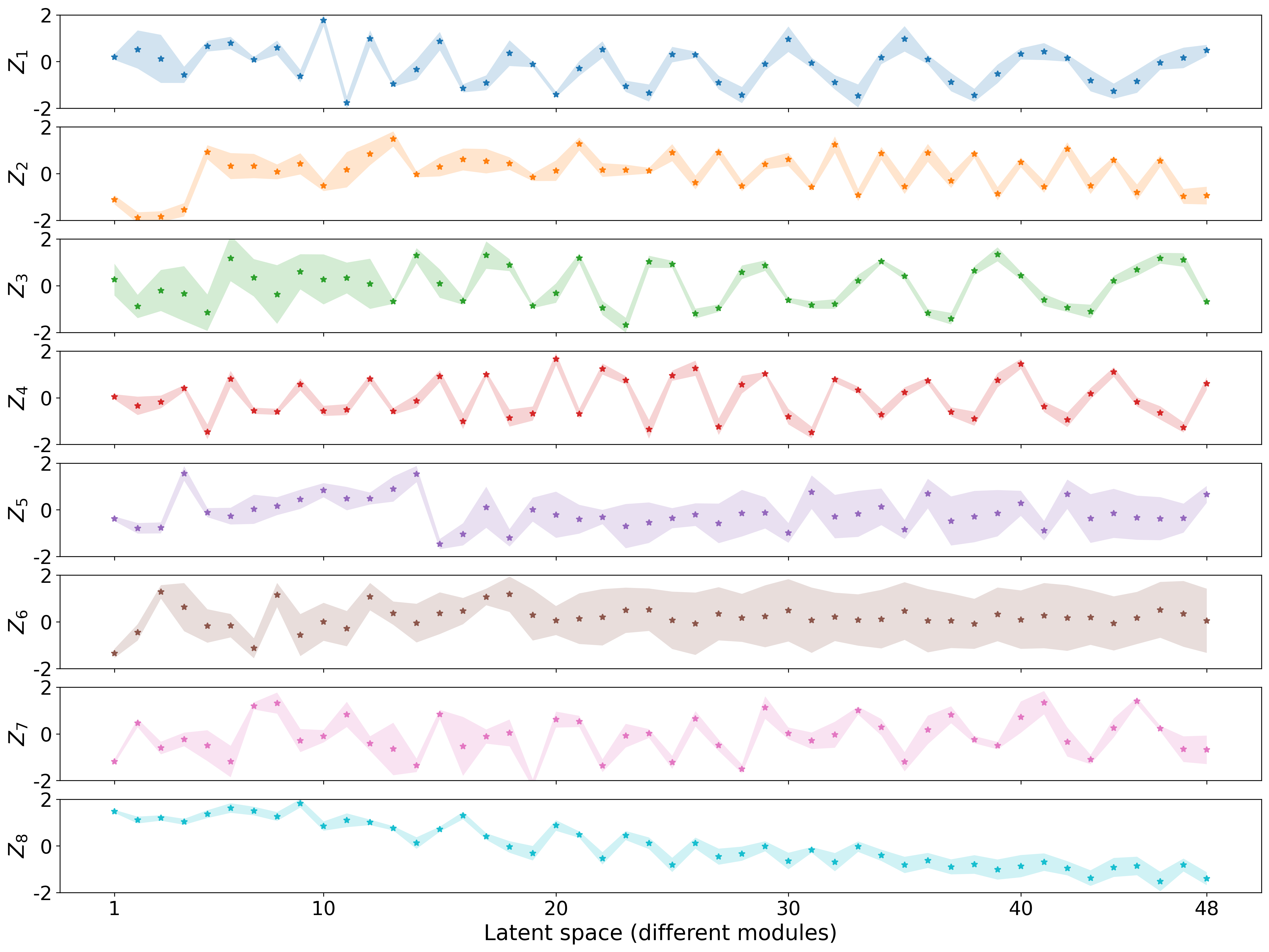}
    \caption{Latent space bounds for all the modules for all the 8 dimensions. The space is conditionally sampled 10k times at different modules followed by calculating the mean and standard deviation for every module and dimension.}
    \label{fig:latentbounds}
\end{figure*}

\begin{figure*}[htbp]
    \centering
    \begin{minipage}[b]{0.45\linewidth}
        \centering
        \includegraphics[trim={0cm 0cm 0cm 0cm},clip, width=1.0\textwidth]{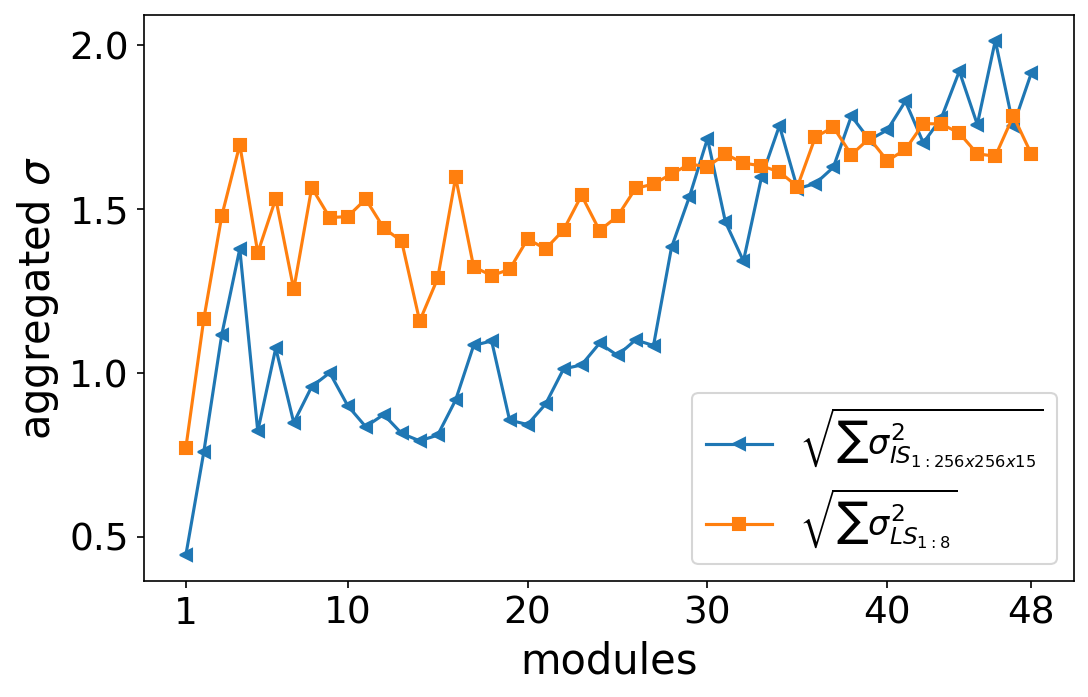}
    \end{minipage}
    \begin{minipage}[b]{0.45\linewidth}
        \centering
        \includegraphics[trim={0cm 0cm 0cm 0cm},clip,width=1.0\textwidth]{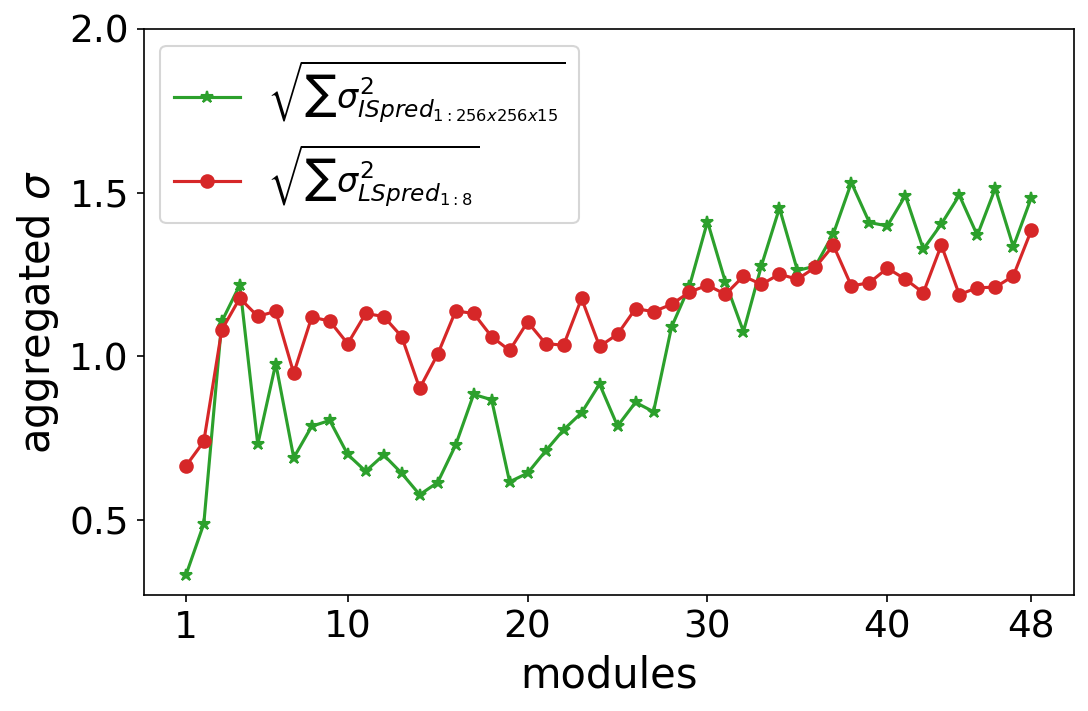}
    \end{minipage}
    \caption{Comparison of aggregated mean and standard deviation of (a) Dataset (Image space (IS): 1400x48x256x256x15) and latent space (LS) across different modules (10k MC samples in 48 modules) (b) LSTM based predictions of the upstream latent space (520 true trajectories) and corresponding decoded images (520x48x256x256x15).}
    \label{fig:comparemeansd_2}
\end{figure*}

The VAE learns a lower-dimensional representation of images through a probabilistic density function, which can be randomly sampled and decoded to generate new, realistic in-distribution (ID) images \cite{rautela2023deep}. This is possible because the VAE's latent space encodes the aleatoric uncertainty of the higher-dimensional input dataset \cite{huang2022evaluating,acharya2023learning}. To quantify this aleatoric uncertainty in the latent space, we collected 10,000 Monte-Carlo samples from all the modules by conditionally sampling the latent space within the bounds of the corresponding modules. The mean and standard deviation are calculated and presented in Fig.~\ref{fig:latentbounds} for all eight dimensions of the latent space across all 48 modules. Notably, the standard deviation varies across different dimensions, being higher in the 4th and 5th dimensions. Additionally, we have plotted the aggregated standard deviation ($\sum{\sigma_{LS_{1:8}}^2}$) in Fig.~\ref{fig:comparemeansd_2}(a), represented by the blue curve. The aggregated standard deviation, which measures the total spread across all dimensions, shows an increasing trend towards the later modules, purely due to the physical nature of the problem with data in later modules simply containing more variance. This is also evident with the orange curve, which demonstrates that the aggregated standard deviation in image space ($\sqrt{\sum{\sigma_{IS_{1:256x256x15}}^2}}$) follows a similar trend to that in latent space (LS). Although a direct comparison of relative differences is challenging due to the different ranges in image space and latent space, the overall trend remains clearly observable. 

Another interesting observation from Fig.~\ref{fig:latentbounds} is that the trajectories are primarily centered and symmetrical around the center of the latent space in all dimensions except $Z_8$. This implies that by moving linearly in the $Z_8$ direction ([-2,2]) while keeping the other dimensions fixed at the origin, we can generate phase space projections for modules from 48 to 1 without the need for conditional sampling in those regions of the latent space. By visualizing the $Z_8$ direction in Fig.~\ref{fig:visualization}(top plot, last image), we can further see the evidence of the aforementioned observation. This demonstrates that a well-structured latent space can effectively generate new samples with customized properties \cite{wang2020deep,rautela2022towards}. Further investigations into this topic can be pursued as part of future research.

In this study, we focus exclusively on encoding aleatoric uncertainty with the CVAE and propagating it using the LSTM network. We assume that epistemic uncertainty is negligible in this context, which is possible with enough training data and sufficient model expressiveness. This assumption is essential for developing a novel approach to aleatoric uncertainty analysis through latent space. Future research will address both epistemic uncertainty and total uncertainty analysis.

\subsubsection{Aleatoric uncertainty propagation}
We also evaluate the robustness of the trained LSTM in predicting upstream states under uncertain downstream measurements. To achieve this, we propagate the aleatoric uncertainty captured in the latent space through the trained LSTM model. Since the CVAE has learned the aleatoric uncertainty of the phase space projections, any uncertain measurement (real observation) at a downstream module can be thought as a random point in the latent space corresponding to that module. This assumption is valid only if the uncertainty of the real observation is well captured by the CVAE, which was trained with the simulated phase space projections corresponding to randomly varied RF settings in a range. In our work, the simulator (HPSim) used to collect the dataset is calibrated every six months with the real accelerator (LANSCE). The drifts within those six months are slow enough to reasonably assume that the measurement will fall within the bounds of our training dataset.

We conditionally sample the latent space corresponding to module 48 multiple times (1000 times) to represent these uncertain measurements. Using these multiple instances as inputs, the trained LSTM predicts the 8D latent points for all upstream modules (47 to 1). Since the LSTM is a deterministic model with no built-in model uncertainty, it provides a single prediction for each upstream module based on the downstream measurements. To capture epistemic uncertainty, future work could incorporate techniques such as a Mixed Density Network (MDN) layer \cite{ha2018world} or Monte Carlo dropout \cite{rautela2023bayesian} into the LSTM framework. These approaches could enable the model to deliver both a mean and a standard deviation for all upstream modules based on a single downstream input.

\begin{figure*}[htbp]
    \centering
    \includegraphics[trim={0cm 0cm 0cm 0cm},clip, width=0.85\textwidth]{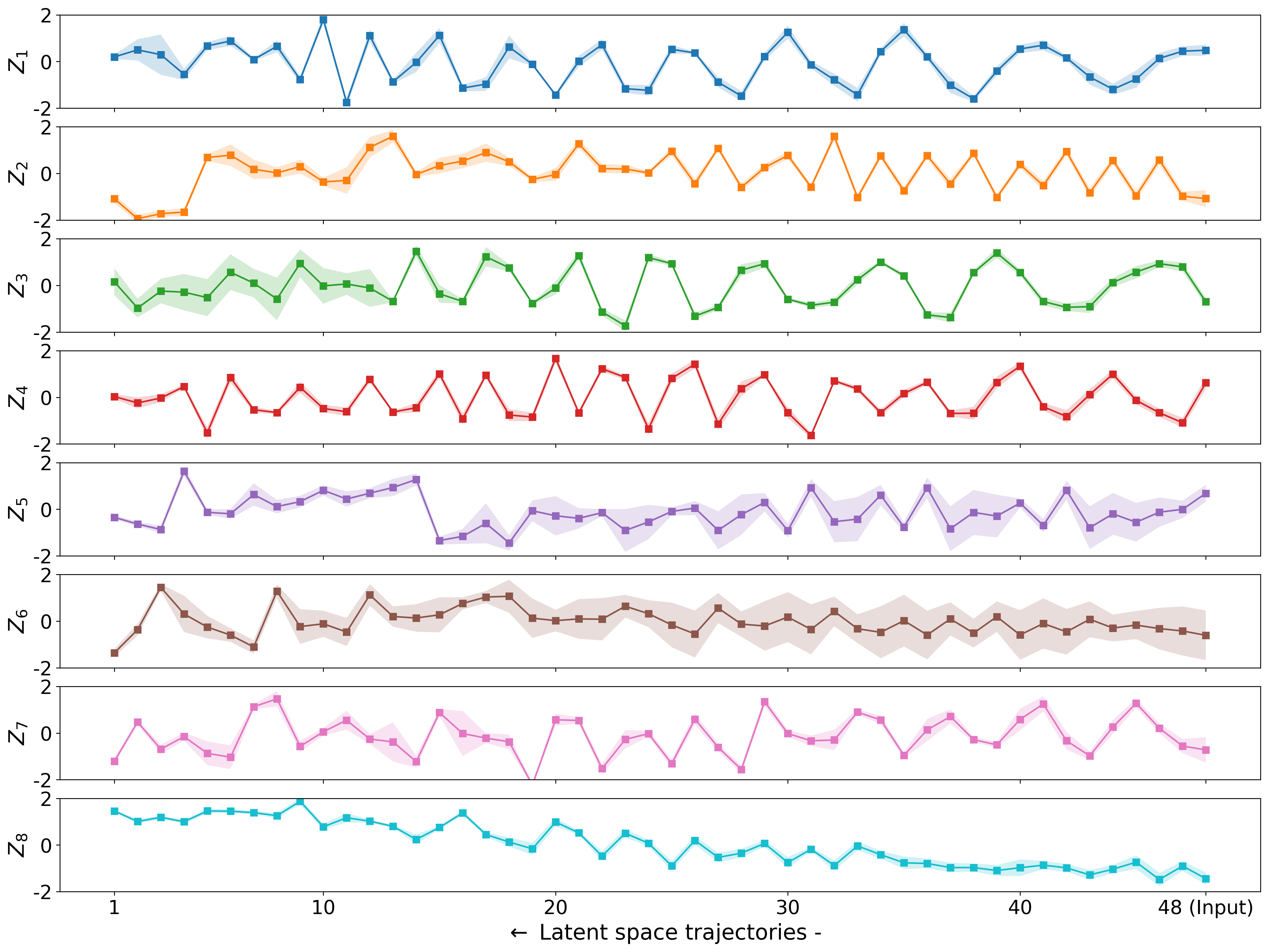}
    \caption{Latent space trajectories for all 8 dimensions with mean and standard deviation bounds. 1000 latent points are randomly sampled from the conditional latent space of module 48 and provided as input to the LSTM to predict the latent points for the upstream modules.}
    \label{fig:trajectories}
\end{figure*}

We decode all the predicted latent points (from LSTM) to generate images using the trained CVAE decoder and then filter out the trajectories corresponding to false positive images. To achieve this, we use a pre-trained ResNet-50 classifier to determine if the generated image belongs to the true class (true module), similar to the method described in \cite{rautela2024conditional}. Through this unsupervised filtering analysis, we determined that nearly 50 percent (approximately 520 out of 1000) of the trajectories are identified as true trajectories. We calculated the mean and standard deviations for these true trajectories and present them in Fig.~\ref{fig:trajectories}. 

The motivation for performing filtering analysis stems from the fact that no generator can consistently produce realistic images every time it is sampled. As demonstrated in \cite{rautela2024conditional}, the precision of the generated images across different modules ranges between 90-100\% -- even with high accuracy, it is still reasonably unlikely that 48 sampled points generate a realistic trajectory. The variability is due to the intermixing areas of latent points from different modules in lower dimensions, as shown in the 2D PCA plots in Fig.~\ref{fig:visualization}. So, we perform filtering to distinguish bad trajectories. 

The bounds of the upstream predictions in Fig.\ref{fig:trajectories}
fall within the bounds of the latent space shown in Fig.\ref{fig:latentbounds}. This is confirmed by Fig.~\ref{fig:comparemeansd_2}(b), which shows that the aggregate standard deviation of the predictions of upstream latent points ($\sqrt{\sum{\sigma_{LSpred_{1:8}}^2}}$ shown by the red curve) is lower than that of the entire latent space ($\sum{\sigma_{LS_{1:8}}^2}$ shown by the orange curve in Fig.~\ref{fig:comparemeansd_2}(a)). This indicates that the predictions from the trained LSTM do not violate the bounds of the latent space, demonstrating the robustness of the trained LSTM when handling uncertain measurements. Additionally, there is a expected similarity in the increasing trend observed across all four curves in Fig.~\ref{fig:comparemeansd_2}. 

\begin{figure*}[htbp]
    \centering
    \begin{minipage}[b]{1.0\linewidth}
        \centering
        \includegraphics[trim={3cm 0.5cm 3cm 0cm},clip, width=1.0\textwidth]{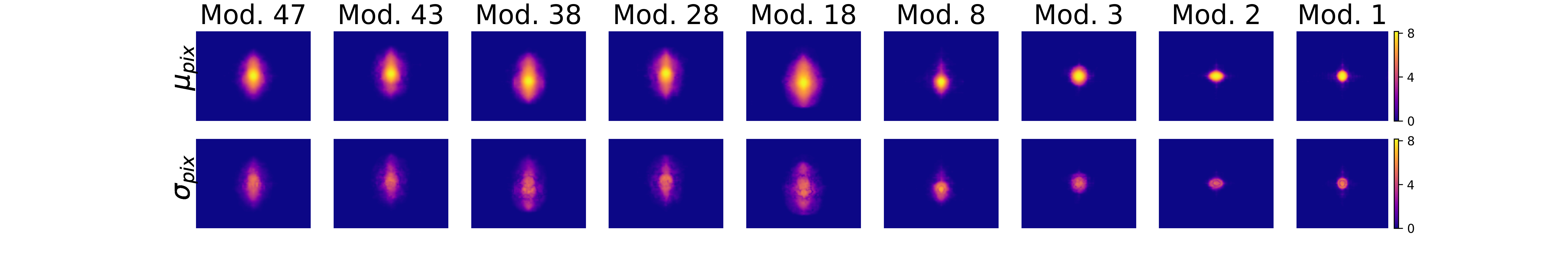}
        \subcaption{$x-y$ projection}
        \hspace{0.25cm}
    \end{minipage}
    \begin{minipage}[b]{1.0\linewidth}
        \centering
        \includegraphics[trim={3cm 0.5cm 3cm 0cm},clip,width=1.0\textwidth]{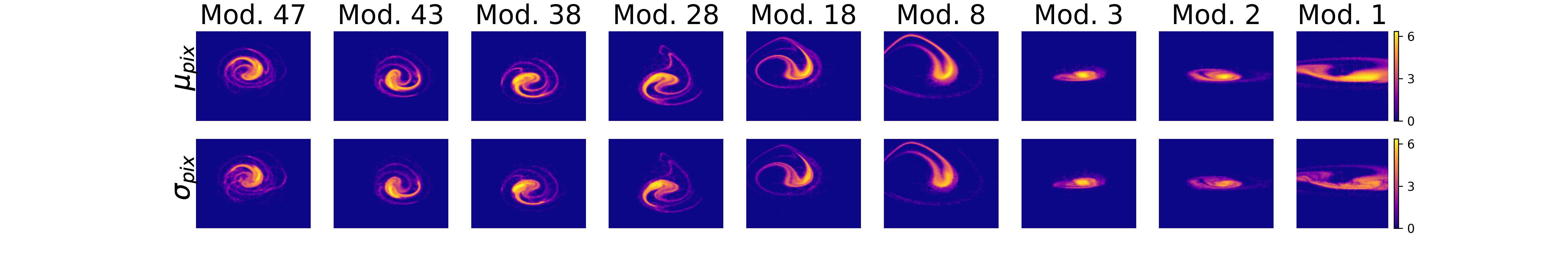}
        \subcaption{$E-\phi$ projection}
        \hspace{0.25cm}
    \end{minipage}
    \begin{minipage}[b]{1.0\linewidth}
        \centering
        \includegraphics[trim={3cm 0.5cm 3cm 0cm},clip,width=1.0\textwidth]{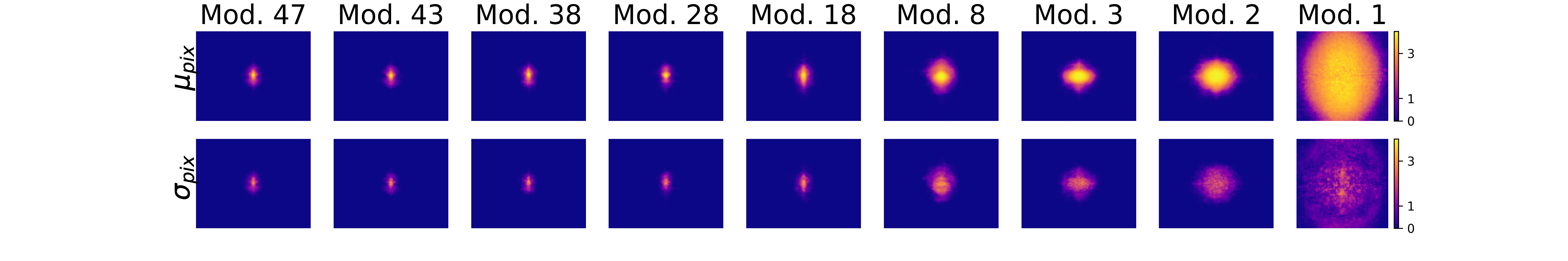}
        \subcaption{$x'-y'$ projection}
        \hspace{0.25cm}
    \end{minipage}
    \caption{Pixel-wise mean and standard deviation of phase space projections decoded from the latent space trajectories.}
    \label{fig:uncertainty_images}
\end{figure*}

We have performed uncertainty analysis in the latent space, whereas measurements and predictions exists in the high-dimensional image space. Thus, the trajectories in the latent space are projected back into the high-dimensional space using the trained decoder to obtain phase space projections in various modules. Pixel-wise mean and standard deviation for the generated projections of all true trajectories (Fig.~\ref{fig:trajectories}) are shown in Fig.~\ref{fig:uncertainty_images}. It can be observed that the pixel-wise standard deviation varies for each projection. Also, the overall pixel-wise aggregate standard deviation can be compared across different modules. Fig.~\ref{fig:comparemeansd_2}(b) (green curve) shows an increasing trend, similar to the corresponding aggregate standard deviation of the latent space trajectories (red curve).

Notable, for better visualization, the images in Fig.~\ref{fig:uncertainty_images} are plotted on a un-normalized logarithmic scale to compress the wider range of pixel intensities and enhance contrast in low-intensity areas. As a result, the standard deviation images appear brighter than they would on a normalized non-logarithmic scale,  but the aforementioned observation still holds.

Four different aggregated standard deviation discussed in this section and shown with are Fig.~\ref{fig:comparemeansd_2} are complied together in a single plot, shown in Fig.~\ref{fig:comparemeansd_all}(b). In Fig.~\ref{fig:comparemeansd_all}(a), we plot four aggregated means (arithmetic mean of means). We see that the pixel-wise aggregated mean of images (dataset images with blue curve and decoded uncertain predictions with red curve) are zero across all the modules due to the normalization of the images between [0,1] and 0 zero regions of the image dominating non-zero regions. In their latent space version, we see that the aggregated mean matches closely, however, it is not zero. It fluctuates between [-0.5, 0.5] symmetrically with the zero line. Since, the latent points for different modules are clustered at different locations in the 8D space, aggregated mean provides the distance of the mean of latent points of a particular module on a real number line. Here, we want to emphasis that the aggregated mean in the latent space gives limited comparison power with the aggregated mean in the image space, contrary to what we discussed for aggregated standard deviation. 

\section{\label{sec:conc} Conclusions}
We have developed a reverse latent evolution model, a two-step self-supervised deep learning framework designed for the temporal inversion of spatiotemporal beam dynamics in particle accelerators. By utilizing an autoregressive CVAE-LSTM architecture, our framework predicts 15 unique projections of 6D phase space across all upstream accelerating sections based on downstream measurements. We enhance interpretability by visualizing the CVAE's latent space using 2D PCA and t-SNE. The CVAE's latent space effectively captures aleatoric uncertainty inherent in high-dimensional phase space projections. Uncertain latent points sampled conditionally from downstream modules are propagated to predict upstream latent points along with their associated uncertainty bounds. We demonstrate that the LSTM is robust against random perturbations of the input. High-dimensional phase space projections corresponding to these predictions are decoded from latent space trajectories, and presented with pixel-wise uncertainty bounds.

\section*{Availability of codes and dataset}
The data used in this article is available on Zenodo at \\ \url{https://zenodo.org/records/10819001}. 

The code used in this article is available on GitHub at \\ \url{https://github.com/mahindrautela/rLEM}.

\section*{Acknowledgements}
Los Alamos National Laboratory (LANL) is operated by Triad National Security, LLC, for the national Nuclear Security Administration of the US Department of Energy (Contract No. 89233218CNA000001). This work was supported by the LANL LDRD Program Directed Research (DR) project 20220074DR.

\bibliography{apssamp}

\begin{thebibliography}{72}%
\makeatletter
\providecommand \@ifxundefined [1]{%
 \@ifx{#1\undefined}
}%
\providecommand \@ifnum [1]{%
 \ifnum #1\expandafter \@firstoftwo
 \else \expandafter \@secondoftwo
 \fi
}%
\providecommand \@ifx [1]{%
 \ifx #1\expandafter \@firstoftwo
 \else \expandafter \@secondoftwo
 \fi
}%
\providecommand \natexlab [1]{#1}%
\providecommand \enquote  [1]{``#1''}%
\providecommand \bibnamefont  [1]{#1}%
\providecommand \bibfnamefont [1]{#1}%
\providecommand \citenamefont [1]{#1}%
\providecommand \href@noop [0]{\@secondoftwo}%
\providecommand \href [0]{\begingroup \@sanitize@url \@href}%
\providecommand \@href[1]{\@@startlink{#1}\@@href}%
\providecommand \@@href[1]{\endgroup#1\@@endlink}%
\providecommand \@sanitize@url [0]{\catcode `\\12\catcode `\$12\catcode `\&12\catcode `\#12\catcode `\^12\catcode `\_12\catcode `\%12\relax}%
\providecommand \@@startlink[1]{}%
\providecommand \@@endlink[0]{}%
\providecommand \url  [0]{\begingroup\@sanitize@url \@url }%
\providecommand \@url [1]{\endgroup\@href {#1}{\urlprefix }}%
\providecommand \urlprefix  [0]{URL }%
\providecommand \Eprint [0]{\href }%
\providecommand \doibase [0]{https://doi.org/}%
\providecommand \selectlanguage [0]{\@gobble}%
\providecommand \bibinfo  [0]{\@secondoftwo}%
\providecommand \bibfield  [0]{\@secondoftwo}%
\providecommand \translation [1]{[#1]}%
\providecommand \BibitemOpen [0]{}%
\providecommand \bibitemStop [0]{}%
\providecommand \bibitemNoStop [0]{.\EOS\space}%
\providecommand \EOS [0]{\spacefactor3000\relax}%
\providecommand \BibitemShut  [1]{\csname bibitem#1\endcsname}%
\let\auto@bib@innerbib\@empty
\bibitem [{\citenamefont {Vinuesa}\ and\ \citenamefont {Brunton}(2022)}]{vinuesa2022enhancing}%
  \BibitemOpen
  \bibfield  {author} {\bibinfo {author} {\bibfnamefont {R.}~\bibnamefont {Vinuesa}}\ and\ \bibinfo {author} {\bibfnamefont {S.~L.}\ \bibnamefont {Brunton}},\ }\bibfield  {title} {\bibinfo {title} {Enhancing computational fluid dynamics with machine learning},\ }\href@noop {} {\bibfield  {journal} {\bibinfo  {journal} {Nature Computational Science}\ }\textbf {\bibinfo {volume} {2}},\ \bibinfo {pages} {358} (\bibinfo {year} {2022})}\BibitemShut {NoStop}%
\bibitem [{\citenamefont {Huerta}\ \emph {et~al.}(2021)\citenamefont {Huerta}, \citenamefont {Khan}, \citenamefont {Huang}, \citenamefont {Tian}, \citenamefont {Levental}, \citenamefont {Chard}, \citenamefont {Wei}, \citenamefont {Heflin}, \citenamefont {Katz}, \citenamefont {Kindratenko} \emph {et~al.}}]{huerta2021accelerated}%
  \BibitemOpen
  \bibfield  {author} {\bibinfo {author} {\bibfnamefont {E.}~\bibnamefont {Huerta}}, \bibinfo {author} {\bibfnamefont {A.}~\bibnamefont {Khan}}, \bibinfo {author} {\bibfnamefont {X.}~\bibnamefont {Huang}}, \bibinfo {author} {\bibfnamefont {M.}~\bibnamefont {Tian}}, \bibinfo {author} {\bibfnamefont {M.}~\bibnamefont {Levental}}, \bibinfo {author} {\bibfnamefont {R.}~\bibnamefont {Chard}}, \bibinfo {author} {\bibfnamefont {W.}~\bibnamefont {Wei}}, \bibinfo {author} {\bibfnamefont {M.}~\bibnamefont {Heflin}}, \bibinfo {author} {\bibfnamefont {D.~S.}\ \bibnamefont {Katz}}, \bibinfo {author} {\bibfnamefont {V.}~\bibnamefont {Kindratenko}}, \emph {et~al.},\ }\bibfield  {title} {\bibinfo {title} {Accelerated, scalable and reproducible ai-driven gravitational wave detection},\ }\href@noop {} {\bibfield  {journal} {\bibinfo  {journal} {Nature Astronomy}\ }\textbf {\bibinfo {volume} {5}},\ \bibinfo {pages} {1062} (\bibinfo {year} {2021})}\BibitemShut {NoStop}%
\bibitem [{\citenamefont {Scheinker}\ \emph {et~al.}(2023)\citenamefont {Scheinker}, \citenamefont {Cropp},\ and\ \citenamefont {Filippetto}}]{scheinker2023adaptive_PRE}%
  \BibitemOpen
  \bibfield  {author} {\bibinfo {author} {\bibfnamefont {A.}~\bibnamefont {Scheinker}}, \bibinfo {author} {\bibfnamefont {F.}~\bibnamefont {Cropp}},\ and\ \bibinfo {author} {\bibfnamefont {D.}~\bibnamefont {Filippetto}},\ }\bibfield  {title} {\bibinfo {title} {Adaptive autoencoder latent space tuning for more robust machine learning beyond the training set for six-dimensional phase space diagnostics of a time-varying ultrafast electron-diffraction compact accelerator},\ }\href@noop {} {\bibfield  {journal} {\bibinfo  {journal} {Physical Review E}\ }\textbf {\bibinfo {volume} {107}},\ \bibinfo {pages} {045302} (\bibinfo {year} {2023})}\BibitemShut {NoStop}%
\bibitem [{\citenamefont {Boehnlein}\ \emph {et~al.}(2022)\citenamefont {Boehnlein}, \citenamefont {Diefenthaler}, \citenamefont {Sato}, \citenamefont {Schram}, \citenamefont {Ziegler}, \citenamefont {Fanelli}, \citenamefont {Hjorth-Jensen}, \citenamefont {Horn}, \citenamefont {Kuchera}, \citenamefont {Lee} \emph {et~al.}}]{boehnlein2022colloquium}%
  \BibitemOpen
  \bibfield  {author} {\bibinfo {author} {\bibfnamefont {A.}~\bibnamefont {Boehnlein}}, \bibinfo {author} {\bibfnamefont {M.}~\bibnamefont {Diefenthaler}}, \bibinfo {author} {\bibfnamefont {N.}~\bibnamefont {Sato}}, \bibinfo {author} {\bibfnamefont {M.}~\bibnamefont {Schram}}, \bibinfo {author} {\bibfnamefont {V.}~\bibnamefont {Ziegler}}, \bibinfo {author} {\bibfnamefont {C.}~\bibnamefont {Fanelli}}, \bibinfo {author} {\bibfnamefont {M.}~\bibnamefont {Hjorth-Jensen}}, \bibinfo {author} {\bibfnamefont {T.}~\bibnamefont {Horn}}, \bibinfo {author} {\bibfnamefont {M.~P.}\ \bibnamefont {Kuchera}}, \bibinfo {author} {\bibfnamefont {D.}~\bibnamefont {Lee}}, \emph {et~al.},\ }\bibfield  {title} {\bibinfo {title} {Colloquium: Machine learning in nuclear physics},\ }\href@noop {} {\bibfield  {journal} {\bibinfo  {journal} {Reviews of Modern Physics}\ }\textbf {\bibinfo {volume} {94}},\ \bibinfo {pages} {031003} (\bibinfo {year} {2022})}\BibitemShut {NoStop}%
\bibitem [{\citenamefont {Bormanis}\ \emph {et~al.}(2024)\citenamefont {Bormanis}, \citenamefont {Leon},\ and\ \citenamefont {Scheinker}}]{bormanis2024solving}%
  \BibitemOpen
  \bibfield  {author} {\bibinfo {author} {\bibfnamefont {A.}~\bibnamefont {Bormanis}}, \bibinfo {author} {\bibfnamefont {C.~A.}\ \bibnamefont {Leon}},\ and\ \bibinfo {author} {\bibfnamefont {A.}~\bibnamefont {Scheinker}},\ }\bibfield  {title} {\bibinfo {title} {Solving the orszag--tang vortex magnetohydrodynamics problem with physics-constrained convolutional neural networks},\ }\href@noop {} {\bibfield  {journal} {\bibinfo  {journal} {Physics of Plasmas}\ }\textbf {\bibinfo {volume} {31}} (\bibinfo {year} {2024})}\BibitemShut {NoStop}%
\bibitem [{\citenamefont {Morison}\ \emph {et~al.}(2024)\citenamefont {Morison}, \citenamefont {Singh}, \citenamefont {Al~Kayed}, \citenamefont {Aadhi}, \citenamefont {Moridsadat}, \citenamefont {Tamura}, \citenamefont {Tait},\ and\ \citenamefont {Shastri}}]{morison2024nonlinear}%
  \BibitemOpen
  \bibfield  {author} {\bibinfo {author} {\bibfnamefont {H.}~\bibnamefont {Morison}}, \bibinfo {author} {\bibfnamefont {J.}~\bibnamefont {Singh}}, \bibinfo {author} {\bibfnamefont {N.}~\bibnamefont {Al~Kayed}}, \bibinfo {author} {\bibfnamefont {A.}~\bibnamefont {Aadhi}}, \bibinfo {author} {\bibfnamefont {M.}~\bibnamefont {Moridsadat}}, \bibinfo {author} {\bibfnamefont {M.}~\bibnamefont {Tamura}}, \bibinfo {author} {\bibfnamefont {A.~N.}\ \bibnamefont {Tait}},\ and\ \bibinfo {author} {\bibfnamefont {B.~J.}\ \bibnamefont {Shastri}},\ }\bibfield  {title} {\bibinfo {title} {Nonlinear dynamics in neuromorphic photonic networks: Physical simulation in verilog-a},\ }\href@noop {} {\bibfield  {journal} {\bibinfo  {journal} {Physical Review Applied}\ }\textbf {\bibinfo {volume} {21}},\ \bibinfo {pages} {034013} (\bibinfo {year} {2024})}\BibitemShut {NoStop}%
\bibitem [{\citenamefont {Kipf}\ and\ \citenamefont {Welling}(2016)}]{kipf2016semi}%
  \BibitemOpen
  \bibfield  {author} {\bibinfo {author} {\bibfnamefont {T.~N.}\ \bibnamefont {Kipf}}\ and\ \bibinfo {author} {\bibfnamefont {M.}~\bibnamefont {Welling}},\ }\bibfield  {title} {\bibinfo {title} {Semi-supervised classification with graph convolutional networks},\ }\href@noop {} {\bibfield  {journal} {\bibinfo  {journal} {arXiv preprint arXiv:1609.02907}\ } (\bibinfo {year} {2016})}\BibitemShut {NoStop}%
\bibitem [{\citenamefont {Shi}\ \emph {et~al.}(2015)\citenamefont {Shi}, \citenamefont {Chen}, \citenamefont {Wang}, \citenamefont {Yeung}, \citenamefont {Wong},\ and\ \citenamefont {Woo}}]{shi2015convolutional}%
  \BibitemOpen
  \bibfield  {author} {\bibinfo {author} {\bibfnamefont {X.}~\bibnamefont {Shi}}, \bibinfo {author} {\bibfnamefont {Z.}~\bibnamefont {Chen}}, \bibinfo {author} {\bibfnamefont {H.}~\bibnamefont {Wang}}, \bibinfo {author} {\bibfnamefont {D.-Y.}\ \bibnamefont {Yeung}}, \bibinfo {author} {\bibfnamefont {W.-K.}\ \bibnamefont {Wong}},\ and\ \bibinfo {author} {\bibfnamefont {W.-c.}\ \bibnamefont {Woo}},\ }\bibfield  {title} {\bibinfo {title} {Convolutional lstm network: A machine learning approach for precipitation nowcasting},\ }\href@noop {} {\bibfield  {journal} {\bibinfo  {journal} {Advances in neural information processing systems}\ }\textbf {\bibinfo {volume} {28}} (\bibinfo {year} {2015})}\BibitemShut {NoStop}%
\bibitem [{\citenamefont {Cheng}\ \emph {et~al.}(2020)\citenamefont {Cheng}, \citenamefont {Fang}, \citenamefont {Pain},\ and\ \citenamefont {Navon}}]{cheng2020data}%
  \BibitemOpen
  \bibfield  {author} {\bibinfo {author} {\bibfnamefont {M.}~\bibnamefont {Cheng}}, \bibinfo {author} {\bibfnamefont {F.}~\bibnamefont {Fang}}, \bibinfo {author} {\bibfnamefont {C.~C.}\ \bibnamefont {Pain}},\ and\ \bibinfo {author} {\bibfnamefont {I.}~\bibnamefont {Navon}},\ }\bibfield  {title} {\bibinfo {title} {Data-driven modelling of nonlinear spatio-temporal fluid flows using a deep convolutional generative adversarial network},\ }\href@noop {} {\bibfield  {journal} {\bibinfo  {journal} {Computer Methods in Applied Mechanics and Engineering}\ }\textbf {\bibinfo {volume} {365}},\ \bibinfo {pages} {113000} (\bibinfo {year} {2020})}\BibitemShut {NoStop}%
\bibitem [{\citenamefont {Wandel}\ \emph {et~al.}(2021)\citenamefont {Wandel}, \citenamefont {Weinmann},\ and\ \citenamefont {Klein}}]{wandel2021teaching}%
  \BibitemOpen
  \bibfield  {author} {\bibinfo {author} {\bibfnamefont {N.}~\bibnamefont {Wandel}}, \bibinfo {author} {\bibfnamefont {M.}~\bibnamefont {Weinmann}},\ and\ \bibinfo {author} {\bibfnamefont {R.}~\bibnamefont {Klein}},\ }\bibfield  {title} {\bibinfo {title} {Teaching the incompressible navier--stokes equations to fast neural surrogate models in three dimensions},\ }\href@noop {} {\bibfield  {journal} {\bibinfo  {journal} {Physics of Fluids}\ }\textbf {\bibinfo {volume} {33}} (\bibinfo {year} {2021})}\BibitemShut {NoStop}%
\bibitem [{\citenamefont {Wiewel}\ \emph {et~al.}(2019)\citenamefont {Wiewel}, \citenamefont {Becher},\ and\ \citenamefont {Thuerey}}]{wiewel2019latent}%
  \BibitemOpen
  \bibfield  {author} {\bibinfo {author} {\bibfnamefont {S.}~\bibnamefont {Wiewel}}, \bibinfo {author} {\bibfnamefont {M.}~\bibnamefont {Becher}},\ and\ \bibinfo {author} {\bibfnamefont {N.}~\bibnamefont {Thuerey}},\ }\bibfield  {title} {\bibinfo {title} {Latent space physics: Towards learning the temporal evolution of fluid flow},\ }in\ \href@noop {} {\emph {\bibinfo {booktitle} {Computer graphics forum}}},\ Vol.~\bibinfo {volume} {38}\ (\bibinfo {organization} {Wiley Online Library},\ \bibinfo {year} {2019})\ pp.\ \bibinfo {pages} {71--82}\BibitemShut {NoStop}%
\bibitem [{\citenamefont {Scheinker}(2021)}]{scheinker2021adaptive_JOI}%
  \BibitemOpen
  \bibfield  {author} {\bibinfo {author} {\bibfnamefont {A.}~\bibnamefont {Scheinker}},\ }\bibfield  {title} {\bibinfo {title} {Adaptive machine learning for time-varying systems: low dimensional latent space tuning},\ }\href@noop {} {\bibfield  {journal} {\bibinfo  {journal} {Journal of Instrumentation}\ }\textbf {\bibinfo {volume} {16}}\bibinfo  {number} { (10)},\ \bibinfo {pages} {P10008}}\BibitemShut {NoStop}%
\bibitem [{\citenamefont {Montes~de Oca~Zapiain}\ \emph {et~al.}(2021)\citenamefont {Montes~de Oca~Zapiain}, \citenamefont {Stewart},\ and\ \citenamefont {Dingreville}}]{montes2021accelerating}%
  \BibitemOpen
\bibfield  {number} {  }\bibfield  {author} {\bibinfo {author} {\bibfnamefont {D.}~\bibnamefont {Montes~de Oca~Zapiain}}, \bibinfo {author} {\bibfnamefont {J.~A.}\ \bibnamefont {Stewart}},\ and\ \bibinfo {author} {\bibfnamefont {R.}~\bibnamefont {Dingreville}},\ }\bibfield  {title} {\bibinfo {title} {Accelerating phase-field-based microstructure evolution predictions via surrogate models trained by machine learning methods},\ }\href@noop {} {\bibfield  {journal} {\bibinfo  {journal} {npj Computational Materials}\ }\textbf {\bibinfo {volume} {7}},\ \bibinfo {pages} {3} (\bibinfo {year} {2021})}\BibitemShut {NoStop}%
\bibitem [{\citenamefont {Nakamura}\ \emph {et~al.}(2021)\citenamefont {Nakamura}, \citenamefont {Fukami}, \citenamefont {Hasegawa}, \citenamefont {Nabae},\ and\ \citenamefont {Fukagata}}]{nakamura2021convolutional}%
  \BibitemOpen
  \bibfield  {author} {\bibinfo {author} {\bibfnamefont {T.}~\bibnamefont {Nakamura}}, \bibinfo {author} {\bibfnamefont {K.}~\bibnamefont {Fukami}}, \bibinfo {author} {\bibfnamefont {K.}~\bibnamefont {Hasegawa}}, \bibinfo {author} {\bibfnamefont {Y.}~\bibnamefont {Nabae}},\ and\ \bibinfo {author} {\bibfnamefont {K.}~\bibnamefont {Fukagata}},\ }\bibfield  {title} {\bibinfo {title} {Convolutional neural network and long short-term memory based reduced order surrogate for minimal turbulent channel flow},\ }\href@noop {} {\bibfield  {journal} {\bibinfo  {journal} {Physics of Fluids}\ }\textbf {\bibinfo {volume} {33}} (\bibinfo {year} {2021})}\BibitemShut {NoStop}%
\bibitem [{\citenamefont {Maulik}\ \emph {et~al.}(2021)\citenamefont {Maulik}, \citenamefont {Lusch},\ and\ \citenamefont {Balaprakash}}]{maulik2021reduced}%
  \BibitemOpen
  \bibfield  {author} {\bibinfo {author} {\bibfnamefont {R.}~\bibnamefont {Maulik}}, \bibinfo {author} {\bibfnamefont {B.}~\bibnamefont {Lusch}},\ and\ \bibinfo {author} {\bibfnamefont {P.}~\bibnamefont {Balaprakash}},\ }\bibfield  {title} {\bibinfo {title} {Reduced-order modeling of advection-dominated systems with recurrent neural networks and convolutional autoencoders},\ }\href@noop {} {\bibfield  {journal} {\bibinfo  {journal} {Physics of Fluids}\ }\textbf {\bibinfo {volume} {33}} (\bibinfo {year} {2021})}\BibitemShut {NoStop}%
\bibitem [{\citenamefont {Vlachas}\ \emph {et~al.}(2022)\citenamefont {Vlachas}, \citenamefont {Arampatzis}, \citenamefont {Uhler},\ and\ \citenamefont {Koumoutsakos}}]{vlachas2022multiscale}%
  \BibitemOpen
  \bibfield  {author} {\bibinfo {author} {\bibfnamefont {P.~R.}\ \bibnamefont {Vlachas}}, \bibinfo {author} {\bibfnamefont {G.}~\bibnamefont {Arampatzis}}, \bibinfo {author} {\bibfnamefont {C.}~\bibnamefont {Uhler}},\ and\ \bibinfo {author} {\bibfnamefont {P.}~\bibnamefont {Koumoutsakos}},\ }\bibfield  {title} {\bibinfo {title} {Multiscale simulations of complex systems by learning their effective dynamics},\ }\href@noop {} {\bibfield  {journal} {\bibinfo  {journal} {Nature Machine Intelligence}\ }\textbf {\bibinfo {volume} {4}},\ \bibinfo {pages} {359} (\bibinfo {year} {2022})}\BibitemShut {NoStop}%
\bibitem [{\citenamefont {Solera-Rico}\ \emph {et~al.}(2024)\citenamefont {Solera-Rico}, \citenamefont {Sanmiguel~Vila}, \citenamefont {G{\'o}mez-L{\'o}pez}, \citenamefont {Wang}, \citenamefont {Almashjary}, \citenamefont {Dawson},\ and\ \citenamefont {Vinuesa}}]{solera2024beta}%
  \BibitemOpen
  \bibfield  {author} {\bibinfo {author} {\bibfnamefont {A.}~\bibnamefont {Solera-Rico}}, \bibinfo {author} {\bibfnamefont {C.}~\bibnamefont {Sanmiguel~Vila}}, \bibinfo {author} {\bibfnamefont {M.}~\bibnamefont {G{\'o}mez-L{\'o}pez}}, \bibinfo {author} {\bibfnamefont {Y.}~\bibnamefont {Wang}}, \bibinfo {author} {\bibfnamefont {A.}~\bibnamefont {Almashjary}}, \bibinfo {author} {\bibfnamefont {S.~T.}\ \bibnamefont {Dawson}},\ and\ \bibinfo {author} {\bibfnamefont {R.}~\bibnamefont {Vinuesa}},\ }\bibfield  {title} {\bibinfo {title} {$\beta$-variational autoencoders and transformers for reduced-order modelling of fluid flows},\ }\href@noop {} {\bibfield  {journal} {\bibinfo  {journal} {Nature Communications}\ }\textbf {\bibinfo {volume} {15}},\ \bibinfo {pages} {1361} (\bibinfo {year} {2024})}\BibitemShut {NoStop}%
\bibitem [{\citenamefont {Rautela}\ \emph {et~al.}(2024{\natexlab{a}})\citenamefont {Rautela}, \citenamefont {Williams},\ and\ \citenamefont {Scheinker}}]{rautela2024conditional}%
  \BibitemOpen
  \bibfield  {author} {\bibinfo {author} {\bibfnamefont {M.}~\bibnamefont {Rautela}}, \bibinfo {author} {\bibfnamefont {A.}~\bibnamefont {Williams}},\ and\ \bibinfo {author} {\bibfnamefont {A.}~\bibnamefont {Scheinker}},\ }\bibfield  {title} {\bibinfo {title} {A conditional latent autoregressive recurrent model for generation and forecasting of beam dynamics in particle accelerators},\ }\href@noop {} {\bibfield  {journal} {\bibinfo  {journal} {Scientific Reports}\ }\textbf {\bibinfo {volume} {14}},\ \bibinfo {pages} {18157} (\bibinfo {year} {2024}{\natexlab{a}})}\BibitemShut {NoStop}%
\bibitem [{\citenamefont {Rautela}\ \emph {et~al.}(2024{\natexlab{b}})\citenamefont {Rautela}, \citenamefont {Scheinker},\ and\ \citenamefont {Williams}}]{rautela24towards}%
  \BibitemOpen
  \bibfield  {author} {\bibinfo {author} {\bibfnamefont {M.}~\bibnamefont {Rautela}}, \bibinfo {author} {\bibfnamefont {A.}~\bibnamefont {Scheinker}},\ and\ \bibinfo {author} {\bibfnamefont {A.}~\bibnamefont {Williams}},\ }\bibfield  {title} {\bibinfo {title} {Towards latent space evolution of spatiotemporal dynamics of six-dimensional phase space of charged particle beams},\ }\href@noop {} {\bibfield  {journal} {\bibinfo  {journal} {Proc. IPAC'24}\ ,\ \bibinfo {pages} {906}} (\bibinfo {year} {2024}{\natexlab{b}})}\BibitemShut {NoStop}%
\bibitem [{\citenamefont {Scheinker}\ \emph {et~al.}(2018)\citenamefont {Scheinker}, \citenamefont {Edelen}, \citenamefont {Bohler}, \citenamefont {Emma},\ and\ \citenamefont {Lutman}}]{scheinker2018demonstration}%
  \BibitemOpen
  \bibfield  {author} {\bibinfo {author} {\bibfnamefont {A.}~\bibnamefont {Scheinker}}, \bibinfo {author} {\bibfnamefont {A.}~\bibnamefont {Edelen}}, \bibinfo {author} {\bibfnamefont {D.}~\bibnamefont {Bohler}}, \bibinfo {author} {\bibfnamefont {C.}~\bibnamefont {Emma}},\ and\ \bibinfo {author} {\bibfnamefont {A.}~\bibnamefont {Lutman}},\ }\bibfield  {title} {\bibinfo {title} {Demonstration of model-independent control of the longitudinal phase space of electron beams in the linac-coherent light source with femtosecond resolution},\ }\href@noop {} {\bibfield  {journal} {\bibinfo  {journal} {Physical review letters}\ }\textbf {\bibinfo {volume} {121}},\ \bibinfo {pages} {044801} (\bibinfo {year} {2018})}\BibitemShut {NoStop}%
\bibitem [{\citenamefont {Emma}\ \emph {et~al.}(2018)\citenamefont {Emma}, \citenamefont {Edelen}, \citenamefont {Hogan}, \citenamefont {O’Shea}, \citenamefont {White},\ and\ \citenamefont {Yakimenko}}]{emma2018machine}%
  \BibitemOpen
  \bibfield  {author} {\bibinfo {author} {\bibfnamefont {C.}~\bibnamefont {Emma}}, \bibinfo {author} {\bibfnamefont {A.}~\bibnamefont {Edelen}}, \bibinfo {author} {\bibfnamefont {M.}~\bibnamefont {Hogan}}, \bibinfo {author} {\bibfnamefont {B.}~\bibnamefont {O’Shea}}, \bibinfo {author} {\bibfnamefont {G.}~\bibnamefont {White}},\ and\ \bibinfo {author} {\bibfnamefont {V.}~\bibnamefont {Yakimenko}},\ }\bibfield  {title} {\bibinfo {title} {Machine learning-based longitudinal phase space prediction of particle accelerators},\ }\href@noop {} {\bibfield  {journal} {\bibinfo  {journal} {Physical Review Accelerators and Beams}\ }\textbf {\bibinfo {volume} {21}},\ \bibinfo {pages} {112802} (\bibinfo {year} {2018})}\BibitemShut {NoStop}%
\bibitem [{\citenamefont {Ivanov}\ and\ \citenamefont {Agapov}(2020)}]{ivanov2020physics}%
  \BibitemOpen
  \bibfield  {author} {\bibinfo {author} {\bibfnamefont {A.}~\bibnamefont {Ivanov}}\ and\ \bibinfo {author} {\bibfnamefont {I.}~\bibnamefont {Agapov}},\ }\bibfield  {title} {\bibinfo {title} {Physics-based deep neural networks for beam dynamics in charged particle accelerators},\ }\href@noop {} {\bibfield  {journal} {\bibinfo  {journal} {Physical review accelerators and beams}\ }\textbf {\bibinfo {volume} {23}},\ \bibinfo {pages} {074601} (\bibinfo {year} {2020})}\BibitemShut {NoStop}%
\bibitem [{\citenamefont {Caliari}\ \emph {et~al.}(2023)\citenamefont {Caliari}, \citenamefont {Oeftiger},\ and\ \citenamefont {Boine-Frankenheim}}]{caliari2023identification}%
  \BibitemOpen
  \bibfield  {author} {\bibinfo {author} {\bibfnamefont {C.}~\bibnamefont {Caliari}}, \bibinfo {author} {\bibfnamefont {A.}~\bibnamefont {Oeftiger}},\ and\ \bibinfo {author} {\bibfnamefont {O.}~\bibnamefont {Boine-Frankenheim}},\ }\bibfield  {title} {\bibinfo {title} {Identification of magnetic field errors in synchrotrons based on deep {L}ie map networks},\ }\href@noop {} {\bibfield  {journal} {\bibinfo  {journal} {Physical Review Accelerators and Beams}\ }\textbf {\bibinfo {volume} {26}},\ \bibinfo {pages} {064601} (\bibinfo {year} {2023})}\BibitemShut {NoStop}%
\bibitem [{\citenamefont {Zhu}\ \emph {et~al.}(2021)\citenamefont {Zhu}, \citenamefont {Chen}, \citenamefont {Brinker}, \citenamefont {Decking}, \citenamefont {Tomin},\ and\ \citenamefont {Schlarb}}]{zhu2021high}%
  \BibitemOpen
  \bibfield  {author} {\bibinfo {author} {\bibfnamefont {J.}~\bibnamefont {Zhu}}, \bibinfo {author} {\bibfnamefont {Y.}~\bibnamefont {Chen}}, \bibinfo {author} {\bibfnamefont {F.}~\bibnamefont {Brinker}}, \bibinfo {author} {\bibfnamefont {W.}~\bibnamefont {Decking}}, \bibinfo {author} {\bibfnamefont {S.}~\bibnamefont {Tomin}},\ and\ \bibinfo {author} {\bibfnamefont {H.}~\bibnamefont {Schlarb}},\ }\bibfield  {title} {\bibinfo {title} {High-fidelity prediction of megapixel longitudinal phase-space images of electron beams using encoder-decoder neural networks},\ }\href@noop {} {\bibfield  {journal} {\bibinfo  {journal} {Physical Review Applied}\ }\textbf {\bibinfo {volume} {16}},\ \bibinfo {pages} {024005} (\bibinfo {year} {2021})}\BibitemShut {NoStop}%
\bibitem [{\citenamefont {Mayet}\ \emph {et~al.}(2022)\citenamefont {Mayet}, \citenamefont {Hachmann}, \citenamefont {Floettmann}, \citenamefont {Burkart}, \citenamefont {Dinter}, \citenamefont {Kuropka}, \citenamefont {Vinatier},\ and\ \citenamefont {Assmann}}]{mayet2022predicting}%
  \BibitemOpen
  \bibfield  {author} {\bibinfo {author} {\bibfnamefont {F.}~\bibnamefont {Mayet}}, \bibinfo {author} {\bibfnamefont {M.}~\bibnamefont {Hachmann}}, \bibinfo {author} {\bibfnamefont {K.}~\bibnamefont {Floettmann}}, \bibinfo {author} {\bibfnamefont {F.}~\bibnamefont {Burkart}}, \bibinfo {author} {\bibfnamefont {H.}~\bibnamefont {Dinter}}, \bibinfo {author} {\bibfnamefont {W.}~\bibnamefont {Kuropka}}, \bibinfo {author} {\bibfnamefont {T.}~\bibnamefont {Vinatier}},\ and\ \bibinfo {author} {\bibfnamefont {R.}~\bibnamefont {Assmann}},\ }\bibfield  {title} {\bibinfo {title} {Predicting the transverse emittance of space charge dominated beams using the phase advance scan technique and a fully connected neural network},\ }\href@noop {} {\bibfield  {journal} {\bibinfo  {journal} {Physical Review Accelerators and Beams}\ }\textbf {\bibinfo {volume} {25}},\ \bibinfo {pages} {094601} (\bibinfo {year} {2022})}\BibitemShut {NoStop}%
\bibitem [{\citenamefont {Cropp}\ \emph {et~al.}(2023)\citenamefont {Cropp}, \citenamefont {Moos}, \citenamefont {Scheinker}, \citenamefont {Gilardi}, \citenamefont {Wang}, \citenamefont {Paiagua}, \citenamefont {Serrano}, \citenamefont {Musumeci},\ and\ \citenamefont {Filippetto}}]{cropp2023virtual}%
  \BibitemOpen
  \bibfield  {author} {\bibinfo {author} {\bibfnamefont {F.}~\bibnamefont {Cropp}}, \bibinfo {author} {\bibfnamefont {L.}~\bibnamefont {Moos}}, \bibinfo {author} {\bibfnamefont {A.}~\bibnamefont {Scheinker}}, \bibinfo {author} {\bibfnamefont {A.}~\bibnamefont {Gilardi}}, \bibinfo {author} {\bibfnamefont {D.}~\bibnamefont {Wang}}, \bibinfo {author} {\bibfnamefont {S.}~\bibnamefont {Paiagua}}, \bibinfo {author} {\bibfnamefont {C.}~\bibnamefont {Serrano}}, \bibinfo {author} {\bibfnamefont {P.}~\bibnamefont {Musumeci}},\ and\ \bibinfo {author} {\bibfnamefont {D.}~\bibnamefont {Filippetto}},\ }\bibfield  {title} {\bibinfo {title} {Virtual-diagnostic-based time stamping for ultrafast electron diffraction},\ }\href@noop {} {\bibfield  {journal} {\bibinfo  {journal} {Physical Review Accelerators and Beams}\ }\textbf {\bibinfo {volume} {26}},\ \bibinfo {pages} {052801} (\bibinfo {year} {2023})}\BibitemShut {NoStop}%
\bibitem [{\citenamefont {Tran}\ \emph {et~al.}(2022)\citenamefont {Tran}, \citenamefont {Hao}, \citenamefont {Mustapha},\ and\ \citenamefont {Martinez~Marin}}]{tran2022predicting}%
  \BibitemOpen
  \bibfield  {author} {\bibinfo {author} {\bibfnamefont {A.}~\bibnamefont {Tran}}, \bibinfo {author} {\bibfnamefont {Y.}~\bibnamefont {Hao}}, \bibinfo {author} {\bibfnamefont {B.}~\bibnamefont {Mustapha}},\ and\ \bibinfo {author} {\bibfnamefont {J.~L.}\ \bibnamefont {Martinez~Marin}},\ }\bibfield  {title} {\bibinfo {title} {Predicting beam transmission using 2-dimensional phase space projections of hadron accelerators},\ }\href@noop {} {\bibfield  {journal} {\bibinfo  {journal} {Frontiers in Physics}\ }\textbf {\bibinfo {volume} {10}},\ \bibinfo {pages} {955555} (\bibinfo {year} {2022})}\BibitemShut {NoStop}%
\bibitem [{\citenamefont {Duris}\ \emph {et~al.}(2020)\citenamefont {Duris}, \citenamefont {Kennedy}, \citenamefont {Hanuka}, \citenamefont {Shtalenkova}, \citenamefont {Edelen}, \citenamefont {Baxevanis}, \citenamefont {Egger}, \citenamefont {Cope}, \citenamefont {McIntire}, \citenamefont {Ermon} \emph {et~al.}}]{duris2020bayesian}%
  \BibitemOpen
  \bibfield  {author} {\bibinfo {author} {\bibfnamefont {J.}~\bibnamefont {Duris}}, \bibinfo {author} {\bibfnamefont {D.}~\bibnamefont {Kennedy}}, \bibinfo {author} {\bibfnamefont {A.}~\bibnamefont {Hanuka}}, \bibinfo {author} {\bibfnamefont {J.}~\bibnamefont {Shtalenkova}}, \bibinfo {author} {\bibfnamefont {A.}~\bibnamefont {Edelen}}, \bibinfo {author} {\bibfnamefont {P.}~\bibnamefont {Baxevanis}}, \bibinfo {author} {\bibfnamefont {A.}~\bibnamefont {Egger}}, \bibinfo {author} {\bibfnamefont {T.}~\bibnamefont {Cope}}, \bibinfo {author} {\bibfnamefont {M.}~\bibnamefont {McIntire}}, \bibinfo {author} {\bibfnamefont {S.}~\bibnamefont {Ermon}}, \emph {et~al.},\ }\bibfield  {title} {\bibinfo {title} {Bayesian optimization of a free-electron laser},\ }\href@noop {} {\bibfield  {journal} {\bibinfo  {journal} {Physical review letters}\ }\textbf {\bibinfo {volume} {124}},\ \bibinfo {pages} {124801} (\bibinfo {year} {2020})}\BibitemShut {NoStop}%
\bibitem [{\citenamefont {Jalas}\ \emph {et~al.}(2023)\citenamefont {Jalas}, \citenamefont {Kirchen}, \citenamefont {Braun}, \citenamefont {Eichner}, \citenamefont {Gonzalez}, \citenamefont {H{\"u}bner}, \citenamefont {H{\"u}lsenbusch}, \citenamefont {Messner}, \citenamefont {Palmer}, \citenamefont {Schnepp} \emph {et~al.}}]{jalas2023tuning}%
  \BibitemOpen
  \bibfield  {author} {\bibinfo {author} {\bibfnamefont {S.}~\bibnamefont {Jalas}}, \bibinfo {author} {\bibfnamefont {M.}~\bibnamefont {Kirchen}}, \bibinfo {author} {\bibfnamefont {C.}~\bibnamefont {Braun}}, \bibinfo {author} {\bibfnamefont {T.}~\bibnamefont {Eichner}}, \bibinfo {author} {\bibfnamefont {J.}~\bibnamefont {Gonzalez}}, \bibinfo {author} {\bibfnamefont {L.}~\bibnamefont {H{\"u}bner}}, \bibinfo {author} {\bibfnamefont {T.}~\bibnamefont {H{\"u}lsenbusch}}, \bibinfo {author} {\bibfnamefont {P.}~\bibnamefont {Messner}}, \bibinfo {author} {\bibfnamefont {G.}~\bibnamefont {Palmer}}, \bibinfo {author} {\bibfnamefont {M.}~\bibnamefont {Schnepp}}, \emph {et~al.},\ }\bibfield  {title} {\bibinfo {title} {Tuning curves for a laser-plasma accelerator},\ }\href@noop {} {\bibfield  {journal} {\bibinfo  {journal} {Physical Review Accelerators and Beams}\ }\textbf {\bibinfo {volume} {26}},\ \bibinfo {pages} {071302} (\bibinfo {year} {2023})}\BibitemShut {NoStop}%
\bibitem [{\citenamefont {Kirschner}\ \emph {et~al.}(2022)\citenamefont {Kirschner}, \citenamefont {Mutn{\`y}}, \citenamefont {Krause}, \citenamefont {de~Portugal}, \citenamefont {Hiller},\ and\ \citenamefont {Snuverink}}]{kirschner2022tuning}%
  \BibitemOpen
  \bibfield  {author} {\bibinfo {author} {\bibfnamefont {J.}~\bibnamefont {Kirschner}}, \bibinfo {author} {\bibfnamefont {M.}~\bibnamefont {Mutn{\`y}}}, \bibinfo {author} {\bibfnamefont {A.}~\bibnamefont {Krause}}, \bibinfo {author} {\bibfnamefont {J.~C.}\ \bibnamefont {de~Portugal}}, \bibinfo {author} {\bibfnamefont {N.}~\bibnamefont {Hiller}},\ and\ \bibinfo {author} {\bibfnamefont {J.}~\bibnamefont {Snuverink}},\ }\bibfield  {title} {\bibinfo {title} {Tuning particle accelerators with safety constraints using bayesian optimization},\ }\href@noop {} {\bibfield  {journal} {\bibinfo  {journal} {Physical Review Accelerators and Beams}\ }\textbf {\bibinfo {volume} {25}},\ \bibinfo {pages} {062802} (\bibinfo {year} {2022})}\BibitemShut {NoStop}%
\bibitem [{\citenamefont {Breckwoldt}\ \emph {et~al.}(2023)\citenamefont {Breckwoldt}, \citenamefont {Son}, \citenamefont {Mazza}, \citenamefont {R{\"o}rig}, \citenamefont {Boll}, \citenamefont {Meyer}, \citenamefont {LaForge}, \citenamefont {Mishra}, \citenamefont {Berrah}, \citenamefont {Santra} \emph {et~al.}}]{breckwoldt2023machine}%
  \BibitemOpen
  \bibfield  {author} {\bibinfo {author} {\bibfnamefont {N.}~\bibnamefont {Breckwoldt}}, \bibinfo {author} {\bibfnamefont {S.-K.}\ \bibnamefont {Son}}, \bibinfo {author} {\bibfnamefont {T.}~\bibnamefont {Mazza}}, \bibinfo {author} {\bibfnamefont {A.}~\bibnamefont {R{\"o}rig}}, \bibinfo {author} {\bibfnamefont {R.}~\bibnamefont {Boll}}, \bibinfo {author} {\bibfnamefont {M.}~\bibnamefont {Meyer}}, \bibinfo {author} {\bibfnamefont {A.~C.}\ \bibnamefont {LaForge}}, \bibinfo {author} {\bibfnamefont {D.}~\bibnamefont {Mishra}}, \bibinfo {author} {\bibfnamefont {N.}~\bibnamefont {Berrah}}, \bibinfo {author} {\bibfnamefont {R.}~\bibnamefont {Santra}}, \emph {et~al.},\ }\bibfield  {title} {\bibinfo {title} {Machine-learning calibration of intense {X}-ray free-electron-laser pulses using {B}ayesian optimization},\ }\href@noop {} {\bibfield  {journal} {\bibinfo  {journal} {Physical Review Research}\ }\textbf {\bibinfo {volume} {5}},\ \bibinfo {pages} {023114} (\bibinfo {year} {2023})}\BibitemShut {NoStop}%
\bibitem [{\citenamefont {Ji}\ \emph {et~al.}(2024)\citenamefont {Ji}, \citenamefont {Edelen}, \citenamefont {Roussel}, \citenamefont {Shen}, \citenamefont {Miskovich}, \citenamefont {Weathersby}, \citenamefont {Luo}, \citenamefont {Mo}, \citenamefont {Kramer}, \citenamefont {Mayes} \emph {et~al.}}]{ji2024multi}%
  \BibitemOpen
  \bibfield  {author} {\bibinfo {author} {\bibfnamefont {F.}~\bibnamefont {Ji}}, \bibinfo {author} {\bibfnamefont {A.}~\bibnamefont {Edelen}}, \bibinfo {author} {\bibfnamefont {R.}~\bibnamefont {Roussel}}, \bibinfo {author} {\bibfnamefont {X.}~\bibnamefont {Shen}}, \bibinfo {author} {\bibfnamefont {S.}~\bibnamefont {Miskovich}}, \bibinfo {author} {\bibfnamefont {S.}~\bibnamefont {Weathersby}}, \bibinfo {author} {\bibfnamefont {D.}~\bibnamefont {Luo}}, \bibinfo {author} {\bibfnamefont {M.}~\bibnamefont {Mo}}, \bibinfo {author} {\bibfnamefont {P.}~\bibnamefont {Kramer}}, \bibinfo {author} {\bibfnamefont {C.}~\bibnamefont {Mayes}}, \emph {et~al.},\ }\bibfield  {title} {\bibinfo {title} {Multi-objective bayesian active learning for mev-ultrafast electron diffraction},\ }\href@noop {} {\bibfield  {journal} {\bibinfo  {journal} {Nature Communications}\ }\textbf {\bibinfo {volume} {15}},\ \bibinfo {pages} {4726} (\bibinfo {year} {2024})}\BibitemShut {NoStop}%
\bibitem [{\citenamefont {Kain}\ \emph {et~al.}(2020)\citenamefont {Kain}, \citenamefont {Hirlander}, \citenamefont {Goddard}, \citenamefont {Velotti}, \citenamefont {Della~Porta}, \citenamefont {Bruchon},\ and\ \citenamefont {Valentino}}]{kain2020sample}%
  \BibitemOpen
  \bibfield  {author} {\bibinfo {author} {\bibfnamefont {V.}~\bibnamefont {Kain}}, \bibinfo {author} {\bibfnamefont {S.}~\bibnamefont {Hirlander}}, \bibinfo {author} {\bibfnamefont {B.}~\bibnamefont {Goddard}}, \bibinfo {author} {\bibfnamefont {F.~M.}\ \bibnamefont {Velotti}}, \bibinfo {author} {\bibfnamefont {G.~Z.}\ \bibnamefont {Della~Porta}}, \bibinfo {author} {\bibfnamefont {N.}~\bibnamefont {Bruchon}},\ and\ \bibinfo {author} {\bibfnamefont {G.}~\bibnamefont {Valentino}},\ }\bibfield  {title} {\bibinfo {title} {Sample-efficient reinforcement learning for cern accelerator control},\ }\href@noop {} {\bibfield  {journal} {\bibinfo  {journal} {Physical Review Accelerators and Beams}\ }\textbf {\bibinfo {volume} {23}},\ \bibinfo {pages} {124801} (\bibinfo {year} {2020})}\BibitemShut {NoStop}%
\bibitem [{\citenamefont {Pang}\ \emph {et~al.}(2020)\citenamefont {Pang}, \citenamefont {Thulasidasan},\ and\ \citenamefont {Rybarcyk}}]{pang2020autonomous}%
  \BibitemOpen
  \bibfield  {author} {\bibinfo {author} {\bibfnamefont {X.}~\bibnamefont {Pang}}, \bibinfo {author} {\bibfnamefont {S.}~\bibnamefont {Thulasidasan}},\ and\ \bibinfo {author} {\bibfnamefont {L.}~\bibnamefont {Rybarcyk}},\ }\bibfield  {title} {\bibinfo {title} {Autonomous control of a particle accelerator using deep reinforcement learning},\ }\href@noop {} {\bibfield  {journal} {\bibinfo  {journal} {arXiv preprint arXiv:2010.08141}\ } (\bibinfo {year} {2020})}\BibitemShut {NoStop}%
\bibitem [{\citenamefont {Meier}\ \emph {et~al.}(2022)\citenamefont {Meier}, \citenamefont {Ramirez}, \citenamefont {V{\"o}lker}, \citenamefont {Viefhaus}, \citenamefont {Sick},\ and\ \citenamefont {Hartmann}}]{meier2022optimizing}%
  \BibitemOpen
  \bibfield  {author} {\bibinfo {author} {\bibfnamefont {D.}~\bibnamefont {Meier}}, \bibinfo {author} {\bibfnamefont {L.~V.}\ \bibnamefont {Ramirez}}, \bibinfo {author} {\bibfnamefont {J.}~\bibnamefont {V{\"o}lker}}, \bibinfo {author} {\bibfnamefont {J.}~\bibnamefont {Viefhaus}}, \bibinfo {author} {\bibfnamefont {B.}~\bibnamefont {Sick}},\ and\ \bibinfo {author} {\bibfnamefont {G.}~\bibnamefont {Hartmann}},\ }\bibfield  {title} {\bibinfo {title} {Optimizing a superconducting radio-frequency gun using deep reinforcement learning},\ }\href@noop {} {\bibfield  {journal} {\bibinfo  {journal} {Physical Review Accelerators and Beams}\ }\textbf {\bibinfo {volume} {25}},\ \bibinfo {pages} {104604} (\bibinfo {year} {2022})}\BibitemShut {NoStop}%
\bibitem [{\citenamefont {Obermair}\ \emph {et~al.}(2022)\citenamefont {Obermair}, \citenamefont {Cartier-Michaud}, \citenamefont {Apollonio}, \citenamefont {Millar}, \citenamefont {Felsberger}, \citenamefont {Fischl}, \citenamefont {Bovbjerg}, \citenamefont {Wollmann}, \citenamefont {Wuensch}, \citenamefont {Catalan-Lasheras} \emph {et~al.}}]{obermair2022explainable}%
  \BibitemOpen
  \bibfield  {author} {\bibinfo {author} {\bibfnamefont {C.}~\bibnamefont {Obermair}}, \bibinfo {author} {\bibfnamefont {T.}~\bibnamefont {Cartier-Michaud}}, \bibinfo {author} {\bibfnamefont {A.}~\bibnamefont {Apollonio}}, \bibinfo {author} {\bibfnamefont {W.}~\bibnamefont {Millar}}, \bibinfo {author} {\bibfnamefont {L.}~\bibnamefont {Felsberger}}, \bibinfo {author} {\bibfnamefont {L.}~\bibnamefont {Fischl}}, \bibinfo {author} {\bibfnamefont {H.~S.}\ \bibnamefont {Bovbjerg}}, \bibinfo {author} {\bibfnamefont {D.}~\bibnamefont {Wollmann}}, \bibinfo {author} {\bibfnamefont {W.}~\bibnamefont {Wuensch}}, \bibinfo {author} {\bibfnamefont {N.}~\bibnamefont {Catalan-Lasheras}}, \emph {et~al.},\ }\bibfield  {title} {\bibinfo {title} {Explainable machine learning for breakdown prediction in high gradient {RF} cavities},\ }\href@noop {} {\bibfield  {journal} {\bibinfo  {journal} {Physical Review Accelerators and Beams}\ }\textbf {\bibinfo {volume} {25}},\ \bibinfo {pages} {104601} (\bibinfo {year} {2022})}\BibitemShut
  {NoStop}%
\bibitem [{\citenamefont {Rajput}\ \emph {et~al.}(2024)\citenamefont {Rajput}, \citenamefont {Schram}, \citenamefont {Blokland}, \citenamefont {Alanazi}, \citenamefont {Ramuhalli}, \citenamefont {Zhukov}, \citenamefont {Peters},\ and\ \citenamefont {Vilalta}}]{rajput2024robust}%
  \BibitemOpen
  \bibfield  {author} {\bibinfo {author} {\bibfnamefont {K.}~\bibnamefont {Rajput}}, \bibinfo {author} {\bibfnamefont {M.}~\bibnamefont {Schram}}, \bibinfo {author} {\bibfnamefont {W.}~\bibnamefont {Blokland}}, \bibinfo {author} {\bibfnamefont {Y.}~\bibnamefont {Alanazi}}, \bibinfo {author} {\bibfnamefont {P.}~\bibnamefont {Ramuhalli}}, \bibinfo {author} {\bibfnamefont {A.}~\bibnamefont {Zhukov}}, \bibinfo {author} {\bibfnamefont {C.}~\bibnamefont {Peters}},\ and\ \bibinfo {author} {\bibfnamefont {R.}~\bibnamefont {Vilalta}},\ }\bibfield  {title} {\bibinfo {title} {Robust errant beam prognostics with conditional modeling for particle accelerators},\ }\href@noop {} {\bibfield  {journal} {\bibinfo  {journal} {Machine Learning: Science and Technology}\ }\textbf {\bibinfo {volume} {5}},\ \bibinfo {pages} {015044} (\bibinfo {year} {2024})}\BibitemShut {NoStop}%
\bibitem [{\citenamefont {Tennant}\ \emph {et~al.}(2020)\citenamefont {Tennant}, \citenamefont {Carpenter}, \citenamefont {Powers}, \citenamefont {Solopova}, \citenamefont {Vidyaratne},\ and\ \citenamefont {Iftekharuddin}}]{tennant2020superconducting}%
  \BibitemOpen
  \bibfield  {author} {\bibinfo {author} {\bibfnamefont {C.}~\bibnamefont {Tennant}}, \bibinfo {author} {\bibfnamefont {A.}~\bibnamefont {Carpenter}}, \bibinfo {author} {\bibfnamefont {T.}~\bibnamefont {Powers}}, \bibinfo {author} {\bibfnamefont {A.~S.}\ \bibnamefont {Solopova}}, \bibinfo {author} {\bibfnamefont {L.}~\bibnamefont {Vidyaratne}},\ and\ \bibinfo {author} {\bibfnamefont {K.}~\bibnamefont {Iftekharuddin}},\ }\bibfield  {title} {\bibinfo {title} {Superconducting radio-frequency cavity fault classification using machine learning at jefferson laboratory},\ }\href@noop {} {\bibfield  {journal} {\bibinfo  {journal} {Physical Review Accelerators and Beams}\ }\textbf {\bibinfo {volume} {23}},\ \bibinfo {pages} {114601} (\bibinfo {year} {2020})}\BibitemShut {NoStop}%
\bibitem [{\citenamefont {Li}\ \emph {et~al.}(2018)\citenamefont {Li}, \citenamefont {Cheng}, \citenamefont {Yu},\ and\ \citenamefont {Rainer}}]{li2018genetic}%
  \BibitemOpen
  \bibfield  {author} {\bibinfo {author} {\bibfnamefont {Y.}~\bibnamefont {Li}}, \bibinfo {author} {\bibfnamefont {W.}~\bibnamefont {Cheng}}, \bibinfo {author} {\bibfnamefont {L.~H.}\ \bibnamefont {Yu}},\ and\ \bibinfo {author} {\bibfnamefont {R.}~\bibnamefont {Rainer}},\ }\bibfield  {title} {\bibinfo {title} {Genetic algorithm enhanced by machine learning in dynamic aperture optimization},\ }\href@noop {} {\bibfield  {journal} {\bibinfo  {journal} {Physical Review Accelerators and Beams}\ }\textbf {\bibinfo {volume} {21}},\ \bibinfo {pages} {054601} (\bibinfo {year} {2018})}\BibitemShut {NoStop}%
\bibitem [{\citenamefont {Wan}\ \emph {et~al.}(2019)\citenamefont {Wan}, \citenamefont {Chu}, \citenamefont {Jiao},\ and\ \citenamefont {Li}}]{wan2019improvement}%
  \BibitemOpen
  \bibfield  {author} {\bibinfo {author} {\bibfnamefont {J.}~\bibnamefont {Wan}}, \bibinfo {author} {\bibfnamefont {P.}~\bibnamefont {Chu}}, \bibinfo {author} {\bibfnamefont {Y.}~\bibnamefont {Jiao}},\ and\ \bibinfo {author} {\bibfnamefont {Y.}~\bibnamefont {Li}},\ }\bibfield  {title} {\bibinfo {title} {Improvement of machine learning enhanced genetic algorithm for nonlinear beam dynamics optimization},\ }\href@noop {} {\bibfield  {journal} {\bibinfo  {journal} {Nuclear Instruments and Methods in Physics Research Section A: Accelerators, Spectrometers, Detectors and Associated Equipment}\ }\textbf {\bibinfo {volume} {946}},\ \bibinfo {pages} {162683} (\bibinfo {year} {2019})}\BibitemShut {NoStop}%
\bibitem [{\citenamefont {Bellotti}\ \emph {et~al.}(2021)\citenamefont {Bellotti}, \citenamefont {Boiger},\ and\ \citenamefont {Adelmann}}]{bellotti2021fast}%
  \BibitemOpen
  \bibfield  {author} {\bibinfo {author} {\bibfnamefont {R.}~\bibnamefont {Bellotti}}, \bibinfo {author} {\bibfnamefont {R.}~\bibnamefont {Boiger}},\ and\ \bibinfo {author} {\bibfnamefont {A.}~\bibnamefont {Adelmann}},\ }\bibfield  {title} {\bibinfo {title} {Fast, efficient and flexible particle accelerator optimisation using densely connected and invertible neural networks},\ }\href@noop {} {\bibfield  {journal} {\bibinfo  {journal} {Information}\ }\textbf {\bibinfo {volume} {12}},\ \bibinfo {pages} {351} (\bibinfo {year} {2021})}\BibitemShut {NoStop}%
\bibitem [{\citenamefont {Li}\ and\ \citenamefont {Adelmann}(2023)}]{li2023time}%
  \BibitemOpen
  \bibfield  {author} {\bibinfo {author} {\bibfnamefont {S.}~\bibnamefont {Li}}\ and\ \bibinfo {author} {\bibfnamefont {A.}~\bibnamefont {Adelmann}},\ }\bibfield  {title} {\bibinfo {title} {Time series forecasting methods and their applications to particle accelerators},\ }\href@noop {} {\bibfield  {journal} {\bibinfo  {journal} {Physical Review Accelerators and Beams}\ }\textbf {\bibinfo {volume} {26}},\ \bibinfo {pages} {024801} (\bibinfo {year} {2023})}\BibitemShut {NoStop}%
\bibitem [{\citenamefont {Scheinker}\ \emph {et~al.}(2021)\citenamefont {Scheinker}, \citenamefont {Cropp}, \citenamefont {Paiagua},\ and\ \citenamefont {Filippetto}}]{scheinker2021adaptive_SciRep}%
  \BibitemOpen
  \bibfield  {author} {\bibinfo {author} {\bibfnamefont {A.}~\bibnamefont {Scheinker}}, \bibinfo {author} {\bibfnamefont {F.}~\bibnamefont {Cropp}}, \bibinfo {author} {\bibfnamefont {S.}~\bibnamefont {Paiagua}},\ and\ \bibinfo {author} {\bibfnamefont {D.}~\bibnamefont {Filippetto}},\ }\bibfield  {title} {\bibinfo {title} {An adaptive approach to machine learning for compact particle accelerators},\ }\href@noop {} {\bibfield  {journal} {\bibinfo  {journal} {Scientific reports}\ }\textbf {\bibinfo {volume} {11}},\ \bibinfo {pages} {19187} (\bibinfo {year} {2021})}\BibitemShut {NoStop}%
\bibitem [{\citenamefont {Wolski}\ \emph {et~al.}(2022)\citenamefont {Wolski}, \citenamefont {Johnson}, \citenamefont {King}, \citenamefont {Militsyn},\ and\ \citenamefont {Williams}}]{wolski2022transverse}%
  \BibitemOpen
  \bibfield  {author} {\bibinfo {author} {\bibfnamefont {A.}~\bibnamefont {Wolski}}, \bibinfo {author} {\bibfnamefont {M.~A.}\ \bibnamefont {Johnson}}, \bibinfo {author} {\bibfnamefont {M.}~\bibnamefont {King}}, \bibinfo {author} {\bibfnamefont {B.~L.}\ \bibnamefont {Militsyn}},\ and\ \bibinfo {author} {\bibfnamefont {P.~H.}\ \bibnamefont {Williams}},\ }\bibfield  {title} {\bibinfo {title} {Transverse phase space tomography in an accelerator test facility using image compression and machine learning},\ }\href@noop {} {\bibfield  {journal} {\bibinfo  {journal} {Physical Review Accelerators and Beams}\ }\textbf {\bibinfo {volume} {25}},\ \bibinfo {pages} {122803} (\bibinfo {year} {2022})}\BibitemShut {NoStop}%
\bibitem [{\citenamefont {Roussel}\ \emph {et~al.}(2023)\citenamefont {Roussel}, \citenamefont {Edelen}, \citenamefont {Mayes}, \citenamefont {Ratner}, \citenamefont {Gonzalez-Aguilera}, \citenamefont {Kim}, \citenamefont {Wisniewski},\ and\ \citenamefont {Power}}]{roussel2023phase}%
  \BibitemOpen
  \bibfield  {author} {\bibinfo {author} {\bibfnamefont {R.}~\bibnamefont {Roussel}}, \bibinfo {author} {\bibfnamefont {A.}~\bibnamefont {Edelen}}, \bibinfo {author} {\bibfnamefont {C.}~\bibnamefont {Mayes}}, \bibinfo {author} {\bibfnamefont {D.}~\bibnamefont {Ratner}}, \bibinfo {author} {\bibfnamefont {J.~P.}\ \bibnamefont {Gonzalez-Aguilera}}, \bibinfo {author} {\bibfnamefont {S.}~\bibnamefont {Kim}}, \bibinfo {author} {\bibfnamefont {E.}~\bibnamefont {Wisniewski}},\ and\ \bibinfo {author} {\bibfnamefont {J.}~\bibnamefont {Power}},\ }\bibfield  {title} {\bibinfo {title} {Phase space reconstruction from accelerator beam measurements using neural networks and differentiable simulations},\ }\href@noop {} {\bibfield  {journal} {\bibinfo  {journal} {Physical Review Letters}\ }\textbf {\bibinfo {volume} {130}},\ \bibinfo {pages} {145001} (\bibinfo {year} {2023})}\BibitemShut {NoStop}%
\bibitem [{\citenamefont {Smith}(2013)}]{smith2013uncertainty}%
  \BibitemOpen
  \bibfield  {author} {\bibinfo {author} {\bibfnamefont {R.~C.}\ \bibnamefont {Smith}},\ }\href@noop {} {\emph {\bibinfo {title} {Uncertainty quantification: theory, implementation, and applications}}}\ (\bibinfo  {publisher} {SIAM},\ \bibinfo {year} {2013})\BibitemShut {NoStop}%
\bibitem [{\citenamefont {Zhang}\ \emph {et~al.}(2019)\citenamefont {Zhang}, \citenamefont {Lu}, \citenamefont {Guo},\ and\ \citenamefont {Karniadakis}}]{zhang2019quantifying}%
  \BibitemOpen
  \bibfield  {author} {\bibinfo {author} {\bibfnamefont {D.}~\bibnamefont {Zhang}}, \bibinfo {author} {\bibfnamefont {L.}~\bibnamefont {Lu}}, \bibinfo {author} {\bibfnamefont {L.}~\bibnamefont {Guo}},\ and\ \bibinfo {author} {\bibfnamefont {G.~E.}\ \bibnamefont {Karniadakis}},\ }\bibfield  {title} {\bibinfo {title} {Quantifying total uncertainty in physics-informed neural networks for solving forward and inverse stochastic problems},\ }\href@noop {} {\bibfield  {journal} {\bibinfo  {journal} {Journal of Computational Physics}\ }\textbf {\bibinfo {volume} {397}},\ \bibinfo {pages} {108850} (\bibinfo {year} {2019})}\BibitemShut {NoStop}%
\bibitem [{\citenamefont {Acharya}\ \emph {et~al.}(2023)\citenamefont {Acharya}, \citenamefont {Russell},\ and\ \citenamefont {Ahmed}}]{acharya2023learning}%
  \BibitemOpen
  \bibfield  {author} {\bibinfo {author} {\bibfnamefont {A.}~\bibnamefont {Acharya}}, \bibinfo {author} {\bibfnamefont {R.}~\bibnamefont {Russell}},\ and\ \bibinfo {author} {\bibfnamefont {N.~R.}\ \bibnamefont {Ahmed}},\ }\bibfield  {title} {\bibinfo {title} {Learning to forecast aleatoric and epistemic uncertainties over long horizon trajectories},\ }\href@noop {} {\bibfield  {journal} {\bibinfo  {journal} {arXiv preprint arXiv:2302.08669}\ } (\bibinfo {year} {2023})}\BibitemShut {NoStop}%
\bibitem [{\citenamefont {Adelmann}(2019)}]{adelmann2019nonintrusive}%
  \BibitemOpen
  \bibfield  {author} {\bibinfo {author} {\bibfnamefont {A.}~\bibnamefont {Adelmann}},\ }\bibfield  {title} {\bibinfo {title} {On nonintrusive uncertainty quantification and surrogate model construction in particle accelerator modeling},\ }\href@noop {} {\bibfield  {journal} {\bibinfo  {journal} {SIAM/ASA Journal on Uncertainty Quantification}\ }\textbf {\bibinfo {volume} {7}},\ \bibinfo {pages} {383} (\bibinfo {year} {2019})}\BibitemShut {NoStop}%
\bibitem [{\citenamefont {Convery}\ \emph {et~al.}(2021)\citenamefont {Convery}, \citenamefont {Smith}, \citenamefont {Gal},\ and\ \citenamefont {Hanuka}}]{convery2021uncertainty}%
  \BibitemOpen
  \bibfield  {author} {\bibinfo {author} {\bibfnamefont {O.}~\bibnamefont {Convery}}, \bibinfo {author} {\bibfnamefont {L.}~\bibnamefont {Smith}}, \bibinfo {author} {\bibfnamefont {Y.}~\bibnamefont {Gal}},\ and\ \bibinfo {author} {\bibfnamefont {A.}~\bibnamefont {Hanuka}},\ }\bibfield  {title} {\bibinfo {title} {Uncertainty quantification for virtual diagnostic of particle accelerators},\ }\href@noop {} {\bibfield  {journal} {\bibinfo  {journal} {Physical Review Accelerators and Beams}\ }\textbf {\bibinfo {volume} {24}},\ \bibinfo {pages} {074602} (\bibinfo {year} {2021})}\BibitemShut {NoStop}%
\bibitem [{\citenamefont {Mishra}\ \emph {et~al.}(2021)\citenamefont {Mishra}, \citenamefont {Edelen}, \citenamefont {Hanuka},\ and\ \citenamefont {Mayes}}]{mishra2021uncertainty}%
  \BibitemOpen
  \bibfield  {author} {\bibinfo {author} {\bibfnamefont {A.~A.}\ \bibnamefont {Mishra}}, \bibinfo {author} {\bibfnamefont {A.}~\bibnamefont {Edelen}}, \bibinfo {author} {\bibfnamefont {A.}~\bibnamefont {Hanuka}},\ and\ \bibinfo {author} {\bibfnamefont {C.}~\bibnamefont {Mayes}},\ }\bibfield  {title} {\bibinfo {title} {Uncertainty quantification for deep learning in particle accelerator applications},\ }\href@noop {} {\bibfield  {journal} {\bibinfo  {journal} {Physical Review Accelerators and Beams}\ }\textbf {\bibinfo {volume} {24}},\ \bibinfo {pages} {114601} (\bibinfo {year} {2021})}\BibitemShut {NoStop}%
\bibitem [{\citenamefont {Blokland}\ \emph {et~al.}(2022)\citenamefont {Blokland}, \citenamefont {Rajput}, \citenamefont {Schram}, \citenamefont {Jeske}, \citenamefont {Ramuhalli}, \citenamefont {Peters}, \citenamefont {Yucesan},\ and\ \citenamefont {Zhukov}}]{blokland2022uncertainty}%
  \BibitemOpen
  \bibfield  {author} {\bibinfo {author} {\bibfnamefont {W.}~\bibnamefont {Blokland}}, \bibinfo {author} {\bibfnamefont {K.}~\bibnamefont {Rajput}}, \bibinfo {author} {\bibfnamefont {M.}~\bibnamefont {Schram}}, \bibinfo {author} {\bibfnamefont {T.}~\bibnamefont {Jeske}}, \bibinfo {author} {\bibfnamefont {P.}~\bibnamefont {Ramuhalli}}, \bibinfo {author} {\bibfnamefont {C.}~\bibnamefont {Peters}}, \bibinfo {author} {\bibfnamefont {Y.}~\bibnamefont {Yucesan}},\ and\ \bibinfo {author} {\bibfnamefont {A.}~\bibnamefont {Zhukov}},\ }\bibfield  {title} {\bibinfo {title} {Uncertainty aware anomaly detection to predict errant beam pulses in the oak ridge spallation neutron source accelerator},\ }\href@noop {} {\bibfield  {journal} {\bibinfo  {journal} {Physical Review Accelerators and Beams}\ }\textbf {\bibinfo {volume} {25}},\ \bibinfo {pages} {122802} (\bibinfo {year} {2022})}\BibitemShut {NoStop}%
\bibitem [{\citenamefont {Schram}\ \emph {et~al.}(2023)\citenamefont {Schram}, \citenamefont {Rajput}, \citenamefont {NS}, \citenamefont {Li}, \citenamefont {John},\ and\ \citenamefont {Sharma}}]{schram2023uncertainty}%
  \BibitemOpen
  \bibfield  {author} {\bibinfo {author} {\bibfnamefont {M.}~\bibnamefont {Schram}}, \bibinfo {author} {\bibfnamefont {K.}~\bibnamefont {Rajput}}, \bibinfo {author} {\bibfnamefont {K.~S.}\ \bibnamefont {NS}}, \bibinfo {author} {\bibfnamefont {P.}~\bibnamefont {Li}}, \bibinfo {author} {\bibfnamefont {J.~S.}\ \bibnamefont {John}},\ and\ \bibinfo {author} {\bibfnamefont {H.}~\bibnamefont {Sharma}},\ }\bibfield  {title} {\bibinfo {title} {Uncertainty aware machine-learning-based surrogate models for particle accelerators: Study at the fermilab booster accelerator complex},\ }\href@noop {} {\bibfield  {journal} {\bibinfo  {journal} {Physical Review Accelerators and Beams}\ }\textbf {\bibinfo {volume} {26}},\ \bibinfo {pages} {044602} (\bibinfo {year} {2023})}\BibitemShut {NoStop}%
\bibitem [{\citenamefont {Garcia-Cardona}\ and\ \citenamefont {Scheinker}(2024)}]{garcia2024machine}%
  \BibitemOpen
  \bibfield  {author} {\bibinfo {author} {\bibfnamefont {C.}~\bibnamefont {Garcia-Cardona}}\ and\ \bibinfo {author} {\bibfnamefont {A.}~\bibnamefont {Scheinker}},\ }\bibfield  {title} {\bibinfo {title} {Machine learning surrogate for charged particle beam dynamics with space charge based on a recurrent neural network with aleatoric uncertainty},\ }\href@noop {} {\bibfield  {journal} {\bibinfo  {journal} {Physical Review Accelerators and Beams}\ }\textbf {\bibinfo {volume} {27}},\ \bibinfo {pages} {024601} (\bibinfo {year} {2024})}\BibitemShut {NoStop}%
\bibitem [{\citenamefont {Uria}\ \emph {et~al.}(2016)\citenamefont {Uria}, \citenamefont {C{\^o}t{\'e}}, \citenamefont {Gregor}, \citenamefont {Murray},\ and\ \citenamefont {Larochelle}}]{uria2016neural}%
  \BibitemOpen
  \bibfield  {author} {\bibinfo {author} {\bibfnamefont {B.}~\bibnamefont {Uria}}, \bibinfo {author} {\bibfnamefont {M.-A.}\ \bibnamefont {C{\^o}t{\'e}}}, \bibinfo {author} {\bibfnamefont {K.}~\bibnamefont {Gregor}}, \bibinfo {author} {\bibfnamefont {I.}~\bibnamefont {Murray}},\ and\ \bibinfo {author} {\bibfnamefont {H.}~\bibnamefont {Larochelle}},\ }\bibfield  {title} {\bibinfo {title} {Neural autoregressive distribution estimation},\ }\href@noop {} {\bibfield  {journal} {\bibinfo  {journal} {The Journal of Machine Learning Research}\ }\textbf {\bibinfo {volume} {17}},\ \bibinfo {pages} {7184} (\bibinfo {year} {2016})}\BibitemShut {NoStop}%
\bibitem [{\citenamefont {Papamakarios}\ \emph {et~al.}(2017)\citenamefont {Papamakarios}, \citenamefont {Pavlakou},\ and\ \citenamefont {Murray}}]{papamakarios2017masked}%
  \BibitemOpen
  \bibfield  {author} {\bibinfo {author} {\bibfnamefont {G.}~\bibnamefont {Papamakarios}}, \bibinfo {author} {\bibfnamefont {T.}~\bibnamefont {Pavlakou}},\ and\ \bibinfo {author} {\bibfnamefont {I.}~\bibnamefont {Murray}},\ }\bibfield  {title} {\bibinfo {title} {Masked autoregressive flow for density estimation},\ }\href@noop {} {\bibfield  {journal} {\bibinfo  {journal} {Advances in neural information processing systems}\ }\textbf {\bibinfo {volume} {30}} (\bibinfo {year} {2017})}\BibitemShut {NoStop}%
\bibitem [{\citenamefont {Papamakarios}\ \emph {et~al.}(2021)\citenamefont {Papamakarios}, \citenamefont {Nalisnick}, \citenamefont {Rezende}, \citenamefont {Mohamed},\ and\ \citenamefont {Lakshminarayanan}}]{papamakarios2021normalizing}%
  \BibitemOpen
  \bibfield  {author} {\bibinfo {author} {\bibfnamefont {G.}~\bibnamefont {Papamakarios}}, \bibinfo {author} {\bibfnamefont {E.}~\bibnamefont {Nalisnick}}, \bibinfo {author} {\bibfnamefont {D.~J.}\ \bibnamefont {Rezende}}, \bibinfo {author} {\bibfnamefont {S.}~\bibnamefont {Mohamed}},\ and\ \bibinfo {author} {\bibfnamefont {B.}~\bibnamefont {Lakshminarayanan}},\ }\bibfield  {title} {\bibinfo {title} {Normalizing flows for probabilistic modeling and inference},\ }\href@noop {} {\bibfield  {journal} {\bibinfo  {journal} {Journal of Machine Learning Research}\ }\textbf {\bibinfo {volume} {22}},\ \bibinfo {pages} {1} (\bibinfo {year} {2021})}\BibitemShut {NoStop}%
\bibitem [{\citenamefont {Ho}\ \emph {et~al.}(2020)\citenamefont {Ho}, \citenamefont {Jain},\ and\ \citenamefont {Abbeel}}]{ho2020denoising}%
  \BibitemOpen
  \bibfield  {author} {\bibinfo {author} {\bibfnamefont {J.}~\bibnamefont {Ho}}, \bibinfo {author} {\bibfnamefont {A.}~\bibnamefont {Jain}},\ and\ \bibinfo {author} {\bibfnamefont {P.}~\bibnamefont {Abbeel}},\ }\bibfield  {title} {\bibinfo {title} {Denoising diffusion probabilistic models},\ }\href@noop {} {\bibfield  {journal} {\bibinfo  {journal} {Advances in neural information processing systems}\ }\textbf {\bibinfo {volume} {33}},\ \bibinfo {pages} {6840} (\bibinfo {year} {2020})}\BibitemShut {NoStop}%
\bibitem [{\citenamefont {Cathey}\ \emph {et~al.}(2018)\citenamefont {Cathey}, \citenamefont {Cousineau}, \citenamefont {Aleksandrov},\ and\ \citenamefont {Zhukov}}]{cathey2018first}%
  \BibitemOpen
  \bibfield  {author} {\bibinfo {author} {\bibfnamefont {B.}~\bibnamefont {Cathey}}, \bibinfo {author} {\bibfnamefont {S.}~\bibnamefont {Cousineau}}, \bibinfo {author} {\bibfnamefont {A.}~\bibnamefont {Aleksandrov}},\ and\ \bibinfo {author} {\bibfnamefont {A.}~\bibnamefont {Zhukov}},\ }\bibfield  {title} {\bibinfo {title} {First six dimensional phase space measurement of an accelerator beam},\ }\href@noop {} {\bibfield  {journal} {\bibinfo  {journal} {Physical Review Letters}\ }\textbf {\bibinfo {volume} {121}},\ \bibinfo {pages} {064804} (\bibinfo {year} {2018})}\BibitemShut {NoStop}%
\bibitem [{\citenamefont {Wiedemann}(1994)}]{wiedemann1994particle}%
  \BibitemOpen
  \bibfield  {author} {\bibinfo {author} {\bibfnamefont {H.}~\bibnamefont {Wiedemann}},\ }\bibfield  {title} {\bibinfo {title} {Particle accelerator physics},\ }\href@noop {} {\bibfield  {journal} {\bibinfo  {journal} {Physics Today}\ }\textbf {\bibinfo {volume} {47}},\ \bibinfo {pages} {61} (\bibinfo {year} {1994})}\BibitemShut {NoStop}%
\bibitem [{\citenamefont {Pang}\ and\ \citenamefont {Rybarcyk}(2014)}]{pang2014gpu}%
  \BibitemOpen
  \bibfield  {author} {\bibinfo {author} {\bibfnamefont {X.}~\bibnamefont {Pang}}\ and\ \bibinfo {author} {\bibfnamefont {L.}~\bibnamefont {Rybarcyk}},\ }\bibfield  {title} {\bibinfo {title} {{GPU} accelerated online multi-particle beam dynamics simulator for ion linear particle accelerators},\ }\href@noop {} {\bibfield  {journal} {\bibinfo  {journal} {Computer Physics Communications}\ }\textbf {\bibinfo {volume} {185}},\ \bibinfo {pages} {744} (\bibinfo {year} {2014})}\BibitemShut {NoStop}%
\bibitem [{\citenamefont {Rautela}\ \emph {et~al.}(2024{\natexlab{c}})\citenamefont {Rautela}, \citenamefont {Williams},\ and\ \citenamefont {Scheinker}}]{rautela_2024_10819001}%
  \BibitemOpen
  \bibfield  {author} {\bibinfo {author} {\bibfnamefont {M.}~\bibnamefont {Rautela}}, \bibinfo {author} {\bibfnamefont {A.}~\bibnamefont {Williams}},\ and\ \bibinfo {author} {\bibfnamefont {A.}~\bibnamefont {Scheinker}},\ }\bibfield  {title} {\bibinfo {title} {{6D phase space of charged beam in particle accelerator}},\ }\href {https://doi.org/10.5281/zenodo.10819001} {10.5281/zenodo.10819001} (\bibinfo {year} {2024}{\natexlab{c}})\BibitemShut {NoStop}%
\bibitem [{\citenamefont {Kingma}\ and\ \citenamefont {Welling}(2013)}]{kingma2013auto}%
  \BibitemOpen
  \bibfield  {author} {\bibinfo {author} {\bibfnamefont {D.~P.}\ \bibnamefont {Kingma}}\ and\ \bibinfo {author} {\bibfnamefont {M.}~\bibnamefont {Welling}},\ }\bibfield  {title} {\bibinfo {title} {Auto-encoding variational bayes},\ }\href@noop {} {\bibfield  {journal} {\bibinfo  {journal} {arXiv preprint arXiv:1312.6114}\ } (\bibinfo {year} {2013})}\BibitemShut {NoStop}%
\bibitem [{\citenamefont {Yin}\ \emph {et~al.}(2021)\citenamefont {Yin}, \citenamefont {Pei},\ and\ \citenamefont {Gao}}]{yin2021neural}%
  \BibitemOpen
  \bibfield  {author} {\bibinfo {author} {\bibfnamefont {J.}~\bibnamefont {Yin}}, \bibinfo {author} {\bibfnamefont {Z.}~\bibnamefont {Pei}},\ and\ \bibinfo {author} {\bibfnamefont {M.~C.}\ \bibnamefont {Gao}},\ }\bibfield  {title} {\bibinfo {title} {Neural network-based order parameter for phase transitions and its applications in high-entropy alloys},\ }\href@noop {} {\bibfield  {journal} {\bibinfo  {journal} {Nature Computational Science}\ }\textbf {\bibinfo {volume} {1}},\ \bibinfo {pages} {686} (\bibinfo {year} {2021})}\BibitemShut {NoStop}%
\bibitem [{\citenamefont {Toneva}\ \emph {et~al.}(2022)\citenamefont {Toneva}, \citenamefont {Mitchell},\ and\ \citenamefont {Wehbe}}]{toneva2022combining}%
  \BibitemOpen
  \bibfield  {author} {\bibinfo {author} {\bibfnamefont {M.}~\bibnamefont {Toneva}}, \bibinfo {author} {\bibfnamefont {T.~M.}\ \bibnamefont {Mitchell}},\ and\ \bibinfo {author} {\bibfnamefont {L.}~\bibnamefont {Wehbe}},\ }\bibfield  {title} {\bibinfo {title} {Combining computational controls with natural text reveals aspects of meaning composition},\ }\href@noop {} {\bibfield  {journal} {\bibinfo  {journal} {Nature computational science}\ }\textbf {\bibinfo {volume} {2}},\ \bibinfo {pages} {745} (\bibinfo {year} {2022})}\BibitemShut {NoStop}%
\bibitem [{\citenamefont {Rautela}\ \emph {et~al.}(2022)\citenamefont {Rautela}, \citenamefont {Senthilnath}, \citenamefont {Huber},\ and\ \citenamefont {Gopalakrishnan}}]{rautela2022towards}%
  \BibitemOpen
  \bibfield  {author} {\bibinfo {author} {\bibfnamefont {M.}~\bibnamefont {Rautela}}, \bibinfo {author} {\bibfnamefont {J.}~\bibnamefont {Senthilnath}}, \bibinfo {author} {\bibfnamefont {A.}~\bibnamefont {Huber}},\ and\ \bibinfo {author} {\bibfnamefont {S.}~\bibnamefont {Gopalakrishnan}},\ }\bibfield  {title} {\bibinfo {title} {Towards deep generation of guided wave representations for composite materials},\ }\href@noop {} {\bibfield  {journal} {\bibinfo  {journal} {IEEE Transactions on Artificial Intelligence}\ } (\bibinfo {year} {2022})}\BibitemShut {NoStop}%
\bibitem [{\citenamefont {Van~der Maaten}\ and\ \citenamefont {Hinton}(2008)}]{van2008visualizing}%
  \BibitemOpen
  \bibfield  {author} {\bibinfo {author} {\bibfnamefont {L.}~\bibnamefont {Van~der Maaten}}\ and\ \bibinfo {author} {\bibfnamefont {G.}~\bibnamefont {Hinton}},\ }\bibfield  {title} {\bibinfo {title} {Visualizing data using t-sne.},\ }\href@noop {} {\bibfield  {journal} {\bibinfo  {journal} {Journal of machine learning research}\ }\textbf {\bibinfo {volume} {9}} (\bibinfo {year} {2008})}\BibitemShut {NoStop}%
\bibitem [{\citenamefont {Rautela}\ \emph {et~al.}(2023{\natexlab{a}})\citenamefont {Rautela}, \citenamefont {Maghareh}, \citenamefont {Dyke},\ and\ \citenamefont {Gopalakrishnan}}]{rautela2023deep}%
  \BibitemOpen
  \bibfield  {author} {\bibinfo {author} {\bibfnamefont {M.}~\bibnamefont {Rautela}}, \bibinfo {author} {\bibfnamefont {A.}~\bibnamefont {Maghareh}}, \bibinfo {author} {\bibfnamefont {S.}~\bibnamefont {Dyke}},\ and\ \bibinfo {author} {\bibfnamefont {S.}~\bibnamefont {Gopalakrishnan}},\ }\bibfield  {title} {\bibinfo {title} {Deep generative models for unsupervised delamination detection using guided waves},\ }\href@noop {} {\bibfield  {journal} {\bibinfo  {journal} {arXiv preprint arXiv:2308.05350}\ } (\bibinfo {year} {2023}{\natexlab{a}})}\BibitemShut {NoStop}%
\bibitem [{\citenamefont {Huang}\ \emph {et~al.}(2022)\citenamefont {Huang}, \citenamefont {Lam},\ and\ \citenamefont {Zhang}}]{huang2022evaluating}%
  \BibitemOpen
  \bibfield  {author} {\bibinfo {author} {\bibfnamefont {Z.}~\bibnamefont {Huang}}, \bibinfo {author} {\bibfnamefont {H.}~\bibnamefont {Lam}},\ and\ \bibinfo {author} {\bibfnamefont {H.}~\bibnamefont {Zhang}},\ }\bibfield  {title} {\bibinfo {title} {Evaluating aleatoric uncertainty via conditional generative models},\ }\href@noop {} {\bibfield  {journal} {\bibinfo  {journal} {arXiv preprint arXiv:2206.04287}\ } (\bibinfo {year} {2022})}\BibitemShut {NoStop}%
\bibitem [{\citenamefont {Wang}\ \emph {et~al.}(2020)\citenamefont {Wang}, \citenamefont {Chan}, \citenamefont {Ahmed}, \citenamefont {Liu}, \citenamefont {Zhu},\ and\ \citenamefont {Chen}}]{wang2020deep}%
  \BibitemOpen
  \bibfield  {author} {\bibinfo {author} {\bibfnamefont {L.}~\bibnamefont {Wang}}, \bibinfo {author} {\bibfnamefont {Y.-C.}\ \bibnamefont {Chan}}, \bibinfo {author} {\bibfnamefont {F.}~\bibnamefont {Ahmed}}, \bibinfo {author} {\bibfnamefont {Z.}~\bibnamefont {Liu}}, \bibinfo {author} {\bibfnamefont {P.}~\bibnamefont {Zhu}},\ and\ \bibinfo {author} {\bibfnamefont {W.}~\bibnamefont {Chen}},\ }\bibfield  {title} {\bibinfo {title} {Deep generative modeling for mechanistic-based learning and design of metamaterial systems},\ }\href@noop {} {\bibfield  {journal} {\bibinfo  {journal} {Computer Methods in Applied Mechanics and Engineering}\ }\textbf {\bibinfo {volume} {372}},\ \bibinfo {pages} {113377} (\bibinfo {year} {2020})}\BibitemShut {NoStop}%
\bibitem [{\citenamefont {Ha}\ and\ \citenamefont {Schmidhuber}(2018)}]{ha2018world}%
  \BibitemOpen
  \bibfield  {author} {\bibinfo {author} {\bibfnamefont {D.}~\bibnamefont {Ha}}\ and\ \bibinfo {author} {\bibfnamefont {J.}~\bibnamefont {Schmidhuber}},\ }\bibfield  {title} {\bibinfo {title} {World models},\ }\href@noop {} {\bibfield  {journal} {\bibinfo  {journal} {arXiv preprint arXiv:1803.10122}\ } (\bibinfo {year} {2018})}\BibitemShut {NoStop}%
\bibitem [{\citenamefont {Rautela}\ \emph {et~al.}(2023{\natexlab{b}})\citenamefont {Rautela}, \citenamefont {Gopalakrishnan},\ and\ \citenamefont {Senthilnath}}]{rautela2023bayesian}%
  \BibitemOpen
  \bibfield  {author} {\bibinfo {author} {\bibfnamefont {M.}~\bibnamefont {Rautela}}, \bibinfo {author} {\bibfnamefont {S.}~\bibnamefont {Gopalakrishnan}},\ and\ \bibinfo {author} {\bibfnamefont {J.}~\bibnamefont {Senthilnath}},\ }\bibfield  {title} {\bibinfo {title} {Bayesian optimized physics-informed neural network for estimating wave propagation velocities},\ }\href@noop {} {\bibfield  {journal} {\bibinfo  {journal} {arXiv preprint arXiv:2312.14064}\ } (\bibinfo {year} {2023}{\natexlab{b}})}\BibitemShut {NoStop}%
\end{thebibliography}%

\section*{Appendix A}
This appendix contains supplementary information in the form of additional results from Sec.~\ref{sec:results}.
\renewcommand{\thefigure}{A.\arabic{figure}}
\setcounter{figure}{0}

\begin{figure*}[htbp]
    \centering
    \begin{minipage}[b]{0.45\linewidth}
        \centering
        \includegraphics[trim={0cm 0cm 0cm 0cm},clip, width=1.0\textwidth]{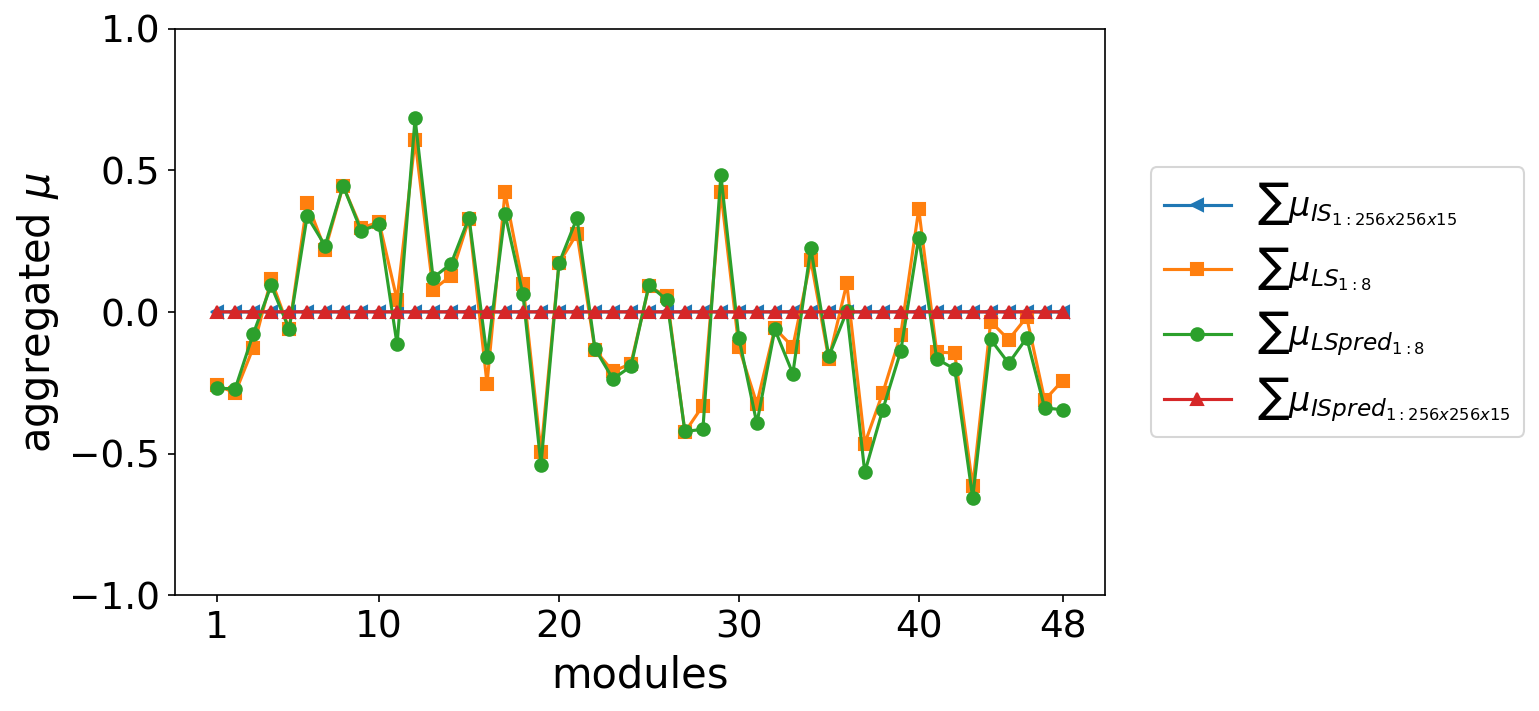}
    \end{minipage}
    \begin{minipage}[b]{0.45\linewidth}
        \centering
        \includegraphics[trim={0cm 0cm 0cm 0cm},clip,width=1.0\textwidth]{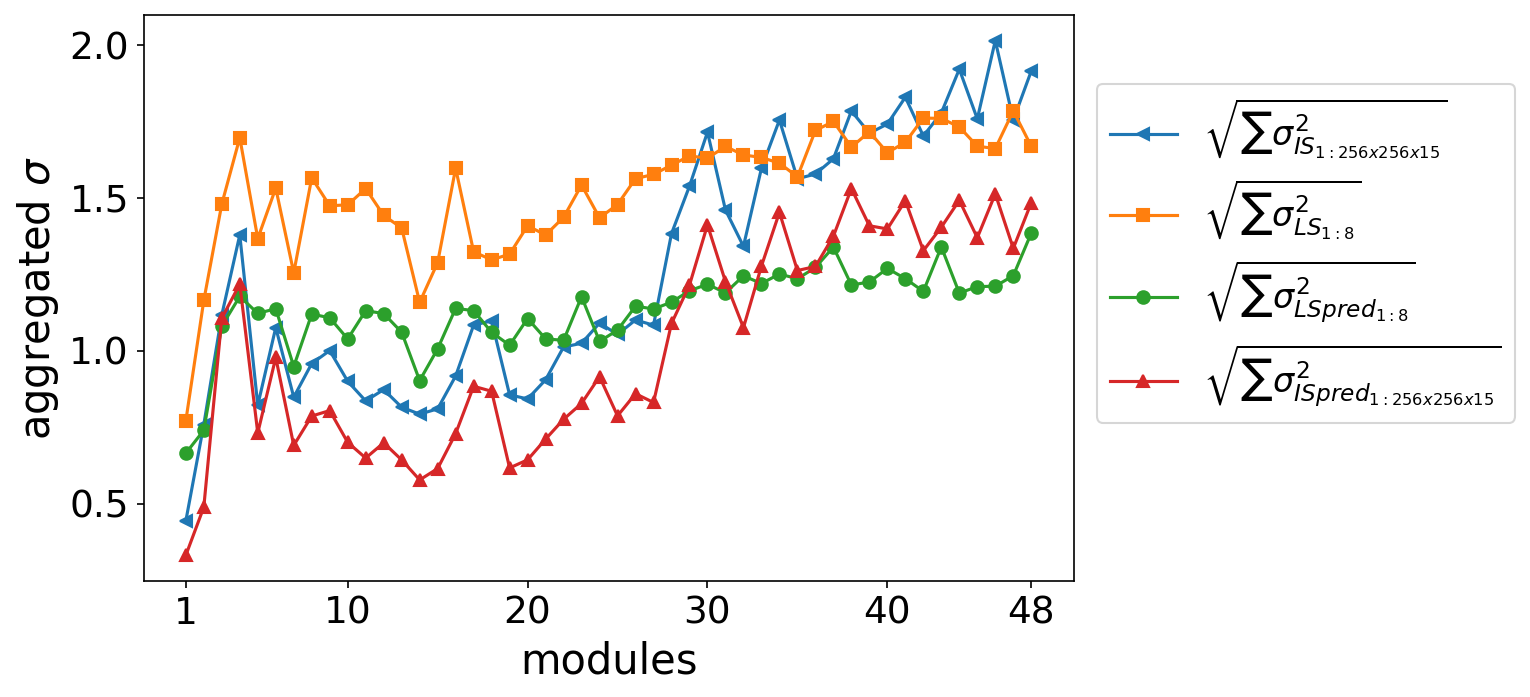}
    \end{minipage}
    \caption{Comparison of (a) aggregated mean and (b) aggregate standard deviation of latent space (orange), Dataset image space (blue), LSTM based latent predictions (green), and decoded latent predictions (red).}
    \label{fig:comparemeansd_all}
\end{figure*}

\begin{figure*}[htbp]
    \centering
    \begin{minipage}[b]{1.0\linewidth}
        \centering
        \includegraphics[trim={0cm 0cm 0cm 0cm},clip, width=1.0\textwidth]{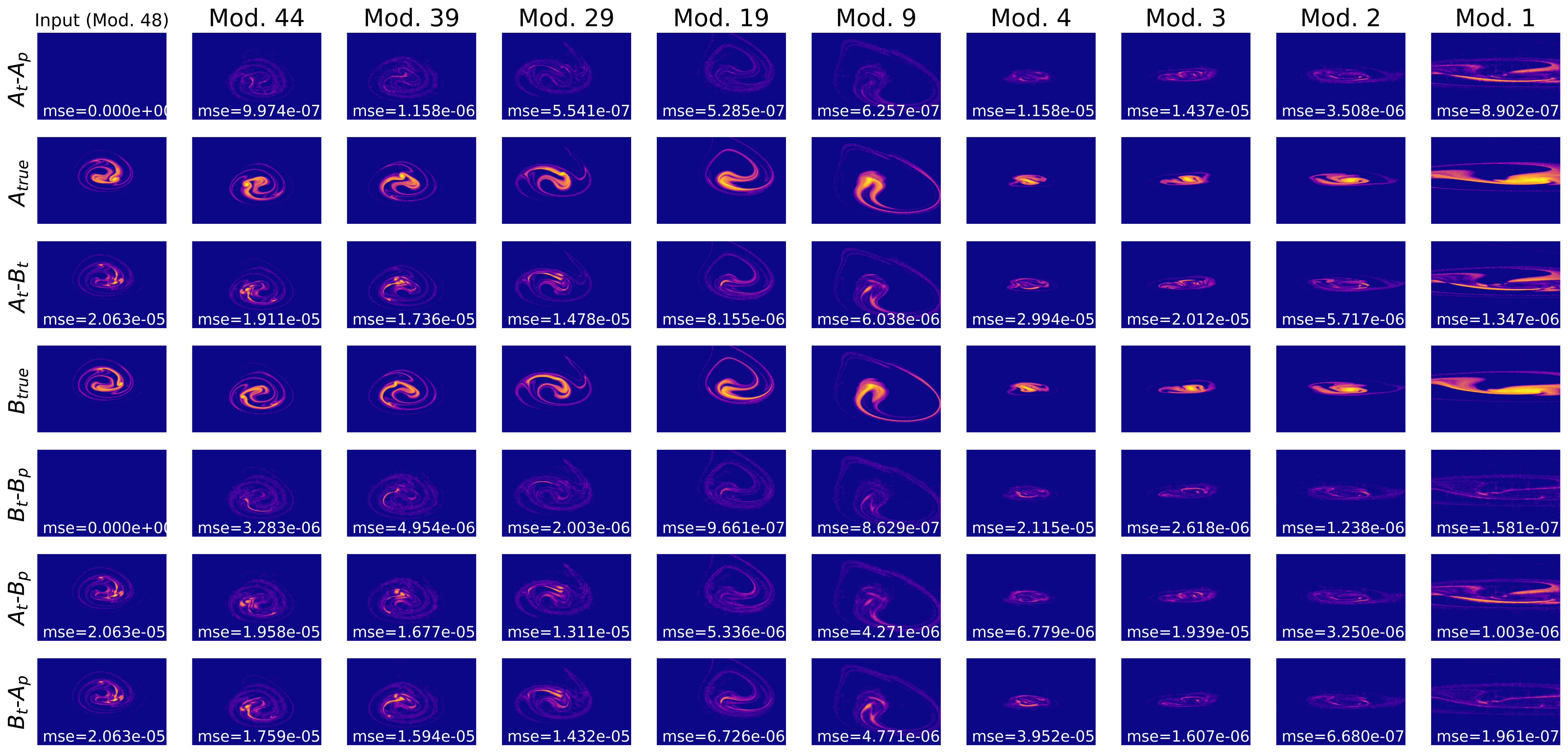}
    \end{minipage}
    \hspace{-10pt}
    \begin{minipage}[b]{1.0\linewidth}
        \centering
        \includegraphics[width=1.0\textwidth]{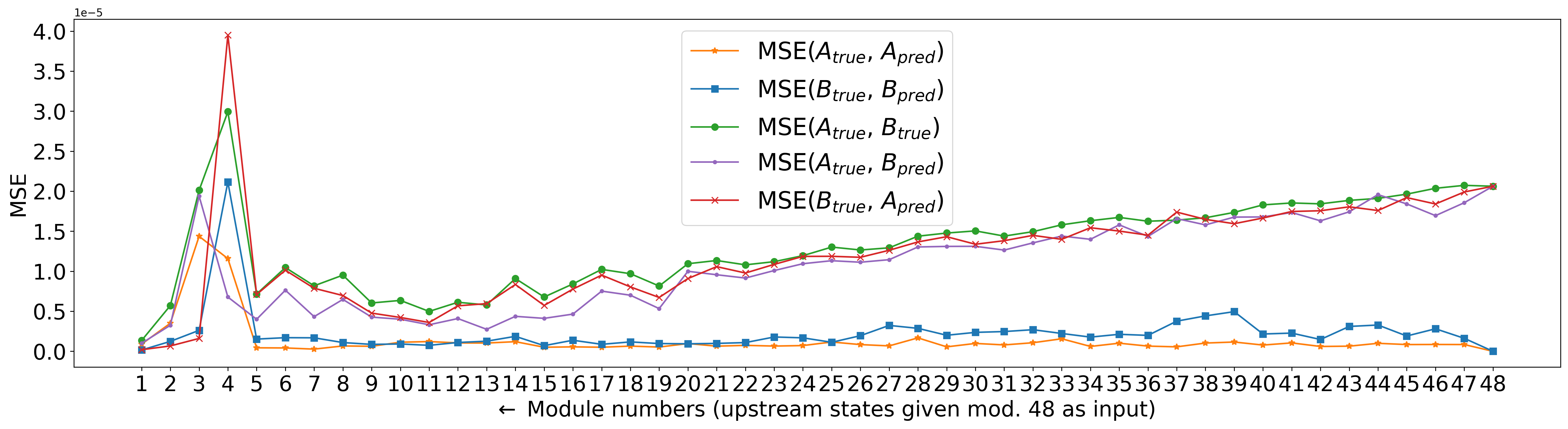}
    \end{minipage}
    \caption{Comparison plot of two different random predictions ($A_{p}$/$A_{pred}$ and $B_{p}$/$B_{pred}$) from the trained RLE at different upstream modules given two different module 48 as the input ($A_t$/$A_{true}$ and $B_t$/$B_{true}$): (a) $E-\phi$ projections (b) MSE vs module numbers.}
    \label{fig:compare_prediction_mse_48}
\end{figure*}

\begin{figure*}[htbp]
    \centering
    \begin{minipage}[b]{1.0\linewidth}
        \centering
        \includegraphics[trim={4cm 1cm 4cm 0cm},clip, width=1.0\textwidth]{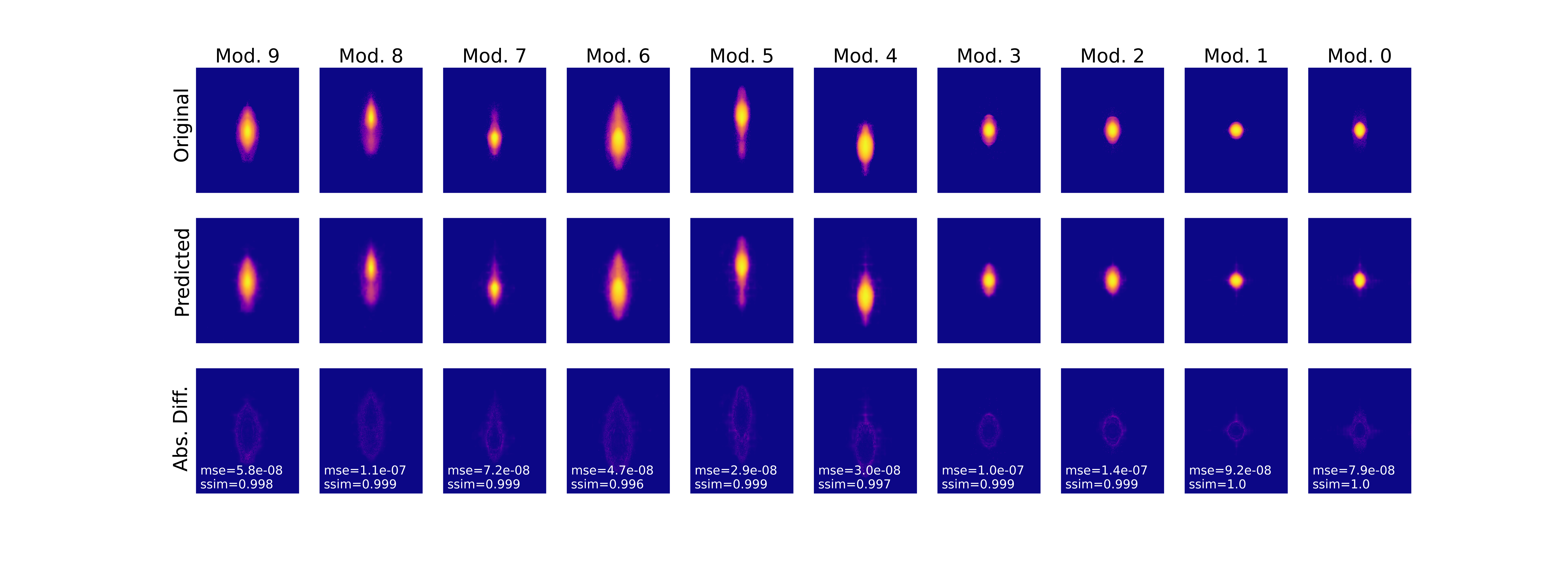}
        \subcaption{$x-y$ projection}
    \end{minipage}
    \begin{minipage}[b]{1.0\linewidth}
        \centering
        \includegraphics[trim={4cm 1cm 4cm 0cm},clip,width=1.0\textwidth]{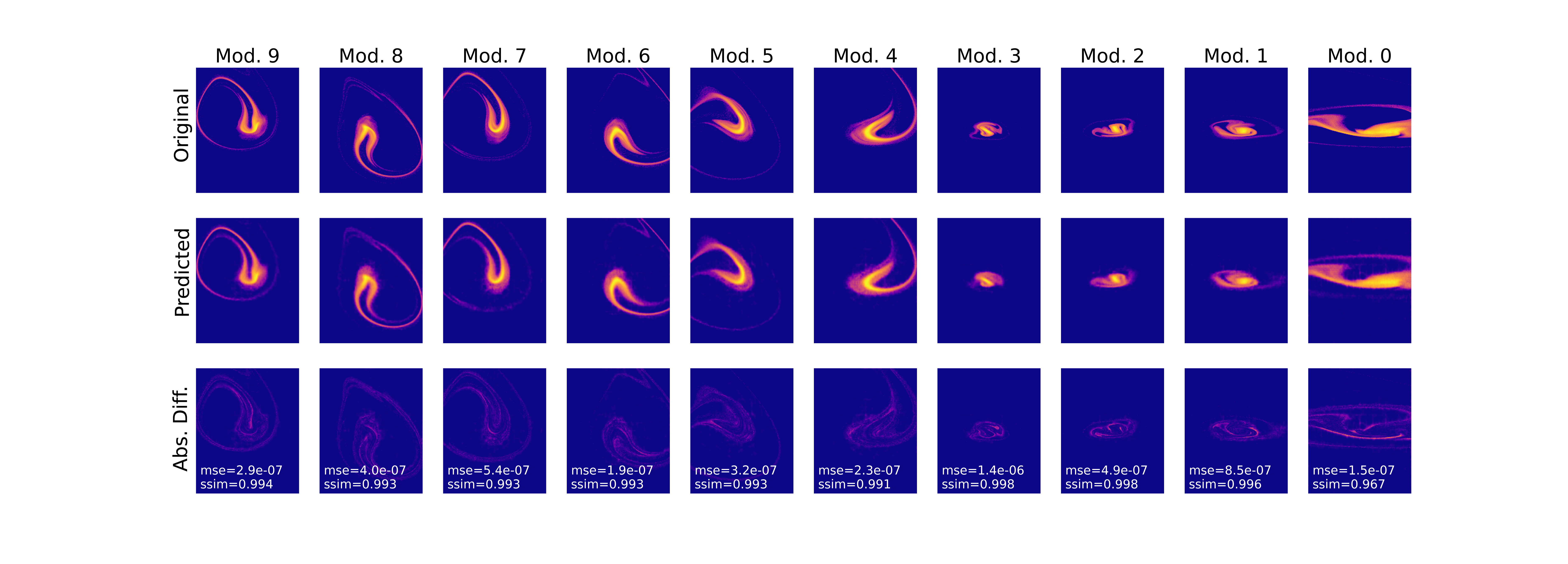}
        \subcaption{$E-\phi$ projection}
    \end{minipage}
    \begin{minipage}[b]{1.0\linewidth}
        \centering
        \includegraphics[trim={4cm 1cm 4cm 0cm},clip,width=1.0\textwidth]{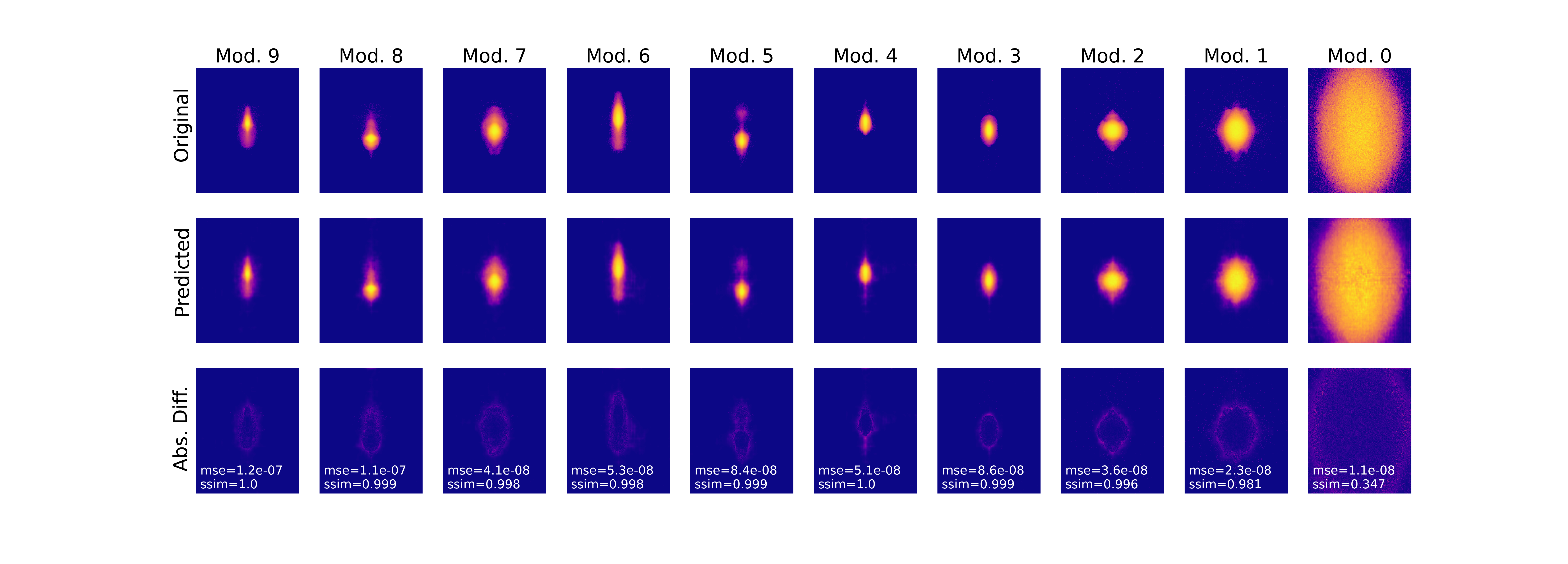}
        \subcaption{$x'-y'$ projection}
    \end{minipage}
    \caption{Predictions from the trained RLE at different upstream modules given module 10 as the input.}
    \label{fig:prediction_10}
\end{figure*}

\begin{figure*}[htbp]
    \centering
    \begin{minipage}[b]{1.0\linewidth}
        \centering
        \includegraphics[trim={4cm 1cm 4cm 0cm},clip, width=1.0\textwidth]{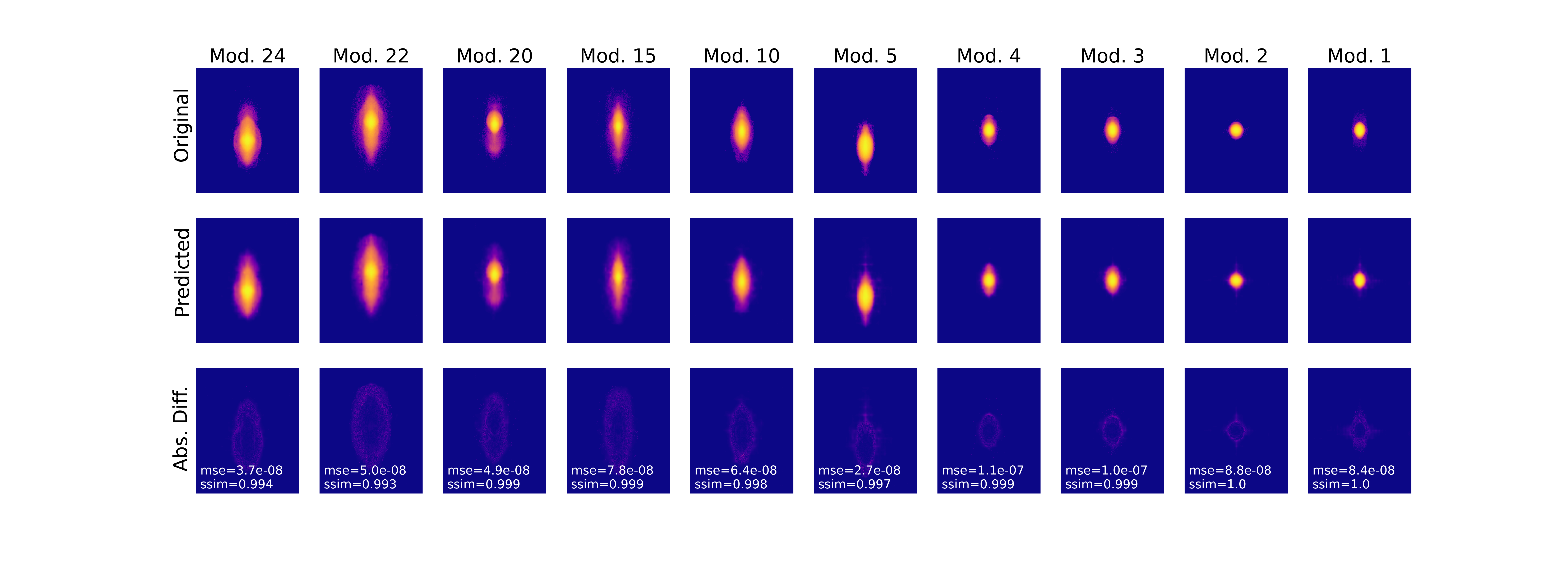}
        \subcaption{$x-y$ projection}
    \end{minipage}
    \begin{minipage}[b]{1.0\linewidth}
        \centering
        \includegraphics[trim={4cm 1cm 4cm 0cm},clip,width=1.0\textwidth]{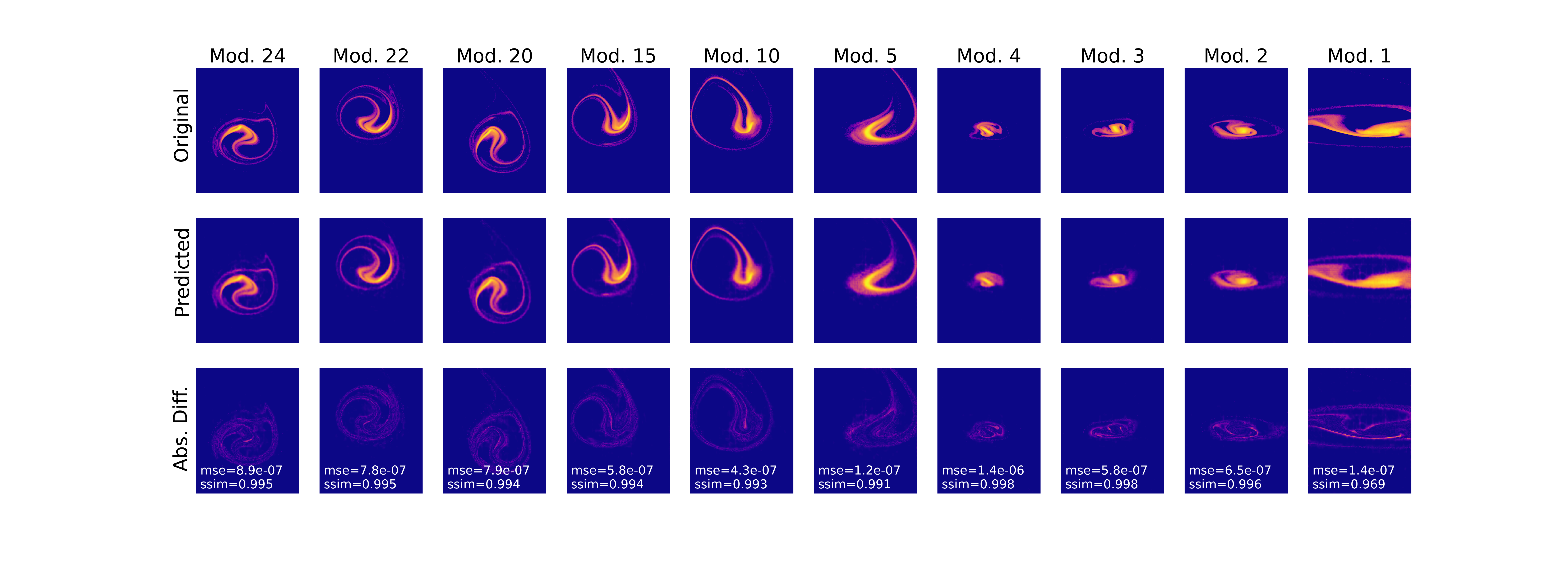}
        \subcaption{$E-\phi$ projection}
    \end{minipage}
    \begin{minipage}[b]{1.0\linewidth}
        \centering
        \includegraphics[trim={4cm 1cm 4cm 0cm},clip,width=1.0\textwidth]{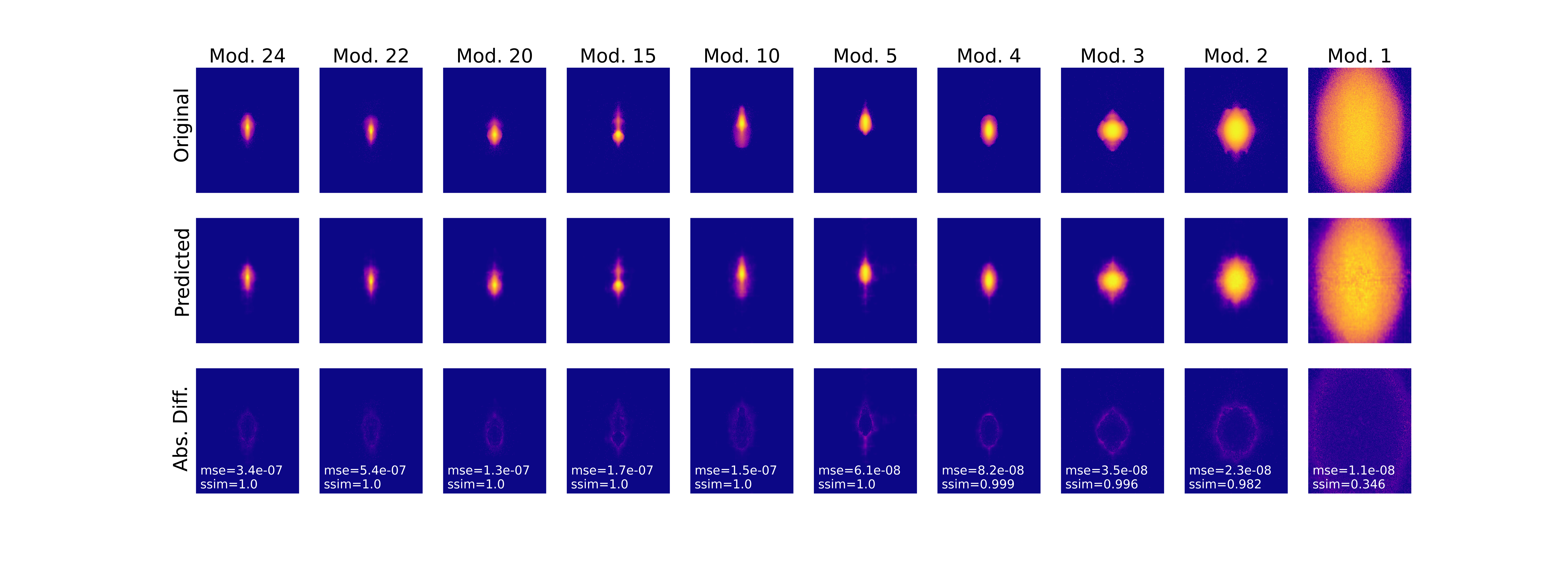}
        \subcaption{$x'-y'$ projection}
    \end{minipage}
    \caption{Predictions from the trained RLE at different upstream modules given module 25 as the input.}
    \label{fig:prediction_25}
\end{figure*}
 
\end{document}